\documentclass{aa}

\usepackage{color}
\usepackage[table,dvipsnames]{xcolor}

\usepackage{epic,eepic}
\usepackage{arydshln}
\usepackage{amstext}
\usepackage{amsmath,bbold,mathrsfs,bm,mathtools}
\usepackage{graphicx}
\usepackage{tabularx}
\usepackage{textcomp}
\usepackage[varg]{txfonts}
\usepackage{bm}
\usepackage{time}

\usepackage[labelformat=simple]{subcaption}
\renewcommand\thesubfigure{(\alph{subfigure})}

\usepackage{url}
\usepackage{upgreek}

\usepackage{natbib,twoopt}

\bibpunct{(}{)}{;}{a}{}{,} %
\makeatletter
\newcommandtwoopt{\citeads}[3][][]{\href{http://adsabs.harvard.edu/abs/#3}%
{\def\hyper@linkstart##1##2{}%
\let\hyper@linkend\@empty\citealp[#1][#2]{#3}}}
\newcommandtwoopt{\citepads}[3][][]{\href{http://adsabs.harvard.edu/abs/#3}%
{\def\hyper@linkstart##1##2{}%
\let\hyper@linkend\@empty\citep[#1][#2]{#3}}}
\newcommandtwoopt{\citetads}[3][][]{\href{http://adsabs.harvard.edu/abs/#3}%
{\def\hyper@linkstart##1##2{}%
\let\hyper@linkend\@empty\citet[#1][#2]{#3}}}
\newcommandtwoopt{\citeyearads}[3][][]%
{\href{http://adsabs.harvard.edu/abs/#3}
{\def\hyper@linkstart##1##2{}%
\let\hyper@linkend\@empty\citeyear[#1][#2]{#3}}}
\newcommandtwoopt{\citeauthorads}[3][][]%
{\href{http://adsabs.harvard.edu/abs/#3}
{\def\hyper@linkstart##1##2{}%
\let\hyper@linkend\@empty\citeauthor[#1][#2]{#3}}}
\makeatother

\usepackage{epic}
\usepackage{fancyvrb}
\usepackage{rotating}
\usepackage{float,afterpage}

\usepackage{multirow}
\usepackage{array}
\usepackage{balance}

\usepackage{overpic}

\usepackage{xifthen}

\usepackage{soulutf8}

\makeatletter
\def\url@leostyle{%
\@ifundefined{selectfont}{\def\UrlFont{\sf}}{\def\UrlFont{\tiny\ttfamily}}}
\makeatother
\title{SSTRED: Data- and metadata-processing pipeline\\ for CHROMIS and CRISP}

\author{Mats~G.~L{\"o}fdahl\inst{1}  \and
Tomas~Hillberg\inst{1} \and
Jaime~de~la~Cruz~Rodr{\'i}guez\inst{1}  \and
Gregal~Vissers\inst{1} \and
Oleksii Andriienko\inst{1} \and
G{\"o}ran~B.~Scharmer\inst{1} \and
Stein~V.~H.~Haugan\inst{2} \and
Terje~Fredvik\inst{2}}
\authorrunning{M.G.~L{\"o}fdahl et al.}

\institute{Institute for Solar Physics, Dept. of Astronomy, Stockholm
University, Albanova University Center, 106\,91 Stockholm, Sweden
\and
Institute of Theoretical Astrophysics, University of Oslo, Postboks
1029, Blindern, 0315 Oslo, Norway
}

\date{Compiled \now\ on \today.}

\abstract
{Data from ground-based, high-resolution solar telescopes can only be
used for science with calibrations and processing, which requires
detailed knowledge about the instrumentation. Space-based solar
telescopes provide science-ready data, which are easier to work with for
researchers whose expertise is in the interpretation of data.
Recently, data-processing pipelines for ground-based instruments
have been constructed.}
{We aim to provide observers with a user-friendly data pipeline for
data from the Swedish 1-meter Solar Telescope (SST) that delivers
science-ready data together with the metadata needed for proper
interpretation and archiving.}
{We briefly describe the CHROMospheric Imaging Spectrometer (CHROMIS)
instrument, including its (pre)filters, as well as recent
upgrades to the CRisp Imaging SpectroPolarimeter (CRISP) prefilters
and polarization optics. We summarize the processing steps from raw
data to science-ready data cubes in FITS files. We report calibrations and compensations for data imperfections in detail.
Misalignment of \ion{Ca}{ii} data due to wavelength-dependent
dispersion is identified, characterized, and compensated for. We
describe intensity calibrations that remove or reduce the effects of
filter transmission profiles as well as solar elevation changes.
We present REDUX, a new version of the MOMFBD image restoration code,
with multiple enhancements and new features. It uses projective
transforms for the registration of multiple detectors. We describe
how image restoration is used with CRISP and CHROMIS data.
The science-ready output is delivered in FITS files, with metadata
compliant with the SOLARNET recommendations. Data cube coordinates
are specified within the World Coordinate System (WCS). Cavity
errors are specified as distortions of the WCS wavelength coordinate
with an extension of existing WCS notation. We establish notation
for specifying the reference system for Stokes vectors with
reference to WCS coordinate directions.
The CRIsp SPectral EXplorer (CRISPEX) data-cube browser has been
extended to accept SSTRED output and to take advantage of the SOLARNET
metadata.}
{SSTRED is a mature data-processing pipeline for imaging instruments,
developed and used for the SST/CHROMIS imaging spectrometer and the
SST/CRISP spectropolarimeter. SSTRED delivers well-characterized,
science-ready, archival-quality FITS files with well-defined
metadata. The SSTRED code, as well as REDUX and CRISPEX, is
freely available through git repositories.}
{}


\usepackage[breaklinks=true]{hyperref} %

\begin{document}

\maketitle

\section{Introduction}
\label{sec:introduction}

Scientists
working with data from ground-based high-resolution solar
telescopes were for many years required to have detailed knowledge
about the telescopes and instruments with which their data were
collected. With the complexity of instruments and observing sequences
developed during the past decades, this has become increasingly
difficult. Only the home institutes of the instruments and a few
other, strong groups were able to maintain the necessary knowledge.

Meanwhile, space-based solar telescopes have come with data pipelines
that deliver well-characterized data, along with metadata that
facilitate their interpretation. Such data have then been made
available to researchers around the world through web-based virtual
observatories with searchable databases, for instance, Hinode
\citepads{2007SoPh..243....3K}, the Interface Region Imaging
Spectrograph (IRIS; \citeads{2014SoPh..289.2733D}), and the Solar
Dynamics Observatory (SDO; \citeads{2016SoPh..291.1887C}). This has
significantly increased the scientific use of these data.

In recent years, data pipelines for the main ground-based
telescopes have also been made available to observers, making the
production of science-ready data a matter of some training and
adequate computer resources. This is a crucial development to realize the scientific potential for solar
data from ground-based telescopes. Some of these pipelines are the IBIS Software
Package \citep{criscouli14IBIS} for the Interferometric BIdimensional
Spectropolarimeter (IBIS); CRISPRED (\citeads{2015A&A...573A..40D},
hereinafter referred to as the CRISPRED paper) for the CRisp Imaging
SPectropolarimeter (CRISP);
the ROSA data reduction pipeline \citep{jess17rosapipeline} for the
Rapid Oscillations in the Solar Atmosphere instrument (ROSA); and sTools
\citepads{2017IAUS..327...20K} for the GREGOR Fabry-P{\'e}rot
Interferometer (GFPI)
and High-resolution Fast Imager (HiFI).

Working within the 2013--2017 EU FP7 SOLARNET project
\citepads{2017psio.confE...1C}, \citet{haugan15metadata} set out to
define the metadata needed for the inclusion of ground-based solar
data in future Solar Virtual Observatories (SVOs). This adds another
set of requirements for the documentation, based on the expectation
that data will be used without the observers being involved.
Within the second SOLARNET project (2019--2022, EU Horizon 2020),
\citetads{2020arXiv201112139H}  further developed these recommendations
as they were confronted with implementations in pipelines,
primarily SSTRED and the pipeline for the Spectral Imaging of the
Coronal Environment instrument (SPICE; \citeads{2020A&A...642A..14S}).

With the August 2016 commissioning of the CHROMospheric Imaging
Spectrometer (CHROMIS; \citeads{2017psio.confE..85S}) at the Swedish 1-meter
Solar Telescope (SST), a data-processing pipeline was needed. Like
CRISP, CHROMIS is based on dual Fabry--P\'erot interferometers (FPIs)
in a telecentric mount and is also similar in the design in other
aspects. It was evident that the CRISPRED code would be an excellent
starting point, but much of the code had to be generalized to remove
CRISP-specific assumptions about cameras, filters, data file formats,
etc.

Inclusion of metadata according to the then recently formulated
SOLARNET recommendations in CRISPRED required a more thorough rewrite.
We decided to work in a new fork, dubbed CHROMISRED, so that observers
reducing CRISP data would not be disturbed by the ongoing
developments. When the new code base was fully operational for
CHROMIS, we re-implemented full support for CRISP data in it.

This paper describes SSTRED: the new, combined pipeline for the two
imaging spectro(polari)meters CRISP and CHROMIS.
With it, we aim to provide the complex ``machinery'' needed to process
multi-instrument, multi-wavelength data from a ground-based solar
telescope, as well as compensate for the known imperfections in these
data and prepare for their use together with space-based data.

SSTRED is being developed in parallel to the development of SOLARNET
recommendations for metadata, and includes multiple calibration and
correction steps beyond the standard pixel bias and gain, alignment
and destretching, and image restoration. SSTRED is therefore well
placed to serve as a model for pipelines of future instruments.

SSTRED was under development 2016--2021, and features were added as
late as in March 2021. A draft version of this paper has been
available since April~2018 %
\citep{lofdahl_data-processing_v1},
with a second version uploaded in July~2019.
We describe the state of SSTRED in June 2021.\footnote{See
Appendix~\ref{sec:pipeline-code} for information about access and
versions of the pipeline code.}

Combining data from multiple instruments with access to different
wavelength bands greatly enhances the scientific analysis. Examples of
this science are published by \citetads{2018A&A...612A..28L}, %
\citetads{2019ApJ...870...88E}, %
\citetads{2019ApJ...876...47B}, %
\citetads{2019ApJ...874..126K}, %
\citetads{2020A&A...637A...1K}, %
\citetads{2020A&A...644A..43P}, %
\citetads{2020A&A...641L...5J}, %
\citetads{2021A&A...648A..54R}, and \citetads{2021A&A...647A.147B}.
This is evidence of the utility of SSTRED-processed data.
\citetads{2017ApJ...851L...6R} and \citetads{2019A&A...627A.101V}
presented data that tapped into an even more complex but also more
powerful application of SSTRED-processed data, which includes the
analysis of CRISP and CHROMIS data together with space-based data from
IRIS. More of the latter can be expected as
\citetads{2020A&A...641A.146R} have released an archive of CRISP and
CHROMIS data that are registered with IRIS data.

The organization of this paper is as follows. We first describe the
instrumentation with relevant details of the CHROMIS and CRISP
instruments (Sect.~\ref{sec:instrumentation}) and give an overview of
the data-processing steps performed by SSTRED
(Sect.~\ref{sec:steps-reduct-chrom}) with references to the literature
(in particular the CRISPRED paper, which is recommended reading for
the interested reader) when possible. We then describe in detail
aspects of the processing that are novel to SSTRED (calibration steps
in Sect.~\ref{sec:calibrations} and image restoration in
Sect.~\ref{sec:image-restoration}). The SSTRED output is rich in
metadata that are stored in accordance with the SOLARNET
recommendations, as described in Sect.~\ref{sec:metadata}. We present
updates to the auxiliary CRIsp SPectral EXplorer (CRISPEX) data
browser, relevant to exploring and analyzing the SSTRED science-ready
output, in Sect.~\ref{sec:crispex}. We conclude with a discussion in
Sect.~\ref{sec:discussion}. Some details about the implementation are
found in the appendices.

\section{Instrumentation}
\label{sec:instrumentation}

\begin{figure}[ht]
\sidecaption
\begin{overpic}[viewport=135 109 549 671,clip,width=\linewidth]{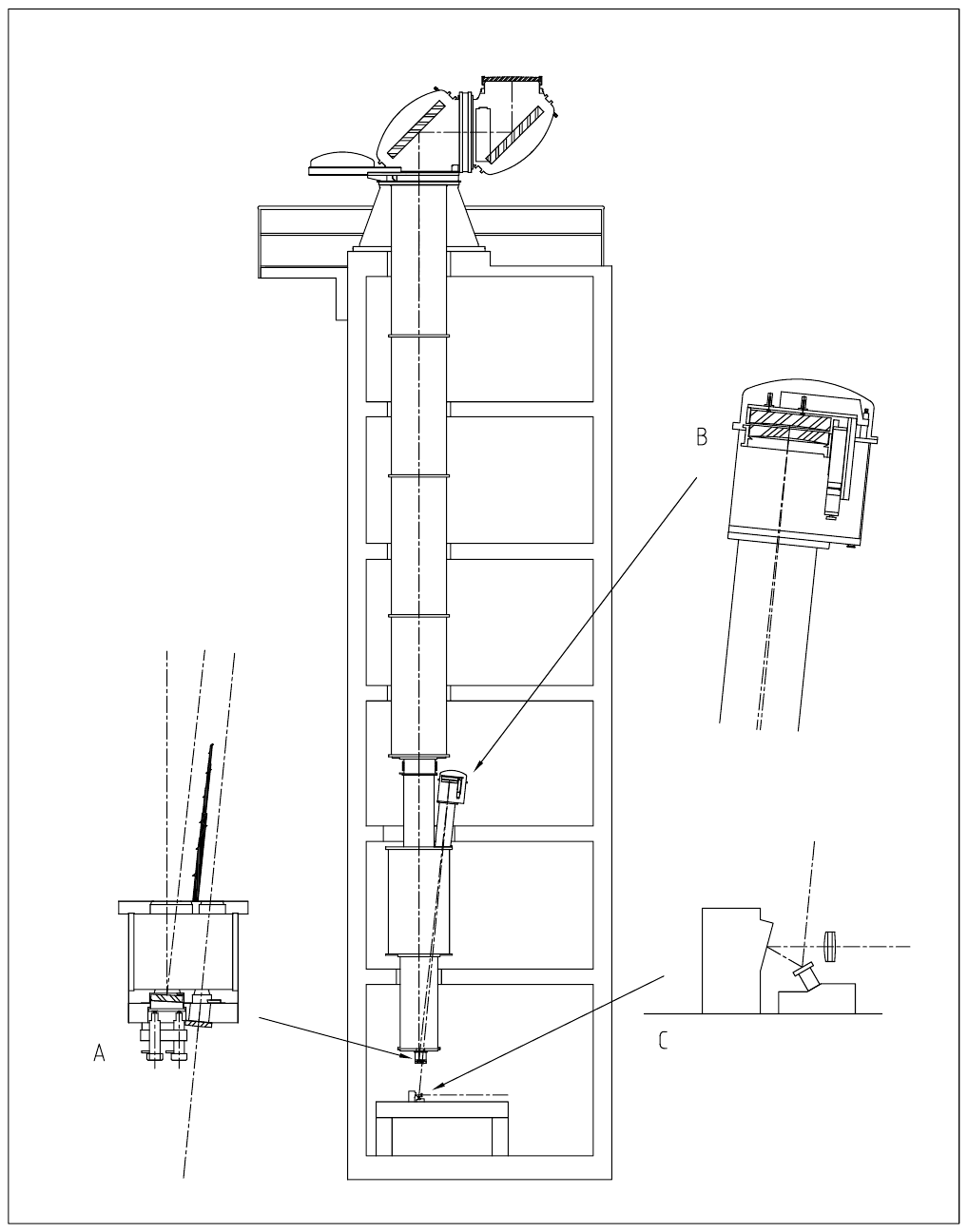}
\put(55.9,21){\tiny{DM}}
\put(64.4,15.7){\tiny{TM}}
\put(66,23.5){\tiny{RL}}
\end{overpic}
\caption{Sketch of the SST (from \citeads{2003SPIE.4853..341S}), from
the 1~m lens, via the two alt-az mirrors, the field mirror on the
bottom plate (inset~\textsf{A}), the Schupmann corrector
(\textsf{B}), the field lens and exit window (\textsf{A}), to the
tip-tilt mirror (TM), deformable mirror (DM), and re-imaging lens
(RL) on the optical table (\textsf{C}). The optical path continues
in Fig.~\ref{fig:setup}.}
\label{fig:sst}
\end{figure}

\begin{figure*}[!tp]
\sidecaption
\includegraphics[viewport=29 459 379 725,clip,width=12cm]{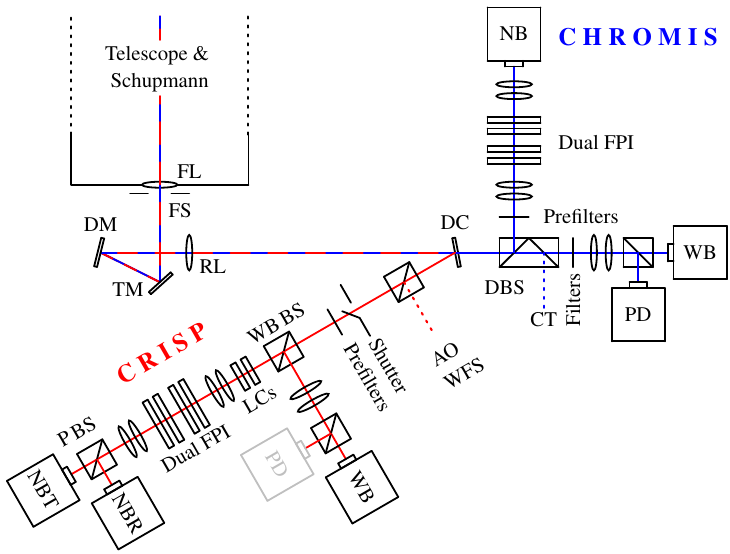}
\caption{Schematics of the setup used from the CHROMIS installation
in August 2016 through the 2020 season. Acronyms: FL = field lens,
FS = field stop, TM = tip-tilt mirror; DM = deformable mirror; RL
= reimaging lens; DC = 500~nm dichroic beamsplitter; DBS = double
beamsplitter, CT = correlation tracker; AO\,WFS = adaptive optics
wavefront sensor; WB\,BS = wide-band beam splitter; FPI =
Fabry--P\'erot interferometer, LCs = liquid crystal modulators;
P\,BS = polarizing beamsplitter, NB = narrowband, WB = wideband,
NBT = NB transmitted, NBR = NB reflected, PD = phase diversity.
White light enters the telescope from the top. It is split into a
red and a blue beam at $\sim$500~nm by the DC. Distances and
angles do not correspond to the physical setup.}
\label{fig:setup}
\end{figure*}

The optics of the SST \citepads{2003SPIE.4853..341S} and its imaging setup are
illustrated in Figs.~\ref{fig:sst} and~\ref{fig:setup}, respectively.
In the caption of Fig. 2, acronyms are defined for many
optical elements that are discussed in the following two subsections,
with details about the CRISP and CHROMIS instruments.

\subsection{CRISP}
\label{sec:crisp}

CRISP operates in light with $\lambda>500$~nm, reflected by the DC
into a red beam. Details about CRISP are given by
\citetads{2008ApJ...689L..69S} and (with an emphasis on data
processing) in the CRISPRED paper. Here follows a brief update.

Before the 2015 season, new ferroelectric LCs were installed as a
polarimetric modulator, replacing the old nematic LCs
\citepads{2021AJ....161...89D}. Because they occupy more space on the
optical table, they were installed directly after the WB\,BS rather
than in the old position before the PBS.

\begin{figure*}[!t]
\begin{subfigure}{0.32\linewidth}
\includegraphics[viewport=43 23 698 529,clip,width=\linewidth]{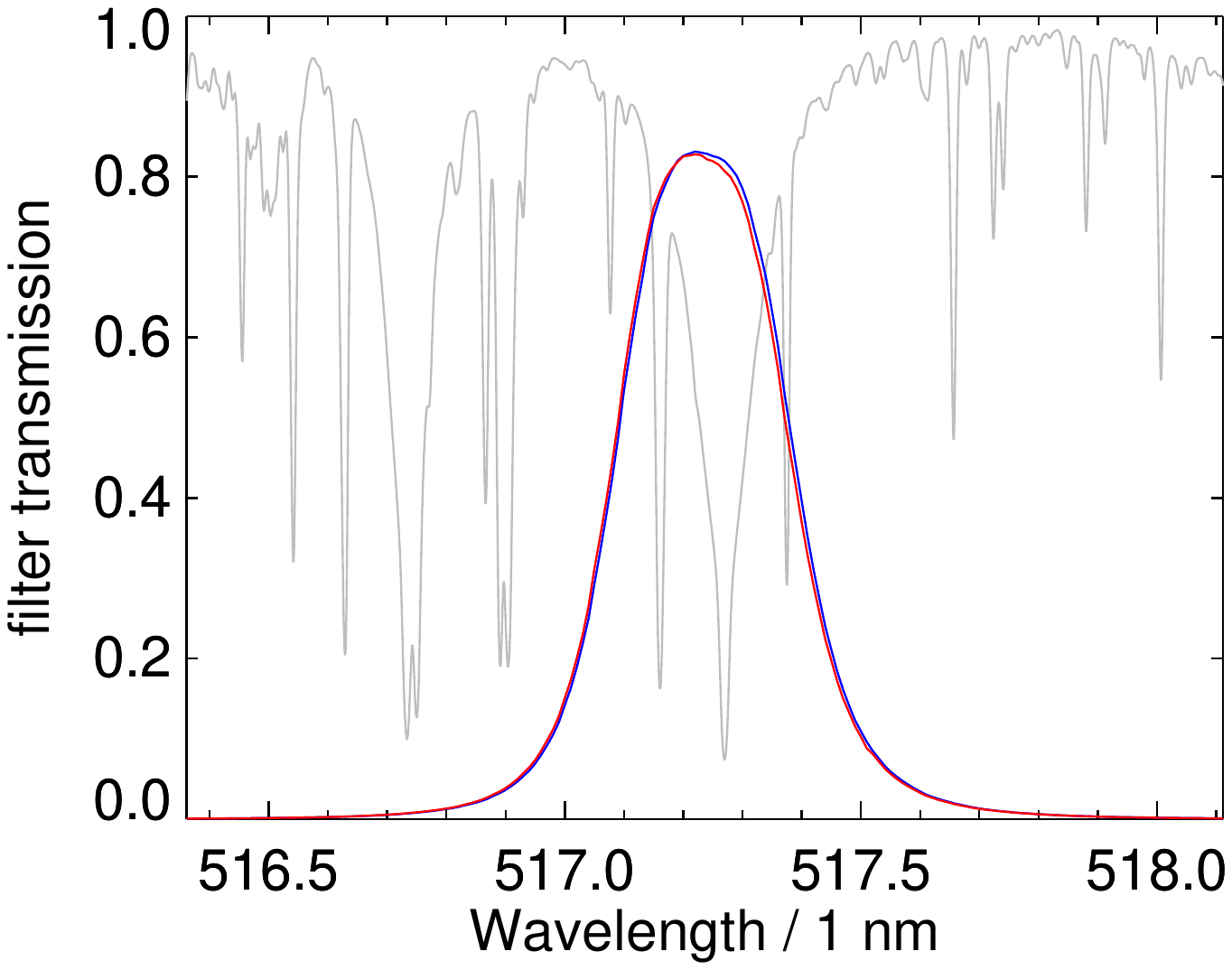}
\caption{$517.2$ nm filter profiles}
\label{fig:CRISP-profiles+spectra_5173}
\end{subfigure}
\hfill
\begin{subfigure}{0.32\linewidth}
\includegraphics[viewport=43 23 698 529,clip,width=\linewidth]{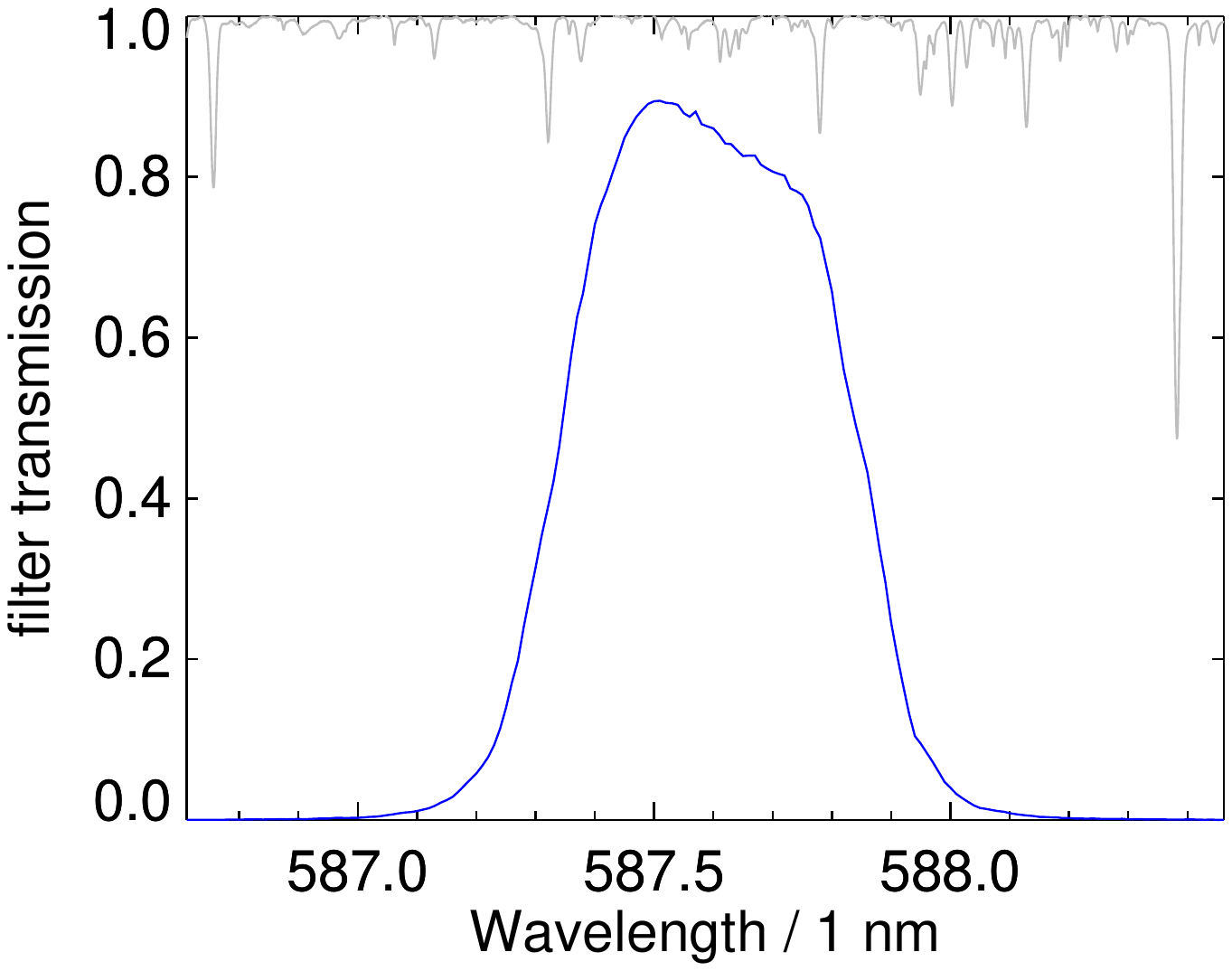}
\caption{$587.6$ nm filter profile}
\label{fig:CRISP-profiles+spectra_5876}
\end{subfigure}
\hfill
\begin{subfigure}{0.32\linewidth}
\includegraphics[viewport=43 23 698 529,clip,width=\linewidth]{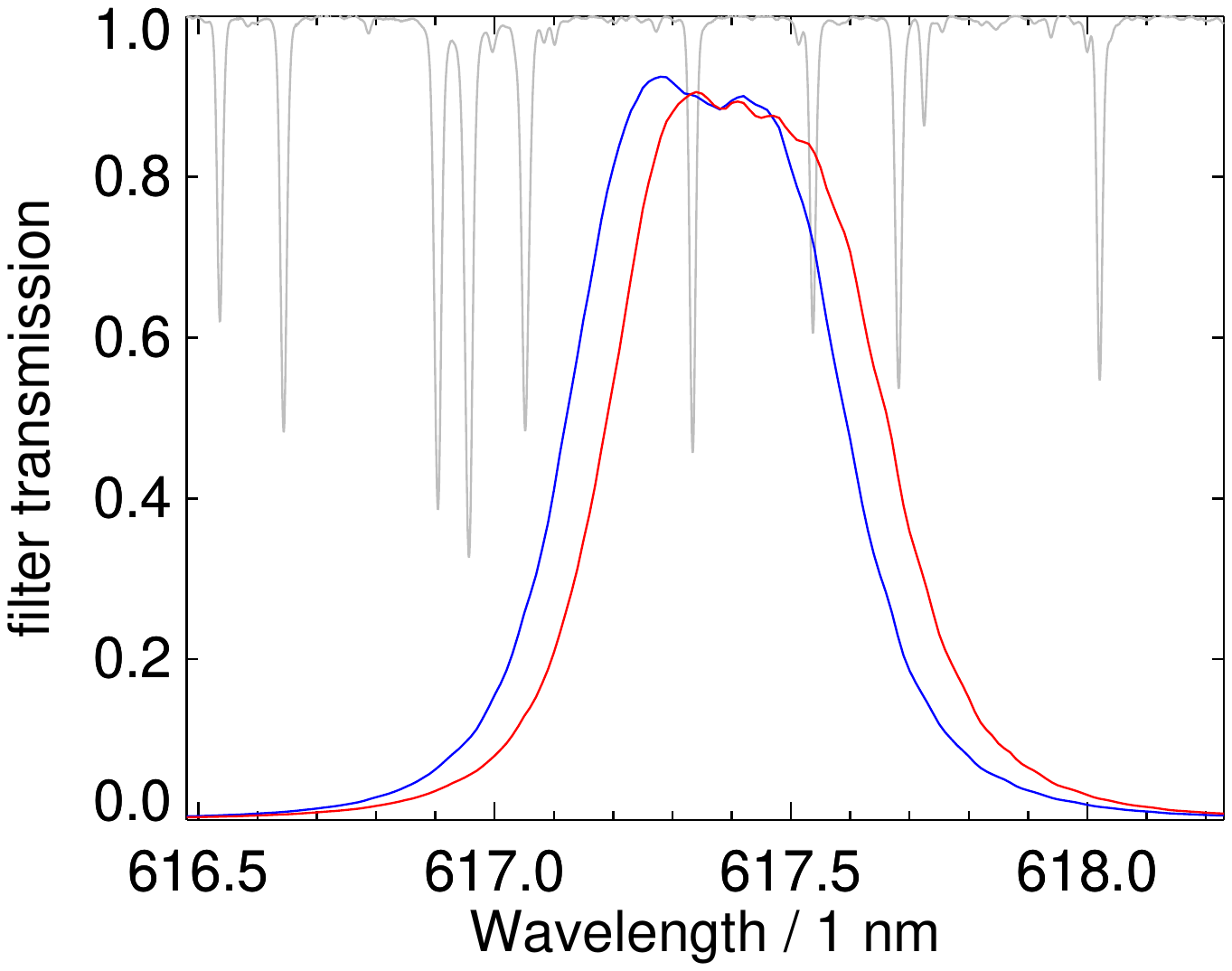}
\caption{$617.4$ nm filter profiles}
\label{fig:CRISP-profiles+spectra_6173}
\end{subfigure}\\[4mm]
\begin{subfigure}{0.32\linewidth}
\includegraphics[viewport=43 23 698 529,clip,width=\linewidth]{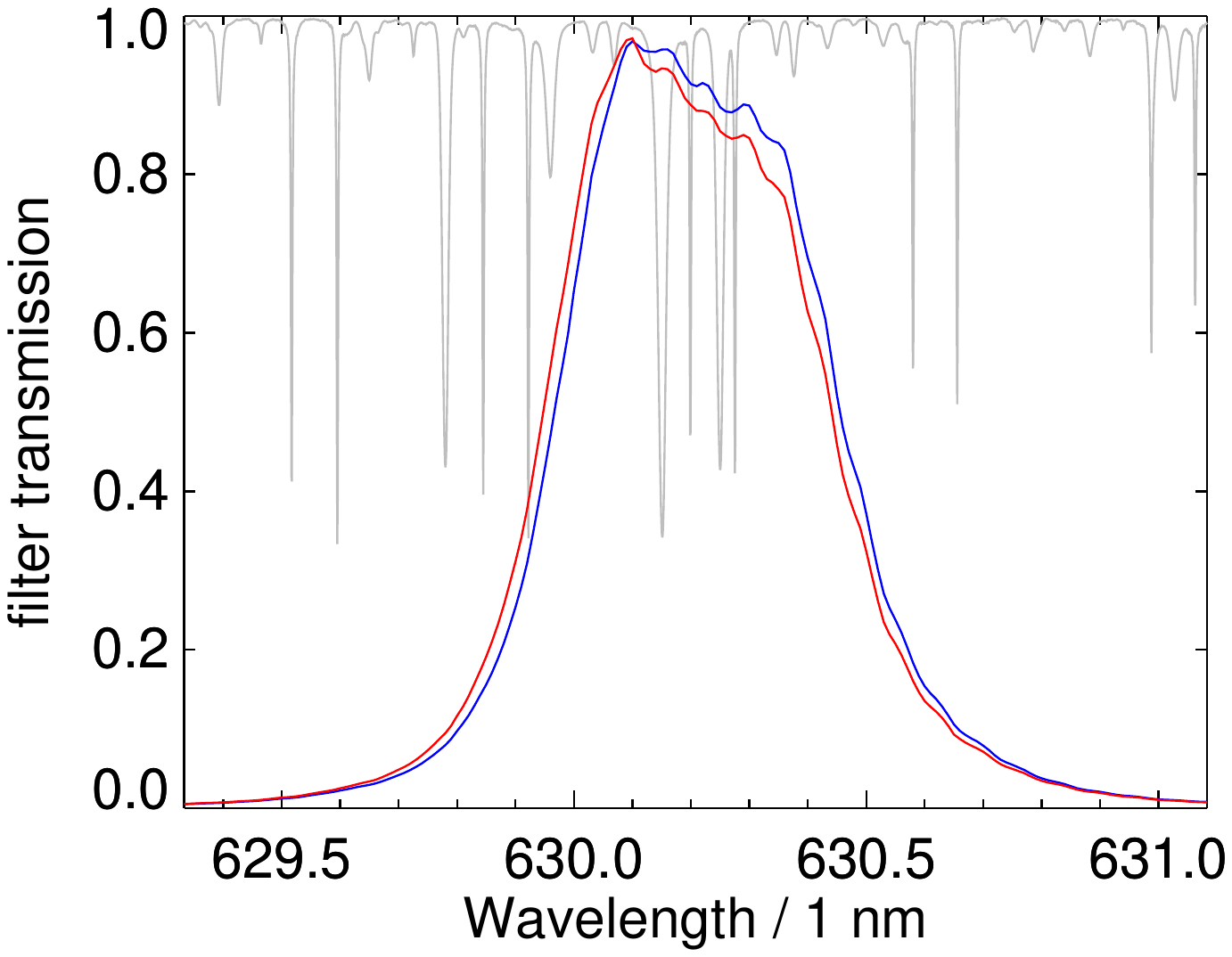}
\caption{$630.2$ nm filter profiles}
\label{fig:CRISP-profiles+spectra_6302}
\end{subfigure}
\hfill
\begin{subfigure}{0.32\linewidth}
\includegraphics[viewport=43 23 698 529,clip,width=\linewidth]{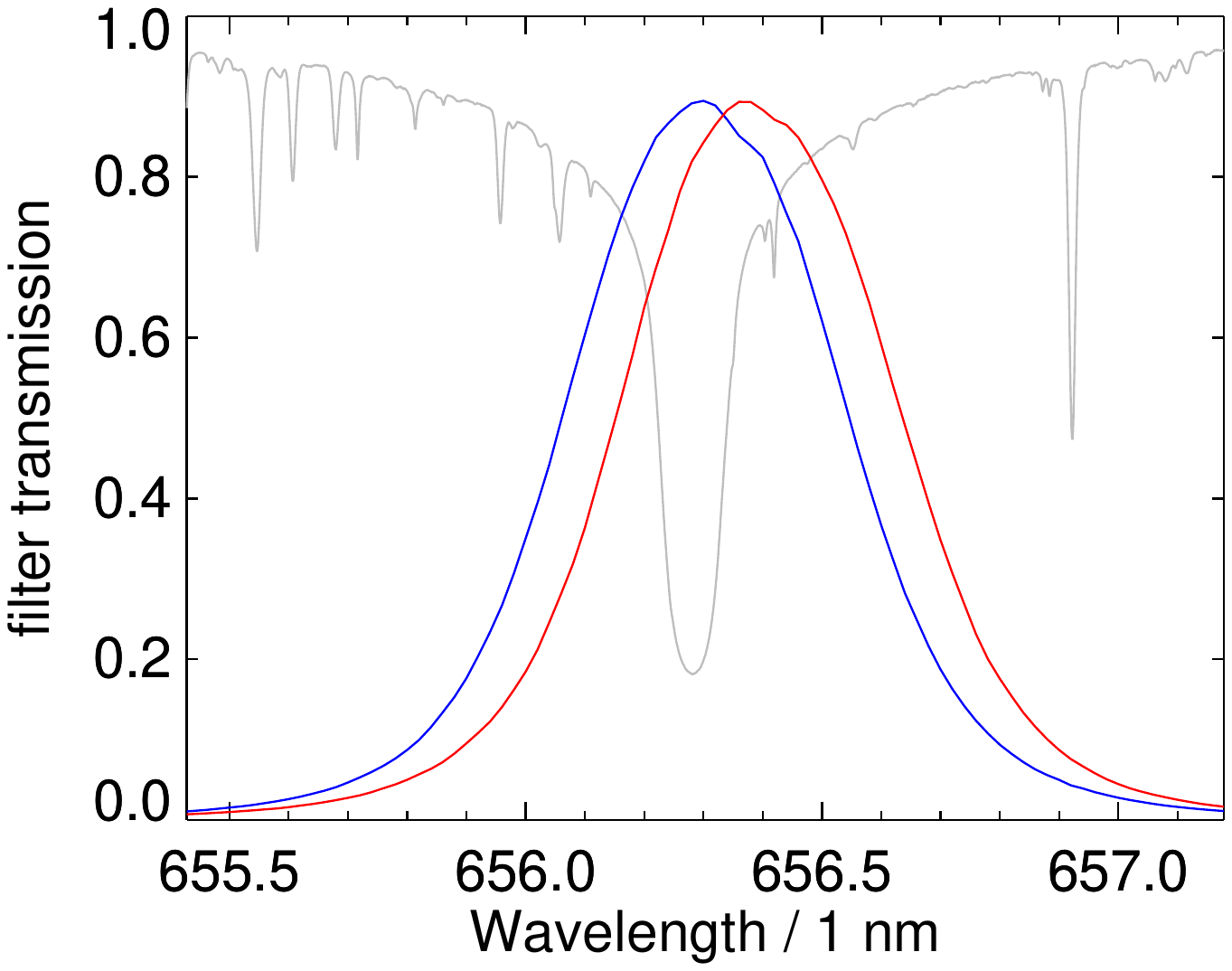}
\caption{$656.3$ nm filter profiles}
\label{fig:CRISP-profiles+spectra_6563}
\end{subfigure}
\hfill
\begin{subfigure}{0.32\linewidth}
\includegraphics[viewport=43 23 698 529,clip,width=\linewidth]{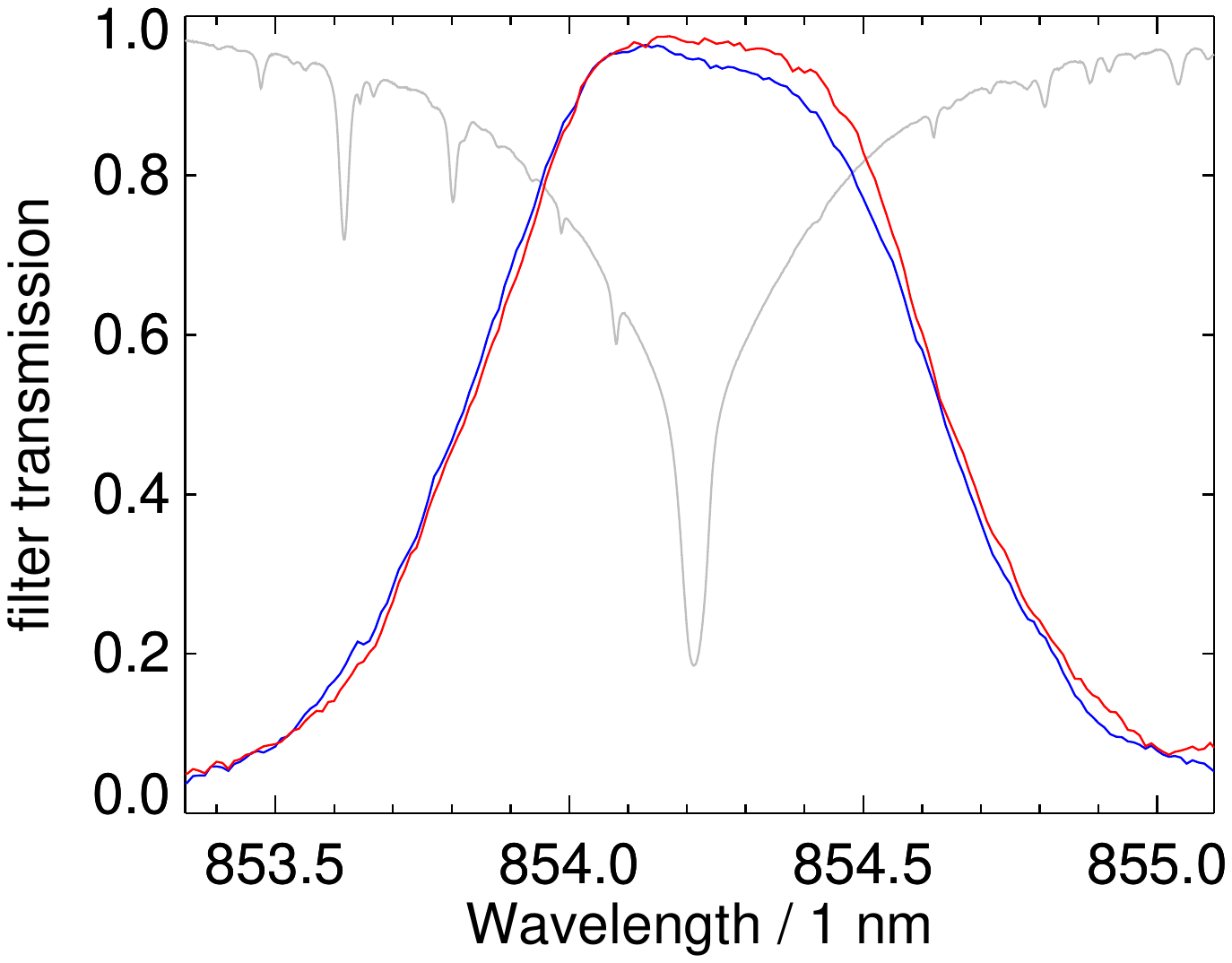}
\caption{$854.3$ nm filter profiles}
\label{fig:CRISP-profiles+spectra_8542}
\end{subfigure}
\caption{CRISP prefilter transmission profiles as measured by the
manufacturer for the filters installed before the 2018 season. The
blue curves correspond to the smaller part number
and (for the wavelength bands where we have spares) red curves the
larger. The gray lines represent the average disk center atlas
spectrum, normalized to the continuum. See also
Table~\ref{tab:CRISP-pref}.}
\label{fig:CRISP-profiles+spectra}
\end{figure*}

We knew since 2013 that the original CRISP prefilters manufactured by
Barr have significant optical power, which causes the focus to vary
between the filters. This was compensated for by use of a variable
focus on the DM, with the unfortunate consequence of leading to focus
errors in the blue beam instead. In practice, this limited most CRISP
observations to using prefilters that happened to have similar optical
power or to not allow simultaneous blue data in all CRISP
observations. New CRISP prefilters without optical power, made by
Alluxa, were installed before the 2018 season. (The 587.66~nm He~D3
filter was installed already in 2016, but it belongs to the new batch
of Alluxa filters with no optical power.) Transmission profiles are
shown in Fig.~\ref{fig:CRISP-profiles+spectra}. See also
Table~\ref{tab:CRISP-pref}, where additional information is given.
Note that there have been problems with fitting the telescope model to
calibration data for the 517~nm filter. Until proper calibrations can
be made, we are using the model for the old 525~nm filter (see the
CRISPRED paper).

\begin{table*}[!t]
\caption{CRISP prefilters installed 2018.}              %
\label{tab:CRISP-pref}      %
\centering                                      %
\begin{tabular}{l@{\quad}c cc c cc c cc c cc c c @{\quad} c}          %
\hline\hline                        %
\noalign{\smallskip}
\multicolumn{2}{c}{Main diagnostic [nm]}%
&&\multicolumn{2}{c}{Part \#}%
&&\multicolumn{2}{c}{CWL [nm]}%
&&\multicolumn{2}{c}{FWHM [nm]}%
&&\multicolumn{2}{c}{$T_\text{peak}$}%
&&PC\\
\cline{1-2}
\cline{4-5}
\cline{7-8}
\cline{10-11}
\cline{13-14}
\noalign{\smallskip}
\hline                                   %
\noalign{\smallskip}
\ion{Mg}{i}  & 517.2 && 1 &(7)&& 517.24 &(517.23)&& 0.32 &(0.32)&& 0.83 &(0.83)&&            \\
He~D3     & 587.6 && 9 &   && 587.66 &        && 0.54 &      && 0.90 &      && \checkmark \\
\ion{Fe}{i}  & 617.3 &&(6)& 9 &&(617.36)& 617.43 &&(0.49)& 0.50 &&(0.92)& 0.91 && \checkmark \\
\ion{Fe}{i}  & 630.2 &&(9)&10 &&(630.21)& 630.19 &&(0.50)& 0.50 &&(0.97)& 0.97 && \checkmark \\
H-$\alpha$& 656.3 &&(1)&10 &&(656.30)& 656.40 &&(0.52)& 0.53 &&(0.89)& 0.89 &&            \\
\ion{Ca}{ii} & 854.2 &&(7)&10 &&(854.22)& 854.23 &&(0.84)& 0.83 &&(0.96)& 0.97 && \checkmark \\
\noalign{\smallskip}
\hline                                             %
\end{tabular}
\tablefoot{The CWL (center wavelength at normal incidence),
$T_\text{peak}$ (peak
transmission), and FWHM values are measured from the profiles
plotted in Fig.~\ref{fig:CRISP-profiles+spectra}. Part numbers
were assigned by the manufacturer. PC: Calibrated telescope
polarization model exists.  The
numbers in parentheses correspond to filters not currently
installed in the CRISP filter wheel. Compare
with Table A.1 in the CRISPRED paper.}

\end{table*}

The varying optical power caused a variation in image scale of up to
several percent between some of the old CRISP prefilters. The image
scale with the new filters agrees to within $\sim$0.1\%.

There are four Sarnoff CAM1M100 cameras for CRISP, but only three at a
time have been used regularly. The fourth has been used as a spare, but
is no longer operational.

\subsection{CHROMIS}
\label{sec:chromis}

CHROMIS (Scharmer et al. in prep.; \citeads{2006A&A...447.1111S}) is
based on a dual FPI mounted in a telecentric setup, similar to CRISP,
but designed for use at wavelengths in the range 380--500 nm and
currently without polarimetry. In particular, CHROMIS is optimized for
use in the \ion{Ca}{ii}~H and~K lines, which are formed in the upper
chromosphere, and for the H-$\upbeta$ line.

The blue light with $\lambda \la 500$~nm is transmitted through the DC
toward the DBS, which reflects most of the light to the narrow-band (NB)
beam and transmits the remainder to the wide-band (WB) beam (where it
is again split between the CT and the two WB cameras). The DBS was
designed to reflect 90\% to the NB path, but the current DBS does not
reflect more than $\sim$60\%. A new DBS with the proper splitting
would improve the signal-to-noise ratio (S/N)
in the NB and should be
installed before a future polarimetry upgrade of the instrument.

\begin{figure*}[t]
\begin{subfigure}{\linewidth}
\includegraphics[viewport=55 316 740 561,clip,scale=0.77]{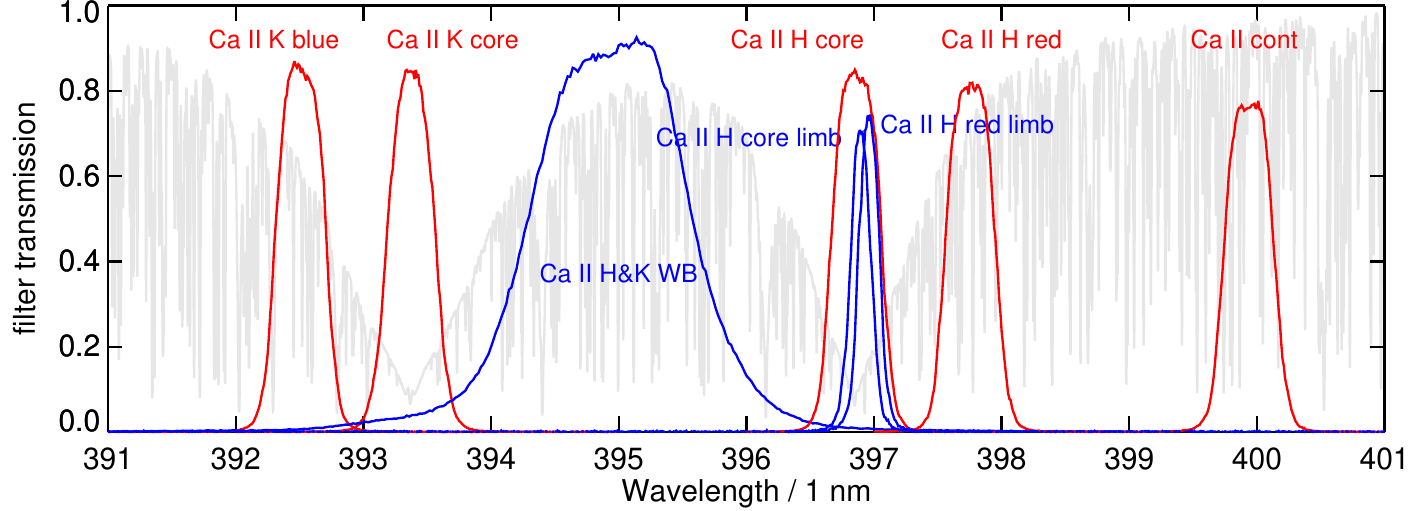}
\caption{\ion{Ca}{ii} filter profiles}
\label{fig:profiles+spectra_ca}
\end{subfigure}\\[4mm]
\begin{subfigure}[b]{12cm}
\includegraphics[viewport=55 52 425 300,clip,scale=0.77]{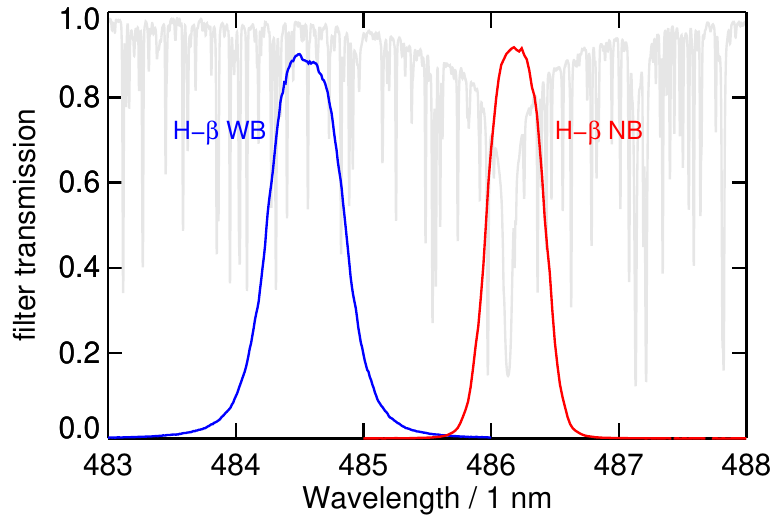}
\caption{H-$\upbeta$ filter profiles}
\label{fig:profiles+spectra_hb}
\end{subfigure}
\begin{minipage}[b]{6cm}
\caption{CHROMIS filter transmission profiles (normal incidence).
Red lines: NB prefilters. Blue lines: WB filters. The filter
profiles shown are measured in the center of the filter by the
manufacturer. The FPI profiles are not wide enough to plot them here.
The gray lines represent the average disk center atlas spectrum,
normalized to the continuum. See also
Table~\ref{tab:chromis-filters}. \bigskip}
\label{fig:profiles+spectra}
\end{minipage}

\end{figure*}

The NB path goes through one of a set of filters mounted in a filter
wheel. The filter characteristics are summarized in
Table~\ref{tab:chromis-filters} and Fig.~\ref{fig:profiles+spectra}.
These filters are all manufactured by Alluxa.

\begin{table*}[t]
\centering
\caption{CHROMIS filters.}
\begin{tabular}{llllcllll}
\hline\hline                        %
\noalign{\smallskip}
\multicolumn{4}{c}{Wide-band filters} && \multicolumn{4}{c}{Narrow-band prefilters}\\
\cline{1-4}
\cline{6-9}
\noalign{\smallskip}

Wavelength band & CWL [nm] & FWHM [nm] & $T_\text{peak}$  && Wavelength band & CWL [nm] & FWHM [nm] & $T_\text{peak}$ \\
\hline \noalign{\smallskip}
&&&&\multirow{5}{*}{$\left.\begin{tabular}{@{}l@{}}
\\  \\ \\
\end{tabular}\right\}
\left\{\begin{tabular}{@{}l@{}}
\\  \\ \\ \\ \\
\end{tabular}\right. $}&   \ion{Ca}{ii} K blue & 392.52& 0.41 & 0.86 \\
\ion{Ca}{ii} H\&K WB &395&1.32&0.9\phantom0&& \ion{Ca}{ii} K core & 393.41 & 0.41 & 0.8\phantom0 \\
\ion{Ca}{ii} H red limb WB &396.96&0.18&0.74&& \ion{Ca}{ii} H core & 396.88 & 0.41 & 0.8\phantom0 \\
\ion{Ca}{ii} H core limb WB &396.90&0.18&0.71&& \ion{Ca}{ii} H red & 397.74& 0.41 & 0.83 \\
&&&&& \ion{Ca}{ii} continuum & 399.04 & 0.42 & 0.76 \\
\noalign{\smallskip}
H-$\upbeta$ WB & 484.55 & 0.65 & 0.9\phantom0
&& H-$\upbeta$ NB & 486.22 & 0.48 & 0.92 \\
\hline
\end{tabular}
\tablefoot{CWL = center wavelength at normal incidence,
$T_\text{peak}$ = peak transmission as given by the manufacturer.
See also the measured transmission profiles in
Fig.~\ref{fig:profiles+spectra}. The curly brackets indicate that
any of the \ion{Ca}{ii} WB filters can be combined with any subset of
the \ion{Ca}{ii} NB prefilters.}
\label{tab:chromis-filters}
\end{table*}

With the present set of prefilters, CHROMIS can be used in two
wavelength regions. One region covers the \ion{Ca}{ii}~H and~K lines.
Scanning through the wide \ion{Ca}{ii} lines and nearby continuum at
399~nm is done through five separate three-cavity NB prefilters, while
simultaneous WB data, used for context and supporting image
restoration, are collected through a single WB filter with a
wavelength between the H \& K lines. The other region covers the
H-$\upbeta$ line, with WB data collected in the continuum to the blue
of the line.

The design FWHM of the FPI NB transmission profile is 8~pm at the
\ion{Ca}{ii}~H and~K lines and 10~pm at H-$\upbeta$. However, the
measured profile width in \ion{Ca}{ii} is estimated to be $\sim$13~pm
\citepads{2017ApJ...851L...6R}, which has been traced to a mismatch in
the etalon reflectivities. (The profile width depends on the
reflectivities in the high-resolution etalon, see Sect.~2.1 and
Fig.~A.1 in the CRISPRED paper.)

The WB re-imaging system provides a telecentric beam with the same
focal ratio as the one of the CHROMIS FPIs. It provides an anchor
object for image restoration, see Sect.~\ref{sec:image-restoration}.
The PD camera collects WB data approximately one~wave out of focus to
further facilitate image restoration with phase diversity.

The \ion{Ca}{ii}~H\&K WB filter has been in use from the start,
providing a photospheric image that works very well for on-disk image
restoration. However, it gives very little signal outside the limb
without saturation on the disk. \citetads{2019ApJ...874..126K} managed to
restore coronal loops, but it has been clear that a WB filter with
better visibility of the fine structure outside the limb would be an
advantage. The \ion{Ca}{ii}~H core WB filters
were installed in August 2020. Details about their use will be
provided in the forthcoming CHROMIS paper.

All three cameras are Grasshopper3 2.3 MP Mono USB3 Vision
(GS3-U3-23S6M-C) cameras, manufactured by PointGrey. They are equipped
with Sony Pregius IMX174 globally shuttered 1920$\times$1200-pixel
CMOS detectors with 5.86~\textmu{}m pitch. These cameras are
synchronized electronically so that there is no need for an external
shutter as in CRISP. The plate scale was $\sim$0\farcs038/pixel during
the 2016--2020 seasons, measured in pinhole array images using a
pinhole spacing of 5\farcs116 (see the CRISPRED paper, Sect.~3.3).

\section{Overview of SSTRED processing}
\label{sec:steps-reduct-chrom}

This section gives an overview of the processing steps for reducing
CRISP and CHROMIS data with SSTRED. The processing divides naturally
into steps performed before image restoration (with multi-object
multi-frame blind deconvolution, MOMFBD), the MOMFBD processing
itself, and post-MOMFBD steps. See Appendix~\ref{sec:pipeline-code}
for notes on the pipeline code.

\subsection{Pre-MOMFBD}
\label{sec:pre-momfbd}

The first step in processing CRISP and CHROMIS data for a particular
observing day is running a setup script that locates and analyzes the
directory tree with the day's observed data, identifying science data
as well as the various kinds of calibration data. A work directory
(one per instrument) is created, where a configuration file and an
IDL\footnote{Interactive Data Language, Harris Geospatial Solutions,
Inc.} script with the recommended processing steps are written.

Most calibration data are collected as bursts of images with the same
settings and exposure times as the science data. These bursts are then
summed, resulting in calibration data where noise and (when relevant)
imprints from granulation are reduced. This summing is performed for dark
frames, flat fields, pinhole array images, and polarimetric calibrations.
While summing, statistics are calculated. This allows checking for
outliers and removal of suspicious frames. Flats and pinholes for the
near-IR wavelength bands 7772 and 8542~\AA{} recorded with CRISP are
corrected for back-scatter. See Sect.~4.2 of the CRISPRED paper for
details.

Based on the summed polarimetric calibration data, modulation matrices
are then calculated for each CRISP pixel. See Sect.~4.1 of the
CRISPRED paper.

The geometrical transform needed to align images from different
cameras is estimated by a calibration step involving pinhole array
images. We describe a procedure for this in
Sect.~\ref{sec:camera-alignment}, which is improved with respect to the
procedure described in Sect.~3.2 of the CRISPRED paper.

The FPI cavity errors are measured from the summed flat fields. See
Sect.~\ref{sec:cavity-errors} below.

The intensities registered by the detectors are affected by the
transmission through the atmosphere, the telescope, other optics
including in particular the prefilters, as well as the quantum
efficiency (QE) of the detectors. These effects depend on both time
and wavelength. We measure the effects from calibration data as
described in Sects.~\ref{sec:fitprefilter} and
\ref{sec:temp-intens-calibr} below.

\subsection{MOMFBD}
\label{sec:momfbd}

The MOMFBD image restoration is made outside of IDL, see details in
Sect.~\ref{sec:image-restoration}. SSTRED uses a recent fork of the
MOMFBD code written by \citetads{2005SoPh..228..191V}, with several
essential new features. The new code, REDUX, is described in
Appendix~\ref{sec:redux-code}.

The MOMFBD processing includes image remapping based on pinhole
calibration. This step aligns images from different cameras, see
Sect.~\ref{sec:camera-alignment}.
Back-scatter correction for CRISP Sarnoff cameras in near IR
wavelengths is performed as described in Sect.~4.2 of the CRISPRED
paper.

\subsection{Post-MOMFBD }
\label{sec:post-momfbd}

The post-MOMFBD steps assemble the MOMFBD-restored images into
science-ready data cubes with metadata.
The cubes have dimensions
$[N_\text{x},N_\text{y},N_\text{tun},N_\text{pol},N_\text{scan}]$,
where $N_\text{x}$ and $N_\text{y}$ are the spatial dimensions,
$N_\text{tun}$ is the number of wavelength-tuning positions,
$N_\text{pol}$ is the number of polarization states (four for Stokes
data, one otherwise), and $N_\text{scan}$ is the number of scans through
the line. The physical coordinates are two spatial coordinates, the
wavelengths, the polarization states, and the temporal coordinates.
They are not strictly equivalent to the pixel coordinates. The spatial
coordinates, solar longitude and latitude, vary with time (due to the
solar rotation), and the time coordinate is advanced both while tuning
and from scan to scan. This is specified in the metadata using the
World Coordinate System (WCS), see Sect.~\ref{sec:wcs}.

The following steps are performed on the restored NB images to
produce a science-ready data cube:
\begin{itemize}
\item For polarimetric CRISP data: demodulation of four LC state
images into Stokes component images. See Sect.~4.1 in the CRISPRED
paper as well as \citetads{2010arXiv1010.4142S} and
\citetads{2008A&A...489..429V}.
\item Removal of periodic image artifacts in polarimetric data. See
Sect.~\ref{sec:periodic-artifacts}
\item For CHROMIS \ion{Ca}{ii} data: Correction for time-variable
misalignment due to chromatic dispersion in the atmosphere and in
the telescope. See Sect.~\ref{sec:time-vari-alignm}.
\item Correction for residual warping from anisoplanatic effects
between NB images with different tuning and/or polarimetric states.
See Sect.~\ref{sec:extra-wb-objects}.
\item Compensation for field rotation caused by the alt-az mount of
the telescope. See Sect.~\ref{sec:spatial-coordinates-1}.
\item For cubes with multiple scans, geometrical alignment and
destretching needed to make the transition from scan to scan smooth.
See Sect.~4.6 in the CRISPRED paper.
\item For polarimetric CRISP data: removal of polarimetric cross-talk.
See Sect.~\ref{sec:polar-cross-talk}.
\item Spectral intensity scaling to compensate for prefilter
transmission profiles and to obtain correct units by use of the
prefilter calibration described in Sect.~\ref{sec:fitprefilter}.
\item Temporal intensity scaling to compensate for changes in solar
elevation. See Sect.~\ref{sec:temp-intens-calibr}
\item Finalize. A final adjustment of metadata to bring cubes made
with earlier versions of SSTRED up to SOLARNET compliance.
Statistics (see Sect.~\ref{sec:statistics}) and checksum
calculations.
\item Archiving and export. Make thumbnail images and context videos.
Upload science data cubes to local disks accessible from the
internet. Upload metadata to databases (local and SVO).
\end{itemize}

\subsection{SVO}
\label{sec:svo}

As the last (optional) steps in the post-MOMFBD processing, we upload
the science-ready data cubes to a web-accessible disk area at
Stockholm University and their metadata to an SVO. At this point, we
specify the SOLARNET keywords \texttt{RELEASE} (a date after which the
data are released to the community) and \texttt{RELEASEC} (a comment
describing the data policy and/or a link to more information about
this policy).

We currently use the prototype SVO, hosted by the Royal Observatory of
Belgium, developed by \citetads{2017psio.confE..91V} in the first
SOLARNET project. We expect to move to the production SVO, developed
by the same group (Mampaey et al., in prep.) in the second SOLARNET
project when it is operational. The SVO facilitates searches in the
metadata and can serve URLs for data downloads from the Stockholm
servers. These search-and-download operations can also be performed by use
of IDL scripts. The server is aware of the \texttt{RELEASE} keyword
and knows to issue a password challenge if the release date is in the
future.

\section{Calibrations}
\label{sec:calibrations}

Here we describe calibrations that were not mentioned in the CRISPRED
paper or are performed differently in SSTRED. See
Sect.~\ref{sec:steps-reduct-chrom} for how they fit within the
pipeline processing as a whole.

\subsection{Camera alignment}
\label{sec:camera-alignment}

Optical setups that use beamsplitters, filters, and lenses are
never perfectly stable in terms of alignment. We therefore need to routinely
calibrate the alignment of the cameras separately for each prefilter.
The tuning of the FPIs and the changes in the polarization states of
the LCs in practice do not have a significant effect, therefore we
average the measurements over those.

The coalignment of the cameras is measured by use of images of an
alignment target mounted in the Schupmann focus, directly after the
FS. This process is described by \citetads{2005SoPh..228..191V} and in
Sect.~3.2 of the CRISPRED paper, although we have changed some aspects
of this calibration. The alignment target, referred to as the pinhole
array, is a rectangular grid of pinholes with a 5\farcs12 spacing,
completely filling the detector area. Three pinholes in an L
configuration are intentionally slightly larger and therefore
brighter. There is also some intensity variation from manufacturing
and dirt. This variation is used to identify the correct orientation
and alignment.

The new method fits a generic transformation matrix to the relative
pinhole positions as measured first by locating the peaks and then by
maximizing the cross-correlation between subimages with the individual
pinhole images. Previously, relative positions were measured as
wavefront tilt components with the MOMFBD program, one hole at a time.

Another change is in how the results of the calibration are
represented. Before, field-dependent shifts in the two axis directions
were stored as detector-pixel maps of offsets in X and Y. This format
can represent very general geometric deformations, but was in practice
only used for global shifts (small ones, large ones represented by
\texttt{ALIGN\_CLIPS}), image-scale differences, skew, and field
rotation. When the MOMFBD program had read a subfield in the anchor
WB image, it would obtain the matching subfield in another
camera or channel by looking up the X and Y shifts for the center
position of the anchor camera and 1) applying the rounded offsets to
the pixels coordinates to read out and 2) adding the sub-pixels
remainders of the offsets to the wavefront tilt terms of the channel.

This scheme has %
problems near the edges of the field of view (FOV). This is mostly due
to the fact that the images are first cropped to the greatest common
FOV before applying the alignment offsets. If there is even a slight
difference in rotation or scale between the cameras, the subfields
near the corners can be shifted outside of the cropped image. The data
that are missed through this operation, if only in a few pixels, will
compromise the restoration. In addition, the shifts are not accurately
described by a quadratic surface near the edges (which the old
calibration assumes). The calibration itself was also not very robust
with respect to a larger rotation misalignment between cameras. The
new method, on the other hand, can handle arbitrary rotations and
scale differences, and it can also be used to align CRISP and CHROMIS
images.

The new method consists of determining the projective transforms
\citep[see, e.g.,][]{hartley00multiple} relating each channel to the
reference channel in the form of a 3$\times$3 matrix,
\begin{equation}
\label{eq:1}
H = \left[
\begin{array}{cc|c}
h_{00} & h_{01} & h_{02} \\
h_{10} & h_{11} & h_{12} \\[1mm]
\hline
&&\\[-3.2mm]
h_{20} & h_{21} & 1
\end{array}
\right],
\end{equation}
acting on the projective coordinate vector, $[x,y,1]^\text{T}$,
followed by a normalization to maintain unity in the third element.
The top left 2$\times$2 block of $H$ encodes rotation, magnification,
and mirroring, $h_{02}$ and $h_{12}$ are translations into $x$ and $y$
(i.e., the top two rows of the matrix make up an affine
transform), and the elements $h_{20}$ and $h_{21}$ are responsible
for perspective skew and keystone effects. A simple sanity-check for
the SST setup is that $h_{00}$ and $h_{11}$ should be close to $+1$ or
$-1$, $h_{02}$ and $h_{12}$ should be within a few tens of pixels from
$0$ (or from the detector size, in case of mirroring), and that
$h_{20}$ and $h_{21}$ should be tiny ($\sim$$10^{-5}$).

In addition to the advantages in accuracy and efficiency of the calibrations
and the alignment during MOMFBD, there are some additional bonuses such as
inverse or composed mappings, which are trivial matrix operations. Many
software libraries (such as OpenCV) can be used to manipulate images
based on these transformation matrices.

\subsection{Cavity errors}
\label{sec:cavity-errors}

CRISP and CHROMIS are based on dual FPIs, mounted in a telecentric
configuration. The central wavelength of the FPI transmission profile
can be tuned to different wavelengths by changes in the separation of
the etalon cavities. Deviations from the nominal tuning
wavelengths are caused by very fine spatial variations, $\delta D$, in the
cavity, $D$, between the two plates of an etalon. Because the etalons
are located near the focal plane, the variation over the etalon
surfaces translates into variations over the FOV: the passband is not
centered at exactly the same wavelength over the entire FOV.

These cavity errors, interpreted as distortions in the wavelength
coordinate, $\lambda$, through
$\delta\lambda=\delta D\cdot \lambda/D$, have to be taken into account
when the data are interpreted, in particular when CRISP and/or
CHROMIS data are fit to models of the solar atmosphere.

\subsubsection{Measurement}
\label{sec:measurement}

The method for measuring the cavity errors in CRISP from flat field
data was developed by \citetads{2011A&A...534A..45S} and is also
described in Sect.~4.3 of the CRISPRED paper.
It fits a pixel map of wavelength shifts, based on the assumption that
the pixels in a scan all show the same spectrum. For this a
sufficiently well sampled line, including slopes in both directions,
is needed. We use the same method for CHROMIS; see
Fig.~\ref{fig:cavity-error} for cavity errors measured for CHROMIS.
See Appendix~\ref{sec:wavel-dist} for how the cavity errors are stored
as distortions to the WCS wavelength coordinate.

\begin{figure}[!ht]
\includegraphics[viewport=76 72 508 731, clip, width=\linewidth]{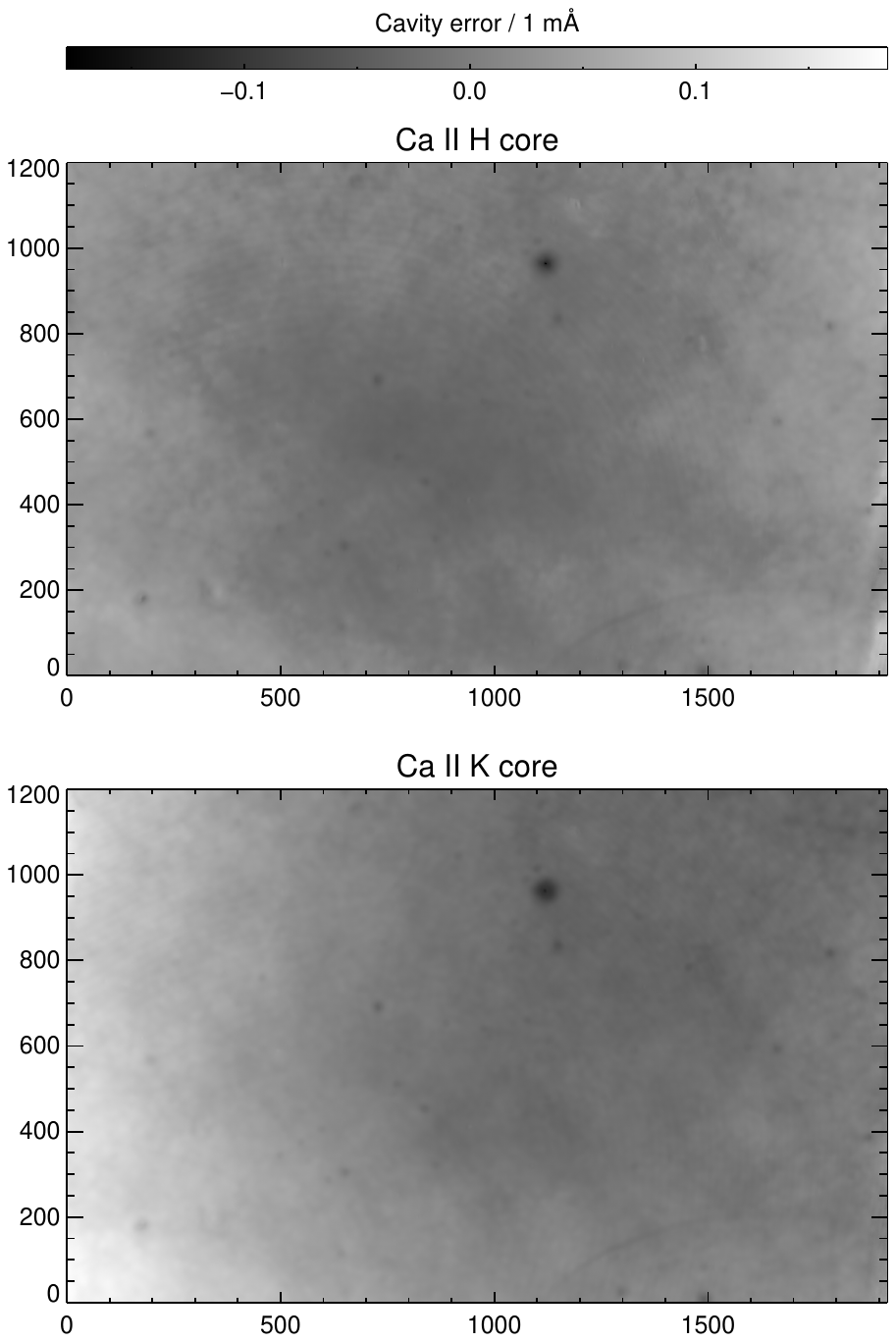}
\caption{Cavity error of the high-resolution CHROMIS etalon through
the \ion{Ca}{ii} H and K core prefilters as measured with data
from 2016 June 19. Mapping the associated wavelength shift over the
1920$\times$1200-pixel NB detector.}
\label{fig:cavity-error}
\end{figure}

The tuning of the CRISP and CHROMIS etalons is controlled digitally,
with step sizes that were calibrated when the instruments were
installed.
The etalon separation might be thought to change with tuning,
as was recently measured by \citetads{2019A&A...626A..43G} in a laboratory
setup. If the three piezo stacks that control the cavity width do not
move by the same amount, we would obtain a time-variable linear component
(tilt) in the cavity separation. However, the accuracy of the tuning
control is estimated to $\pm 0.0005$~m\AA{}/du for CRISP and
$\pm0.01$~m\AA{}/du for CHROMIS (P.~S\"utterlin 2020, priv. comm.).

While the calibration is good enough that tilt variations with tuning
are negligible within the tuning range of a prefilter, the cavity
errors still do have a tilt component. The tilt component depends on
the settings of parameters related to the piezos that control the gap
when tuning, and these parameters are set by daily calibrations that
are independent for each prefilter. For CHROMIS \ion{Ca}{ii}, where
multiple prefilters are used in a single scan, the tilt (and possibly
other low-order components) of the gap therefore varies from one
prefilter to another. Compare the cavity maps for the two \ion{Ca}{ii}
core filters in Fig.~\ref{fig:cavity-error}. This effect could have
been reduced by sending control signals to each piezo stack
individually. Sending a single control signal to all three is needed
for fast scans.

The \ion{Ca}{ii}~H and~K wing filters have so far hardly been used, but
it should be possible to calibrate them using some of the blends in
the wing filter passbands. It should be possible also for the
\ion{Ca}{ii}~continuum filter, although it is not needed as long as it
is not used to scan through any of the lines within its passband.

\subsubsection{Flat fielding}
\label{sec:flat-fielding}

When observations are made in the parts of spectral lines in which the
intensity changes rapidly, the shifts in cavity error wavelength cause a
spatial imprint on the intensities of both data and flats. The
wavelength shifts cause intensity changes that are not accounted for
in the monochromatic image-formation model assumed by MOMFBD. This can
cause suboptimal solutions for the wavefronts (i.e., the PSFs) in the
image restoration step because of mismatches with the image formation
model, which assumes quasi-monochromatic data. For polarimetric
observations, the time-dependent telescope polarization also has an
effect on the flat fields. Moreover, deconvolving large discontinuities in
the cavity errors can cause artifacts in the MOMFBD output. For these
reasons, \citetads{2011A&A...534A..45S} developed a method for taking
all this into account.

Science data are often collected in areas in which the spectrum varies
with both time and position in the FOV and is quite different from
that of the quiet Sun, where flat-field data are collected. The
CRISPRED paper (see their Sect.~3.1) improved upon the procedure by
\citetads{2011A&A...534A..45S} to take this into account as well. This is
what is used for CRISP data in SSTRED.

The cavity errors of CHROMIS are both smoother and smaller than those
of CRISP. CHROMIS cavity errors are $\sim$0.1~m\AA{} (see
Fig.~\ref{fig:cavity-error}), while CRISP cavity errors are
$\sim$10~m\AA{} (see Fig.~4 of the CRISPRED paper).
For CHROMIS, we have therefore found that the faster method by
\citetads{2011A&A...534A..45S} is sufficient.

\subsection{Periodic artifacts }
\label{sec:periodic-artifacts}

\begin{figure*}[!tp]
\def\tilewidth{0.715\linewidth}
\def\tilewidth{0.32\linewidth}
\begin{subfigure}[t]{\tilewidth}
\includegraphics[width=\linewidth]{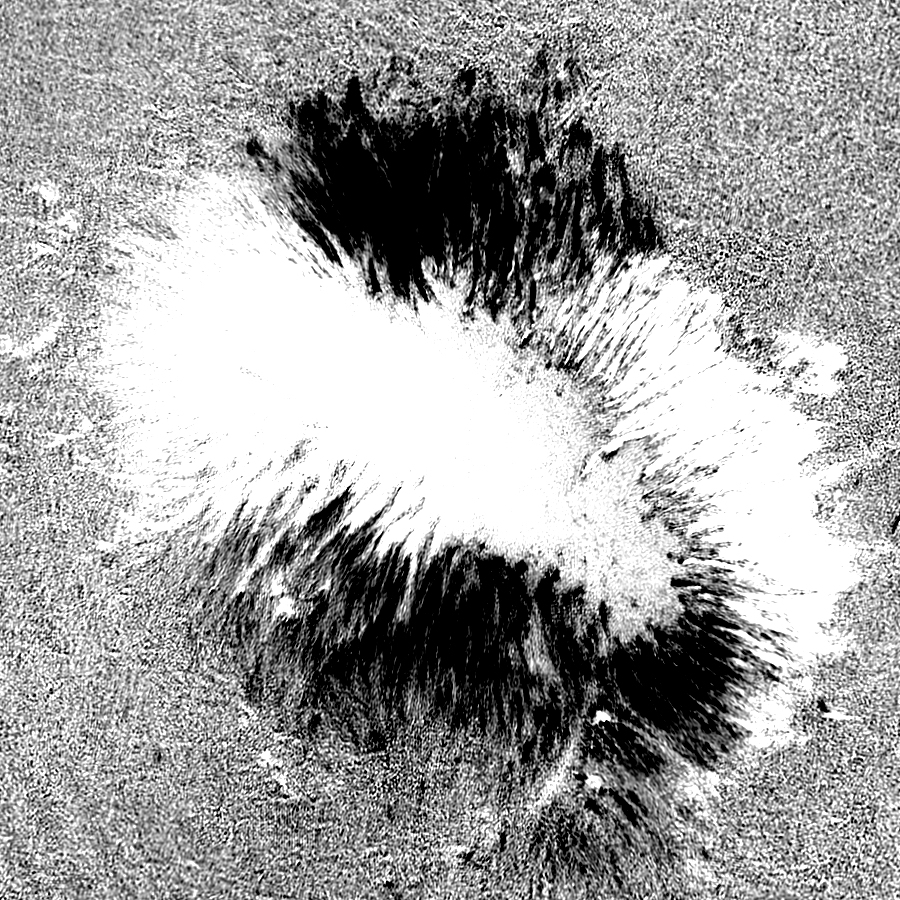}
\caption{Stokes $U$ component, sunspot saturated to emphasize
artifacts in the lower left corner.}
\label{fig:periodicartifacts_orig}
\end{subfigure}\hfill
\begin{subfigure}[t]{\tilewidth}
\includegraphics[width=\linewidth]{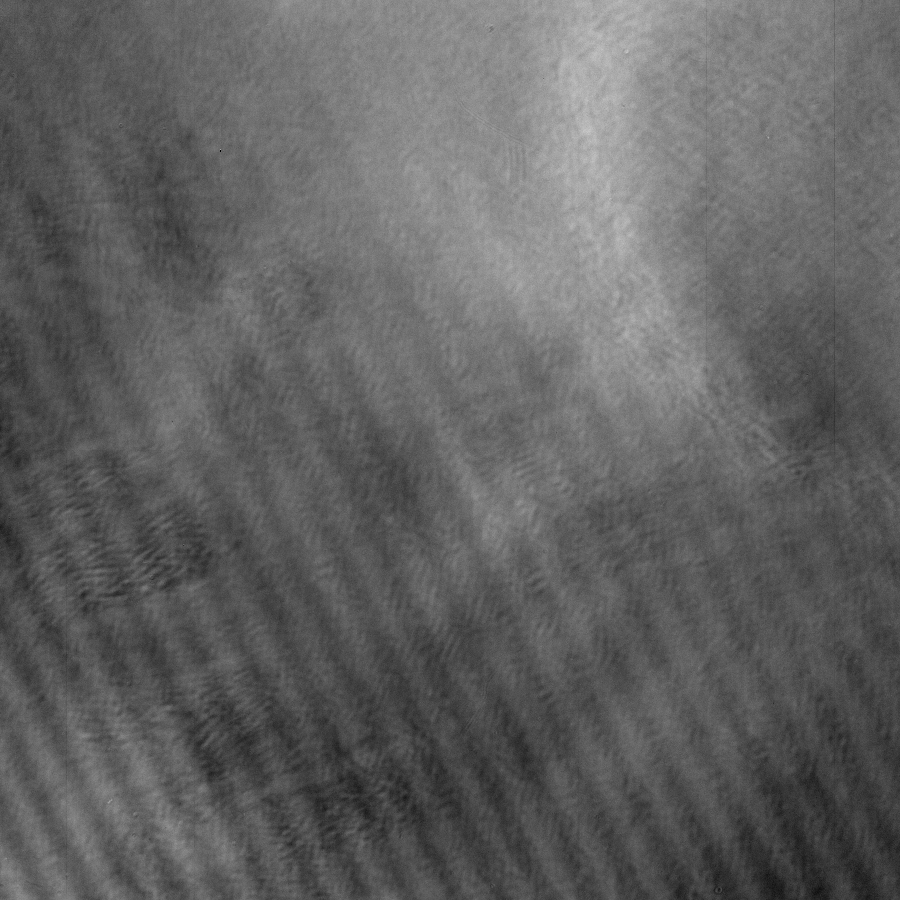}
\caption{The polarization calibration image on which the Fourier
filter is based.}
\label{fig:periodicartifacts_polcal}
\end{subfigure}\hfill
\begin{subfigure}[t]{\tilewidth}
\includegraphics[width=\linewidth]{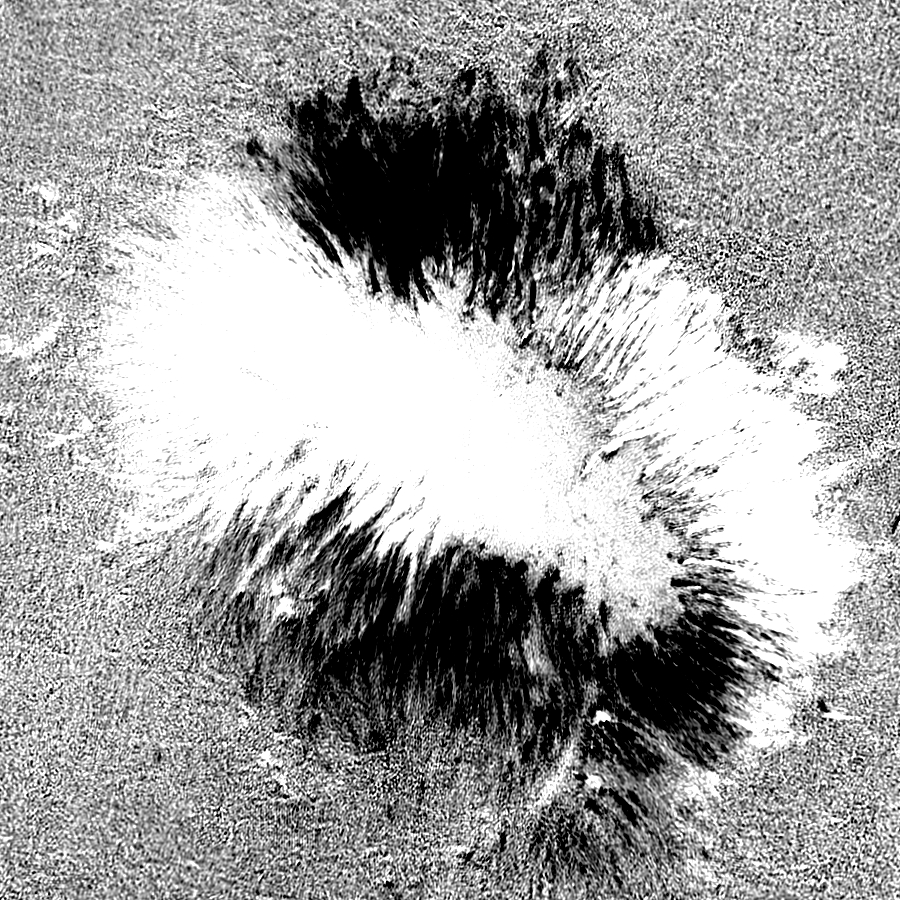}
\caption{Fourier filtered $U$ component without the artifacts.}
\label{fig:periodicartifacts_filt}
\end{subfigure}
\caption{Removal of periodic artifacts in polarimetric data. The
sample CRISP \ion{Fe}{i} 6302~\AA{}, $-210$~m\AA{} from line core.
Data were collected on 2019 May 10 at 09:07~UT with AR 2741 in the
FOV. The image shown is 53\arcsec{} squared.}
\label{fig:periodicartifacts}
\end{figure*}

Polarimetric CRISP datasets
often show periodic artifacts in the
Stokes~$Q$, $U$, and $V$ components, in particular $U$ and $V$. A
sample $U$ image with such artifacts is shown in
Fig.~\ref{fig:periodicartifacts_orig}. With dominating high-contrast
features such as sunspots in the FOV, it is difficult to separate the
signal from the artifact.
However, the same artifact is visible in the polarization calibration
(polcal) image data. As polcal data are always collected in regions
with quiet Sun, the offending spatial frequency can quite easily be
found manually in the 2D power spectrum of sums of polcal images. A
sample image is shown in Fig.~\ref{fig:periodicartifacts_polcal}.
This spatial frequency is then removed from the Stokes images by use
of a 2D notch filter, the result of which is shown in
Fig.~\ref{fig:periodicartifacts_filt}.

Similar Fourier filtering was implemented in CRISPRED, but it was left
to the user to actually construct the filter. \citet[his
Fig.~2.19]{diaz_baso18analysis} and \citeauthorads{2020A&A...644A..43P}
(\citeyearads[][their Fig.~2]{2020A&A...644A..43P})
implemented their own similar corrections
before the filter described here was included in SSTRED.

The artifact pattern seems stable over years of observations, but is
measured independently for each new day of observations.
A pattern with the same period and direction is visible also
in the modulation matrix components. Our first thought was to apply
the notch filter to this matrix before demodulating the data. However,
this exacerbated the artifacts.

A close inspection of Fig.~\ref{fig:periodicartifacts_filt} shows
cross-talk from Stokes~$I$ as a weak granulation pattern imprint
outside the spot (in particular in the top part). See
Sect.~\ref{sec:polar-cross-talk} for how this is removed.

\subsection{Intensity calibration}
\label{sec:intens-calibr}

\subsubsection{Spectral intensity calibration}
\label{sec:fitprefilter}

In order to characterize the wavelength-dependent instrument response
in our observations, we use the Hamburg disk center atlas spectrum
(\citeads{1999SoPh..184..421.}; \citealp{brault87spectal}) as a
reference. We degrade the atlas spectrum by convolution with a
theoretical FPI transmission profile. We then estimate the parameters
in a model prefilter profile by fitting the degraded atlas spectrum to
quiet-Sun data collected at disk center. The prefilter model is a
Lorentzian multiplied with an antisymmetric polynomial,
\begin{equation}
\bar{P}(\lambda,p_0,\ldots,p_6) =
\frac{p_0}
{\displaystyle
1 + \left(
2 \frac{\lambda_\text{c}-p_2}
{p_3}
\right)^{2p_4}
}
\cdot
( 1 + p_5\lambda_\text{c} + p_6\lambda_\text{c}^3 )
\label{eq:pref}
,\end{equation}
where
\begin{equation}
\lambda_\text{c}= p_1 + (\lambda-\text{CWL})\cdot p_7 .
\end{equation}
The parameters have the following effects on the fit: $p_0$ is a scale
factor from counts to the units of the atlas; $p_1$ shifts the
wavelength scale to align the spectral lines in the data with those in
the atlas spectrum; $p_2$ is a shift of the CWL of the prefilter;
$p_3$ is the FWHM of its transmission profile; $p_4$ is the number of
cavities in the filter; the polynomial with parameters $p_5$ and $p_6$
represent asymmetries in the profile as well as in other
wavelength-dependent effects (such as the QE of the detector as well
as the transmission of other optics and the atmosphere) on the number
of collected photons; and the seldom used $p_7$ stretches the
wavelength grid. The Lorentzian represents the prefilter transmission
profile, where not constraining the number of cavities to be an
integer accounts for some deviations from an idealized filter.

\begin{figure*}[!tbp]
\begin{subfigure}{.49\linewidth}
\includegraphics[viewport=90 47 720 527,clip,width=\linewidth]{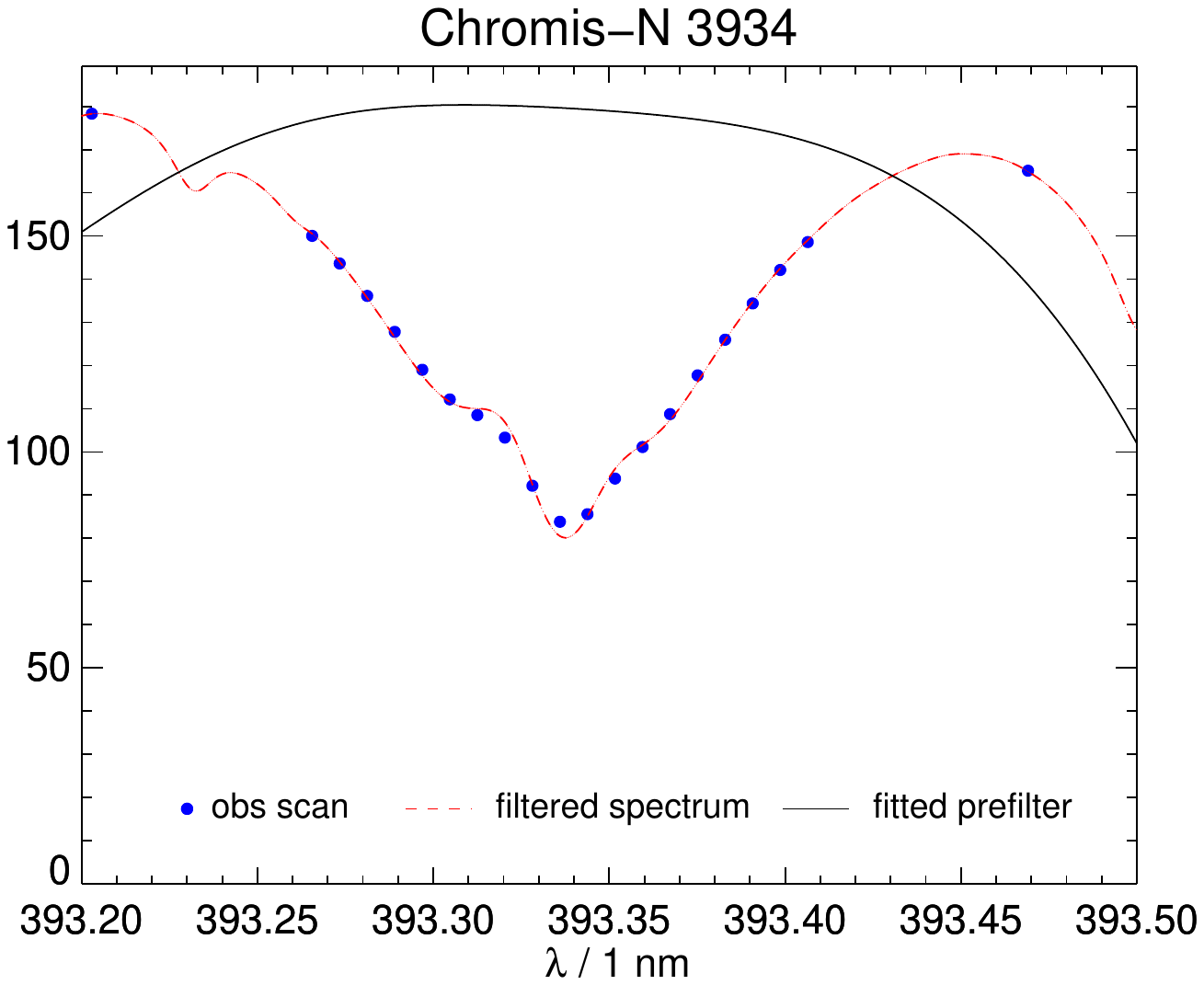}
\caption{\ion{Ca}{ii}~K core}
\end{subfigure} \hfill
\begin{subfigure}{.49\linewidth}
\includegraphics[viewport=90 47 720 527,clip,width=\linewidth]{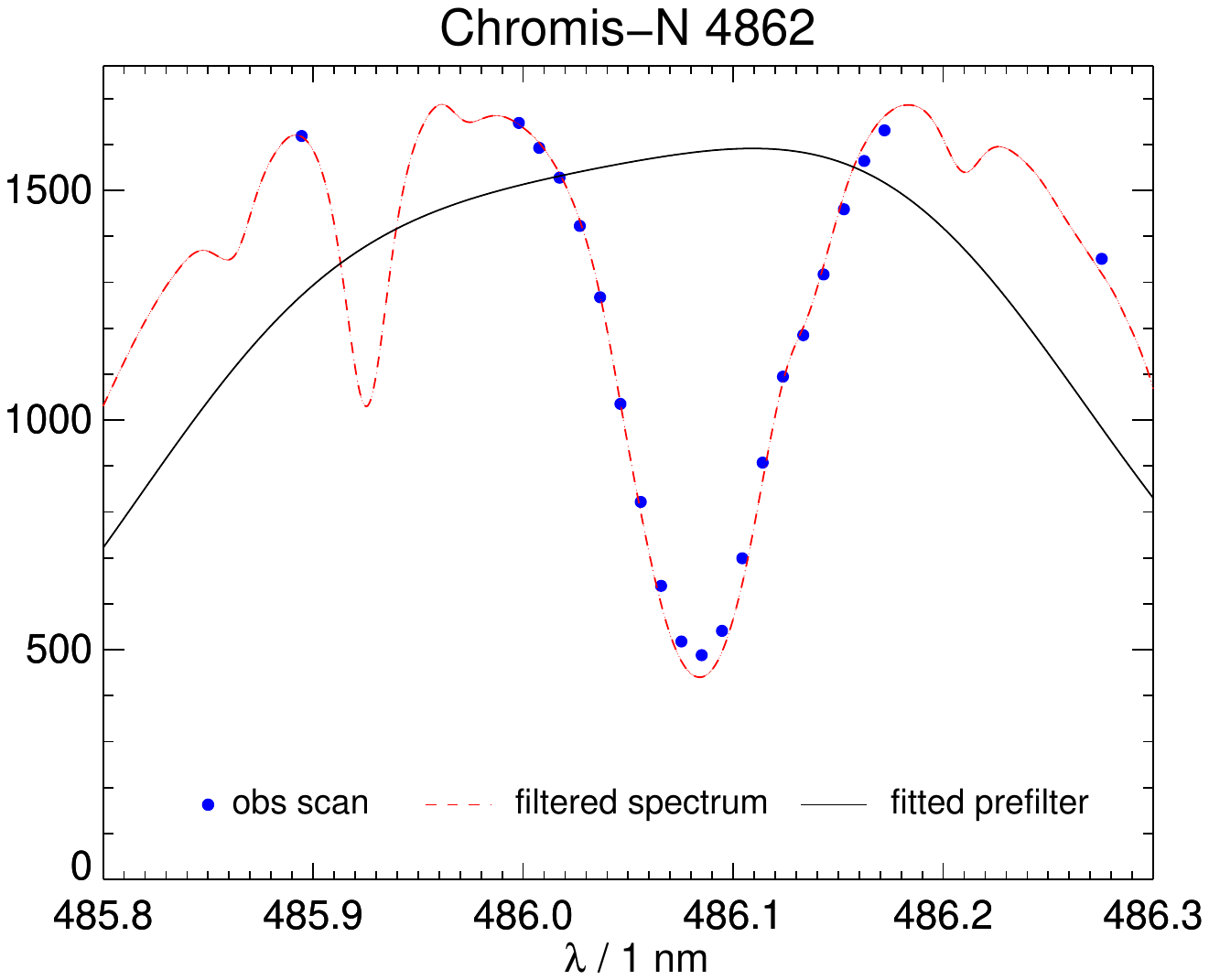}
\caption{H-$\upbeta$}
\end{subfigure}
\caption{Sample diagnostic plots from the spectral
intensity calibration. The scan intensities are plotted in counts,
while the fitted prefilter transmission is plotted in
arbitrary units.}
\label{fig:fitprefilter-plots}
\end{figure*}

Fig.~\ref{fig:fitprefilter-plots} shows examples of these fits for
\ion{Ca}{ii}~K and H-$\beta$. The derived prefilter curves are also
plotted along with the fits in this figure.

As described in Sect.~4.5 of the CRISPRED paper, the fitted profile
(defined by parameters $p_0$ and $p_2$--$p_6$) is used to compensate
science data for the wavelength-dependent response of the instrument
and to convert the camera counts into intensity in SI units
($\rm W \, m^{-2} \, Hz^{-1} \, sr^{-1}$). In addition, $p_1$ is used
to adjust the wavelength coordinate from the one set by the instrument
calibration made each morning at the telescope.

\subsubsection{Temporal intensity calibration}
\label{sec:temp-intens-calibr}

The spectral intensity calibration in the previous subsection is used
to remove the effect of the prefilter transmission profiles and scale
intensity counts to SI units. However, because of variations in solar
elevation and the amount of dust in the line of sight, this conversion
is really only valid for the time the calibration data were collected.

Our previous approach to correcting for this was to apply a correction
factor that is the ratio of median WB intensities of the individual
scan and that of the calibration data. This compensates for changes in
elevation and dust, but has the drawback that it also removes the
inherent intensity variations from solar disk center to limb. A better
intensity calibration would need a similarly measured intensity ratio,
but with both WB intensities obtained at disk center.

\begin{figure*}[!tp]
\begin{subfigure}{0.49\linewidth}
\includegraphics[viewport=61 44 700  531,clip,width=\linewidth]{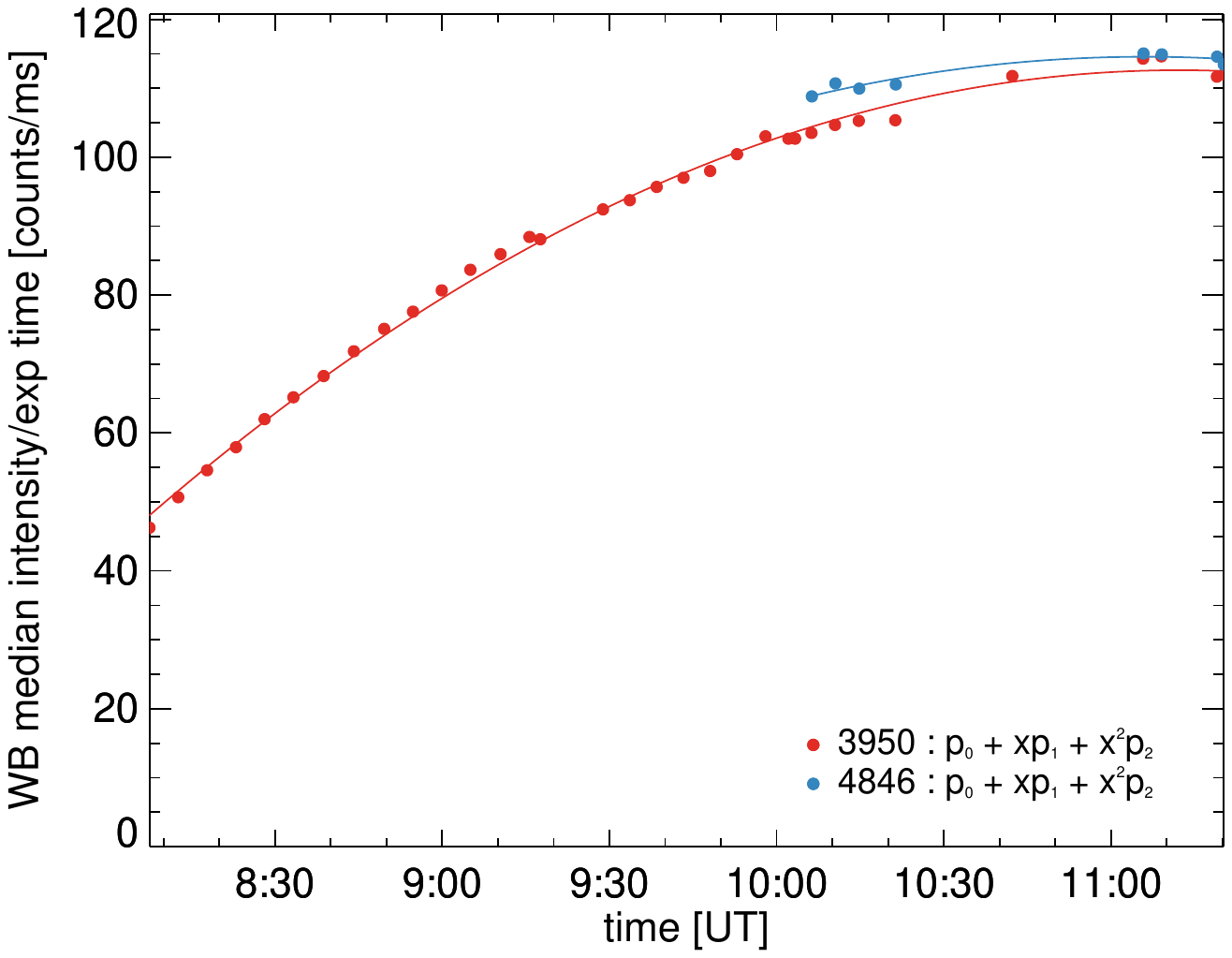}
\caption{CHROMIS}
\label{fig:wb-intensities-chromis}
\end{subfigure}
\hfill
\begin{subfigure}{0.49\linewidth}
\includegraphics[viewport=61 44 700 531,clip,width=\linewidth]{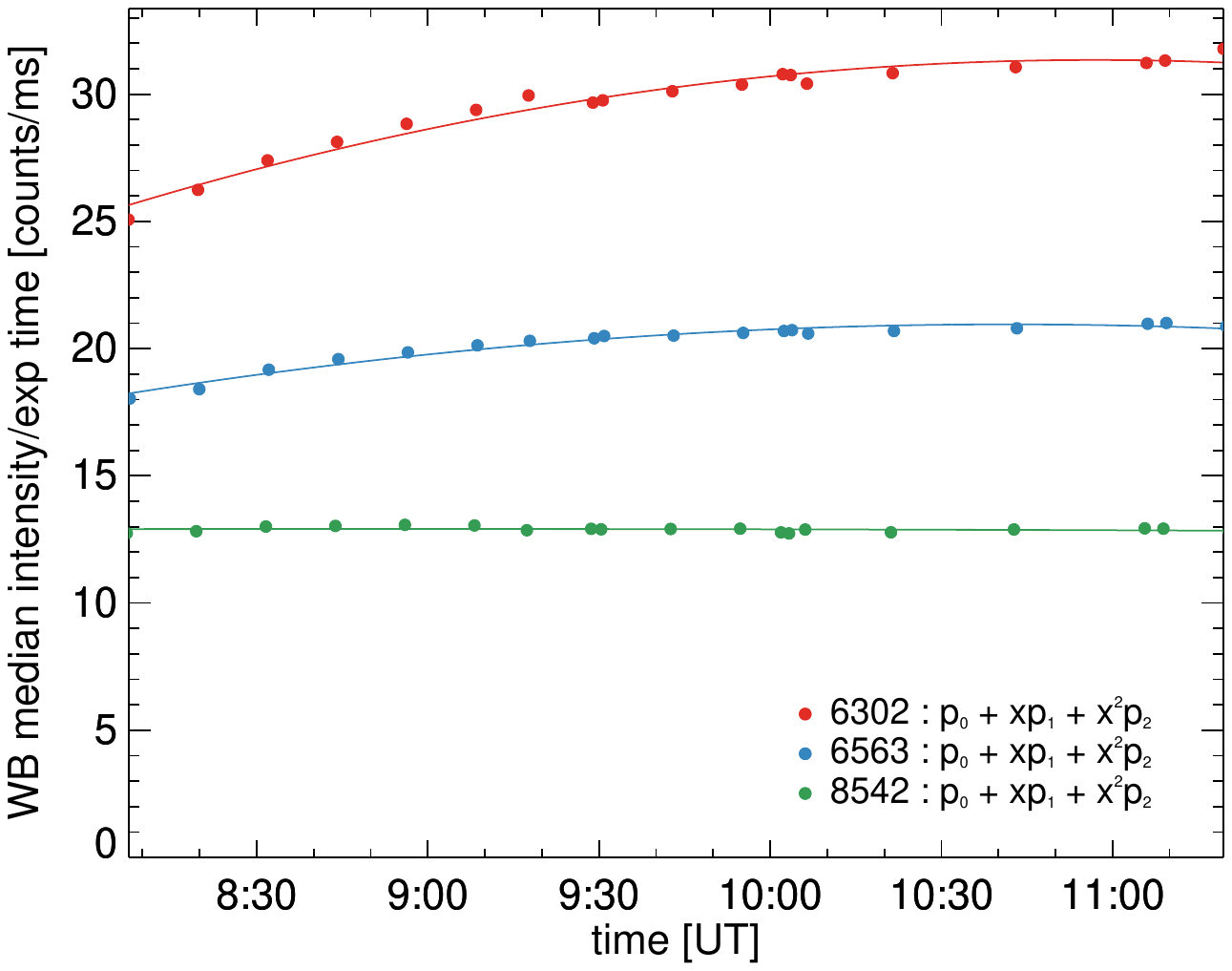}
\caption{CRISP}
\label{fig:wb-intensities-crisp}
\end{subfigure}
\caption{Disk center median WB calibration intensities from
2016 September 19. Prefilter wavelength bands in Å as indicated in the
legends. The solid lines are fits to second-order polynomials as
shown in the legends.}
\label{fig:wb-intensities}
\end{figure*}

The present default calibration procedure is based on a polynomial fit
to WB intensity data collected throughout the observing day. It
does not matter if the data are science data or flats, as long as they
are collected near disk center and temporally span
the entire period of observations. We use the median intensity
so that the procedure can tolerate science data obtained from active
regions, as long as the FOV is not dominated by large sunspots or
faculae. Sample data for both CRISP and CHROMIS are shown in
Fig.~\ref{fig:wb-intensities}. The data illustrate the decrease in
extinction toward the red wavelengths, to essentially no effect in the
IR. By default, second-order polynomials are used for the fit (first
order if there are only two data points). Other fit functions can be
specified if needed.

It might be possible to relax the requirement that the calibration
data are collected near disk center by correcting the intensities for
limb darkening using measurements by, e.g.,
\citetads{1994SoPh..153...91N} or \citetads{2005SoPh..229...13N}.
However, this is not implemented at this point.

\begin{figure*}[!tp]
\begin{subfigure}{0.49\linewidth}
\includegraphics[viewport=61 44 720 531,clip,width=\linewidth]{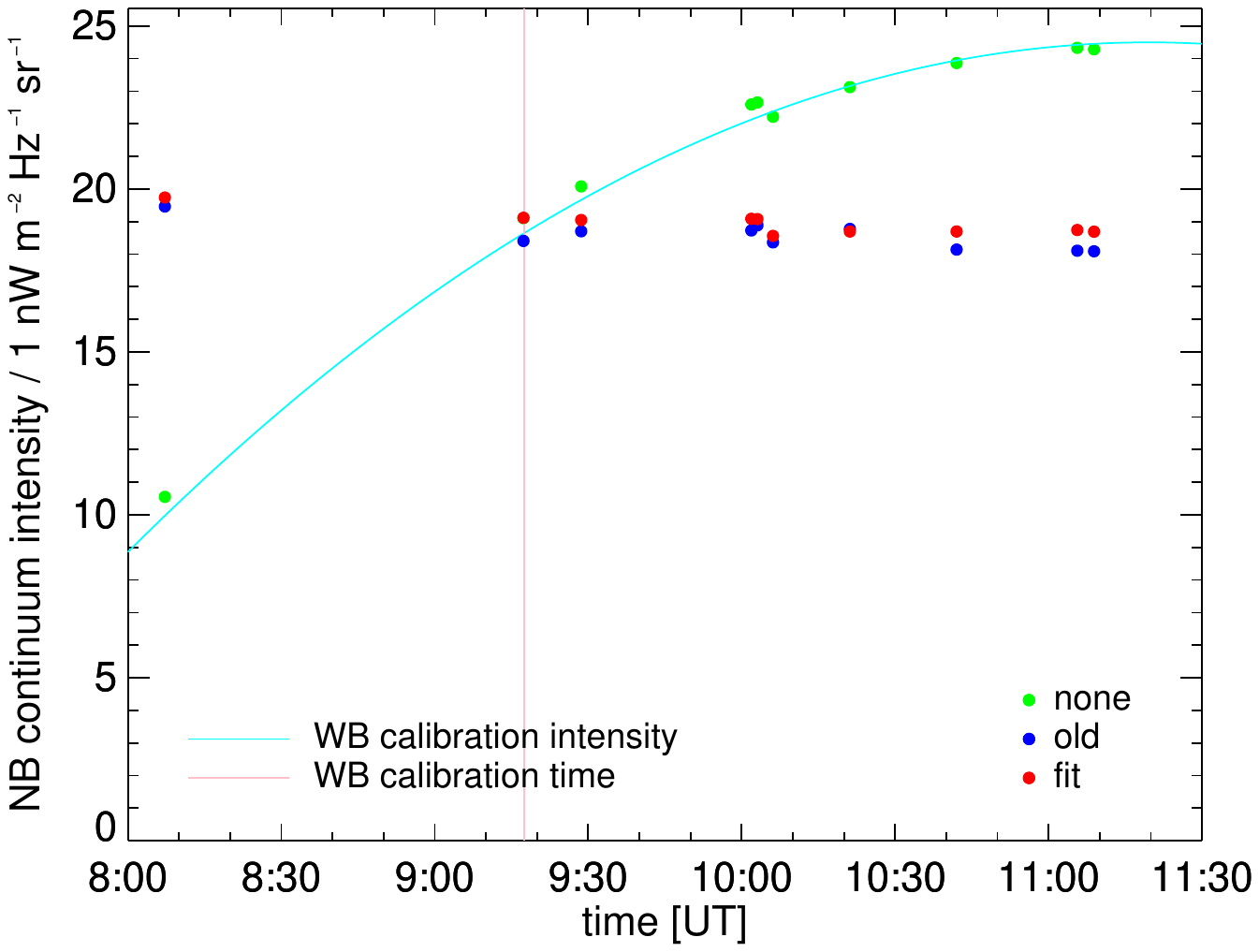}
\caption{Data from 2016-09-19, $\mu\approx 1$ (near disk center).}
\label{fig:nb-intensities-dc}
\end{subfigure}
\hfill
\begin{subfigure}{0.49\linewidth}
\includegraphics[viewport=61 44 720 531,clip,width=\linewidth]{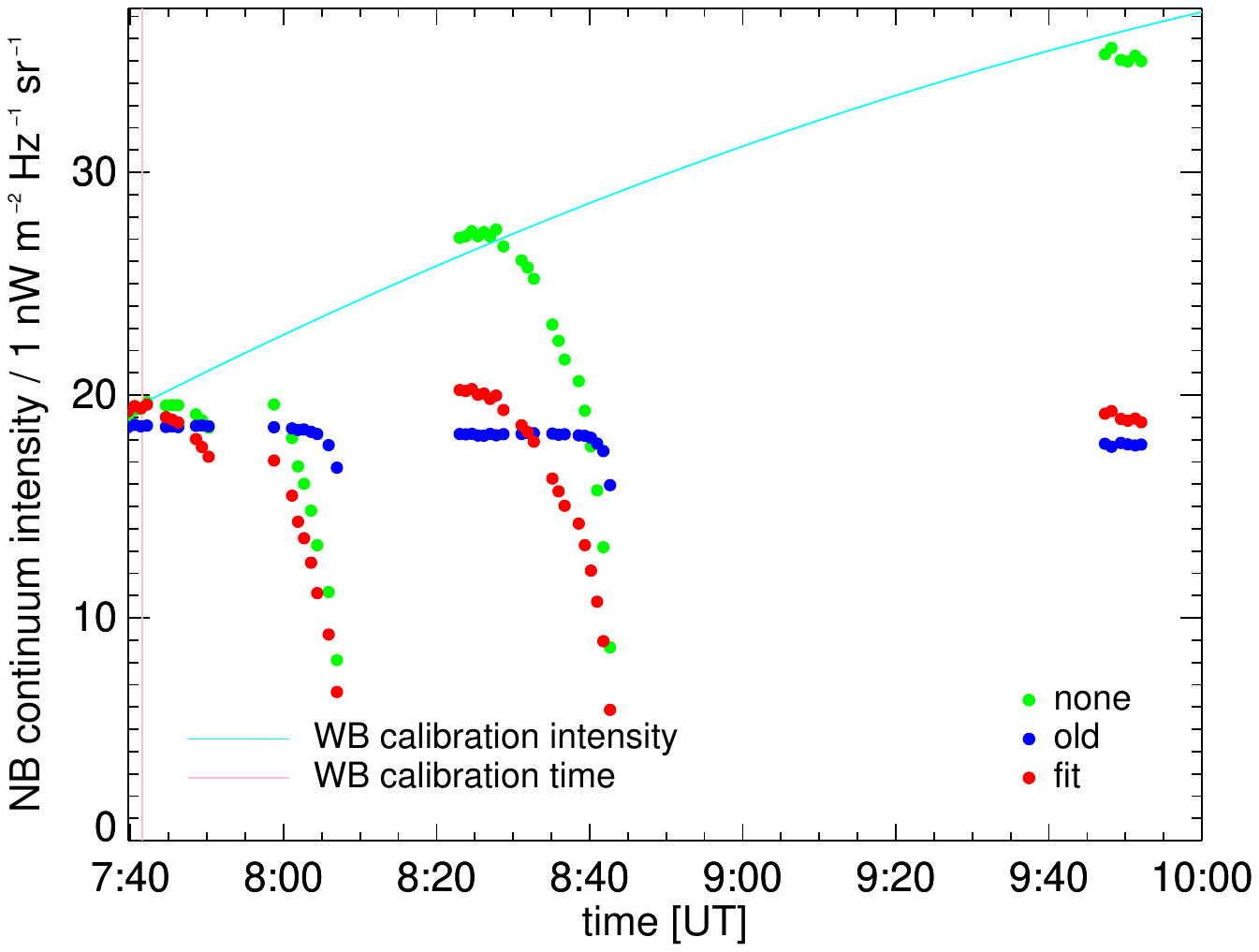}
\caption{Data from 2019-06-21, varying $\mu$.}
\label{fig:nb-intensities-mu}
\end{subfigure}
\caption{CHROMIS Ca II NB continuum (median) intensities of restored
data (individual scans), with and without temporal correction as
indicated in the legends. The WB calibration curves are the fit
functions (shown in Fig.~\ref{fig:wb-intensities-chromis} for the
2016 September 19 data), scaled to approximately match the disk center
continuum intensities. The dot colors refer to data with no
correction (green), the old correction based on the WB intensity
in the data set itself, and the correction based on a fit to disk
center data (red).}
\label{fig:nb-intensities}
\end{figure*}

Figure~\ref{fig:nb-intensities} shows the effect of the corrections
over two morning periods of observations. For the disk center data in
Fig.~\ref{fig:nb-intensities-dc}, the uncorrected intensities show the
variation with solar elevation, similar to the WB calibration curve.
The corrected intensities show much smaller temporal variation,
crossing the uncorrected curve at the time the spectral
intensity calibration data were collected. This holds for both the old
and the new correction. In Fig.~\ref{fig:nb-intensities-mu} we show a
similar plot where data were collected during two center-to-limb
scans. The disk center intensities are like those in
Fig.~\ref{fig:nb-intensities-dc}: following the increase with
elevation for the uncorrected data, and are essentially identical for
the two correction methods. The variations with $\mu$ are removed from
the old method data (except very near the limb, an effect of the
FOV including a gradient toward the limb as well as off-limb
darkness), but preserved with the fit method.

In the new procedure, we also implemented a correction for varying
exposure time. This was not necessary in CRISPRED as CRISP is
operated with a constant rotating-shutter speed. The exposure time of
CHROMIS can easily be changed, for instance, when the target is moved from
disk center to limb.

\begin{figure}[!htp]
\includegraphics[viewport=61 44 720 531,clip,width=\linewidth]{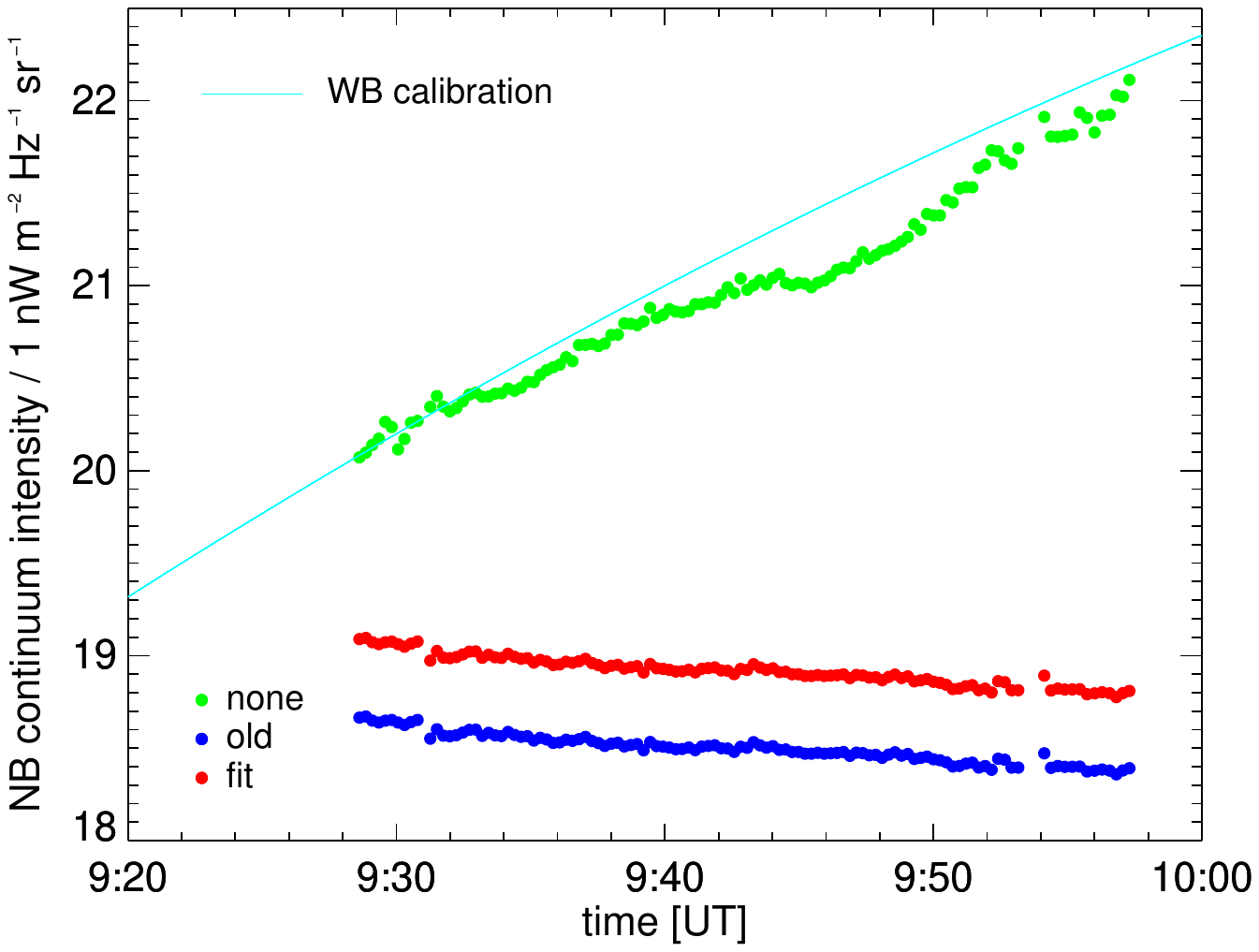}
\caption{CHROMIS Ca II NB continuum (median) intensities of restored
data (multiple scans), with and without temporal correction, as
indicated in the legends. See also Fig.~\ref{fig:nb-intensities}.}
\label{fig:nb-intensities2}
\end{figure}

For datasets consisting of multiple scans, the old method also
corrects for rapid variations in intensity that would be unaccounted
for if the ratio of fitted WB intensities alone were applied as for
the individual scans. Therefore the fit correction takes the
scan-to-scan variations in WB mean values into account. This results
in corrections that remove the rapid variations; see the sample
results in Fig.~\ref{fig:nb-intensities2}.

We have implemented this step so that the new procedure is used by
default, but for data not temporally spanned by WB disk center data, it
will fall back to the old method. See
Sect.~\ref{sec:temp-intens-calibr} for a description that the metadata of the cube
include information about which version of this step was actually
performed.

\subsubsection{Residual intensity variation}
\label{sec:residual-variation}

After the two intensity calibrations described in
Sects.~\ref{sec:fitprefilter} and ~\ref{sec:temp-intens-calibr},
\ion{Ca}{ii} continuum data still show systematic residual intensity
variations. The 1\% drop in intensity over the 30~min data collection
interval in Fig.~\ref{fig:nb-intensities2} (common to both methods) is
consistent with the $\sim$5\% intensity drop over 2.5~h shown in
Fig.~\ref{fig:nb-intensities}.

The WB and NB continuum passbands are separated by $\sim$5~nm and the
gradient in the extinction wavelength-dependence is large around
400~nm. The effect can be replicated qualitatively with Rayleigh
scattering and the semi-empirical expression for relative air mass as
a function of zenith angle by \citetads{1989ApOpt..28.4735K}, although not
quantitatively well enough for an accurate correction.

We also found a systematic residual time-dependent variation in the
ratio of intensities in the K line core and the continuum (separated
even more in wavelength) in the same datasets (not shown).

\subsection{Time-variable alignment}
\label{sec:time-vari-alignm}

MOMFBD restoration, together with the pinhole calibration described in
Sect.~\ref{sec:camera-alignment}, usually delivers images recorded at
different wavelengths that are very well aligned. However, the
wavelength range of scans involving multiple \ion{Ca}{ii} prefilters
can extend over more than 7~nm (see Fig.~\ref{fig:profiles+spectra}),
which makes misalignment caused by wavelength-dependent dispersion
significant. As a comparison, the wavelength ranges of scans through
H-$\upbeta$ and any of the CRISP lines are typically less than 1~nm.

\begin{figure*}[!tp]
\begin{subfigure}{0.49\linewidth}
\includegraphics[viewport=64 161 750 431,clip,width=\linewidth]{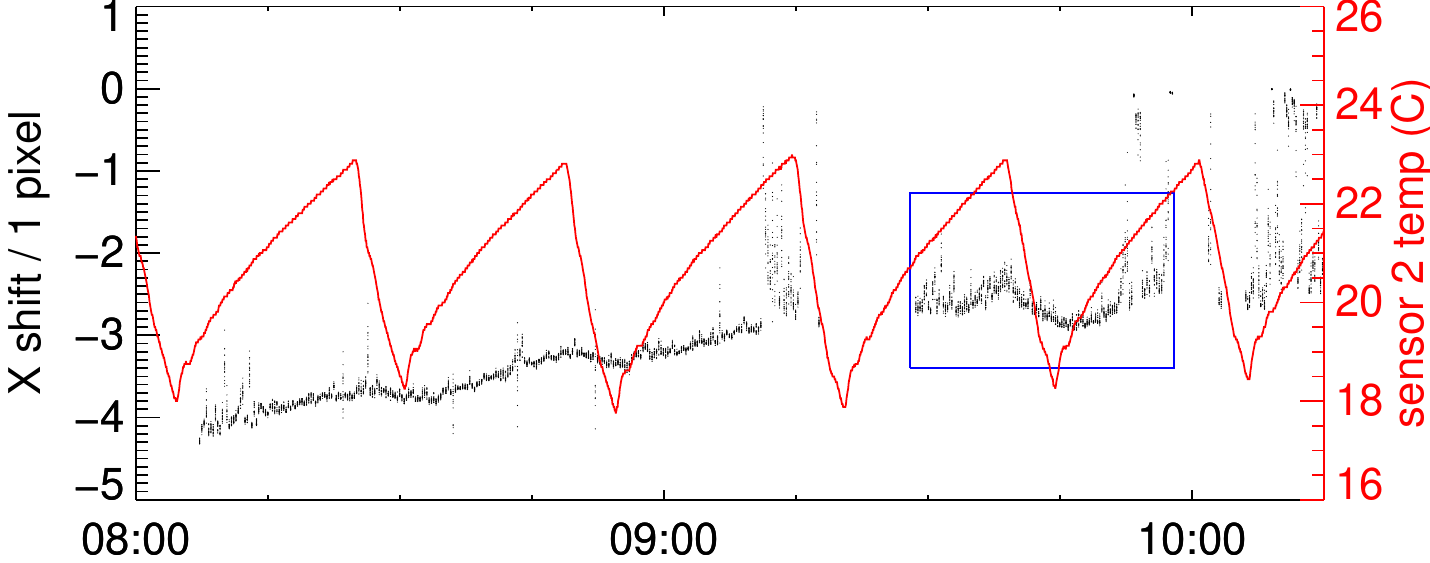}
\caption{X misalignment}
\end{subfigure}
\quad
\begin{subfigure}{0.49\linewidth}
\includegraphics[viewport=64 161 750 431,clip,width=\linewidth]{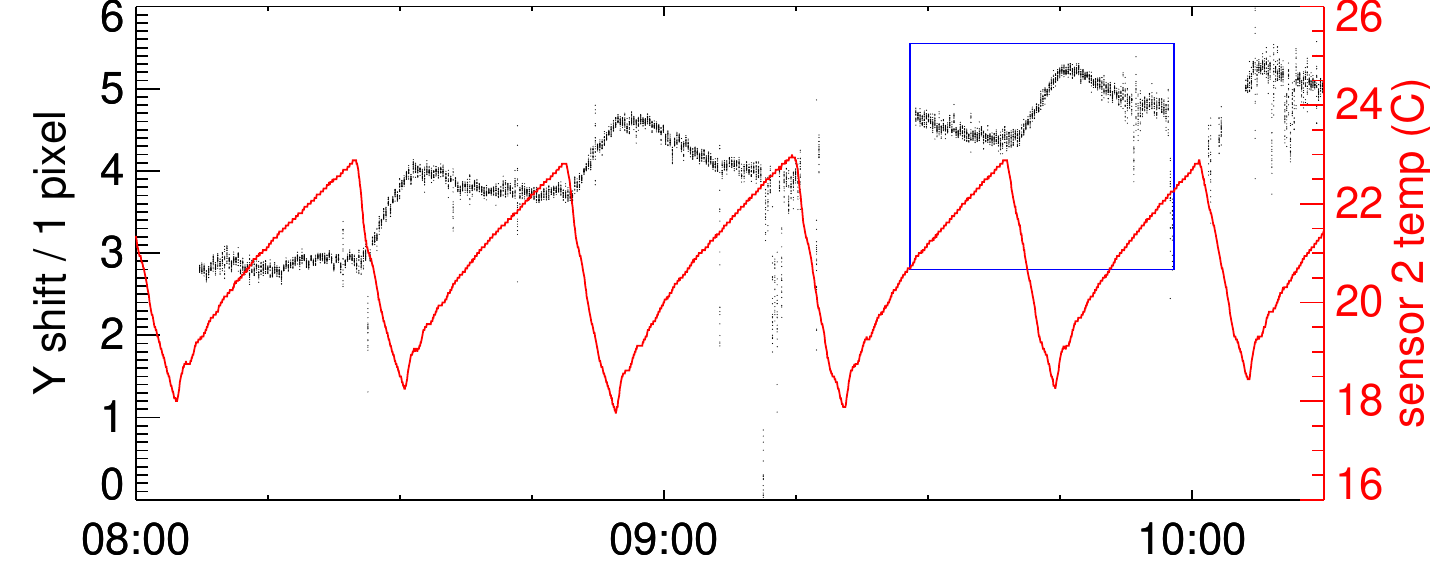}
\caption{Y misalignment}
\end{subfigure}
\caption{Covariation of the misalignment in continuum and WB measured with raw
data and telescope bottom plate temperature. These data were
collected during the morning hours of 2016 June 19. In bad seeing,
measurements most often tend toward zero. The blue boxes
correspond approximately to the plot ranges of
Fig.~\ref{fig:contalignX} and \subref{fig:contalignY}.}
\label{fig:contalign-temp}
\end{figure*}

The misalignment can be measured by use of cross-correlation of H\&K
WB and continuum data, which are both formed mostly in the
photosphere. Figure~\ref{fig:contalign-temp} shows such measurements
made with raw data collected during two morning hours.
The misalignment is time-dependent, and there are two components, one
of which varies with telescope elevation in a way that suggests
atmospheric dispersion. The other component is periodic, and we found
that the period matches that of the temperature of the telescope
bottom plate,
also shown in Fig.~\ref{fig:contalign-temp}. The periodic variation
was traced to an air-conditioning unit that switched on and off at regular
time intervals.
Our working hypothesis for the periodic component is that the
temperature variations have a minute effect on the alignment of the
field mirror that reflects the SST beam to the Schupmann corrector
(see Fig.~\ref{fig:sst}, inset A) or within the corrector itself. The
alignment of the Schupmann corrector has a wavelength-dependent
dispersion effect \citepads{2003SPIE.4853..341S}.

\begin{figure*}[!tbp]
\begin{subfigure}{0.32\linewidth}
\includegraphics[viewport=64 47 720 529,clip,width=\linewidth]{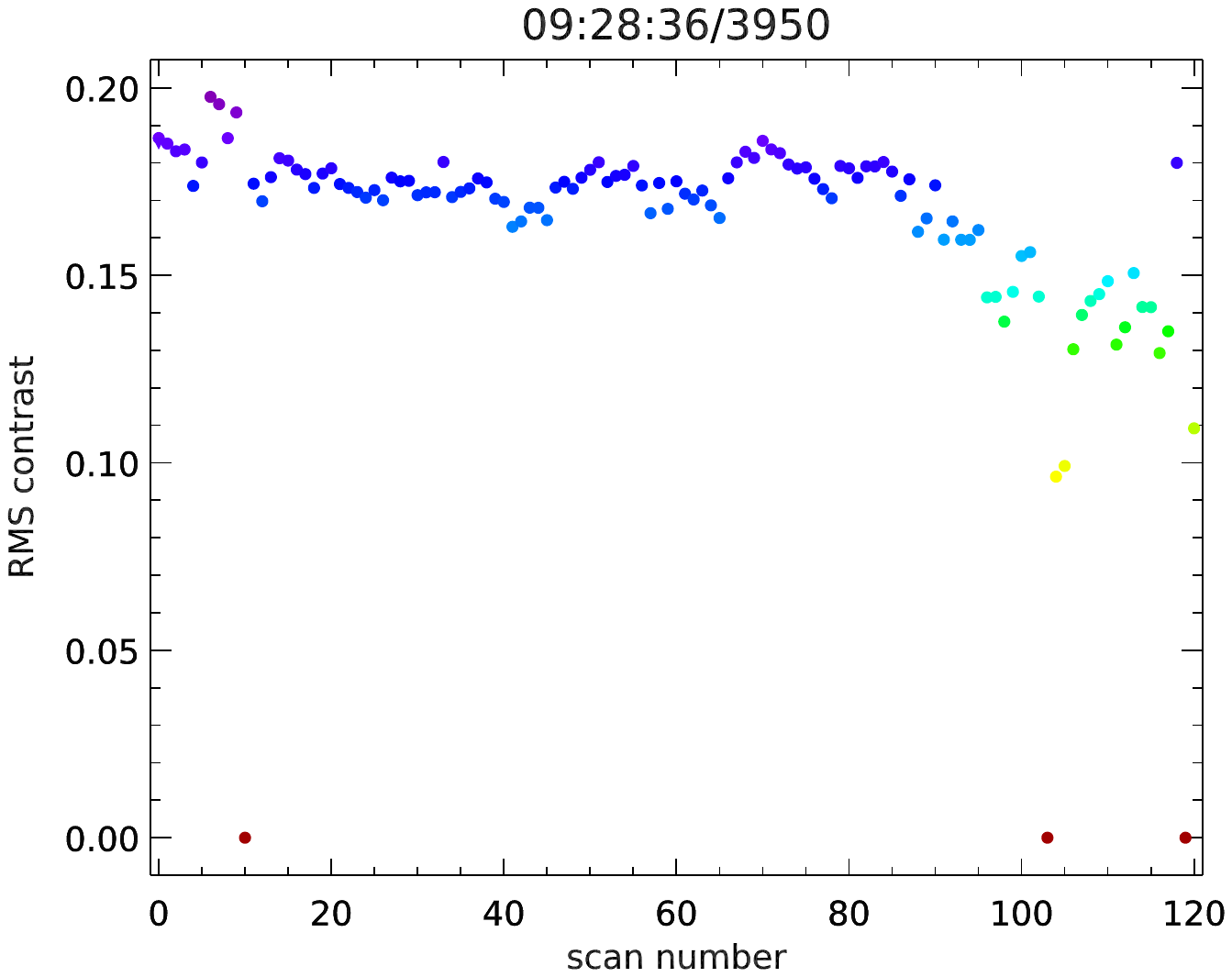}\
\caption{RMS contrast}
\label{fig:contalignR}
\end{subfigure}
\
\begin{subfigure}{0.32\linewidth}
\includegraphics[viewport=64 47 720 529,clip,width=\linewidth]{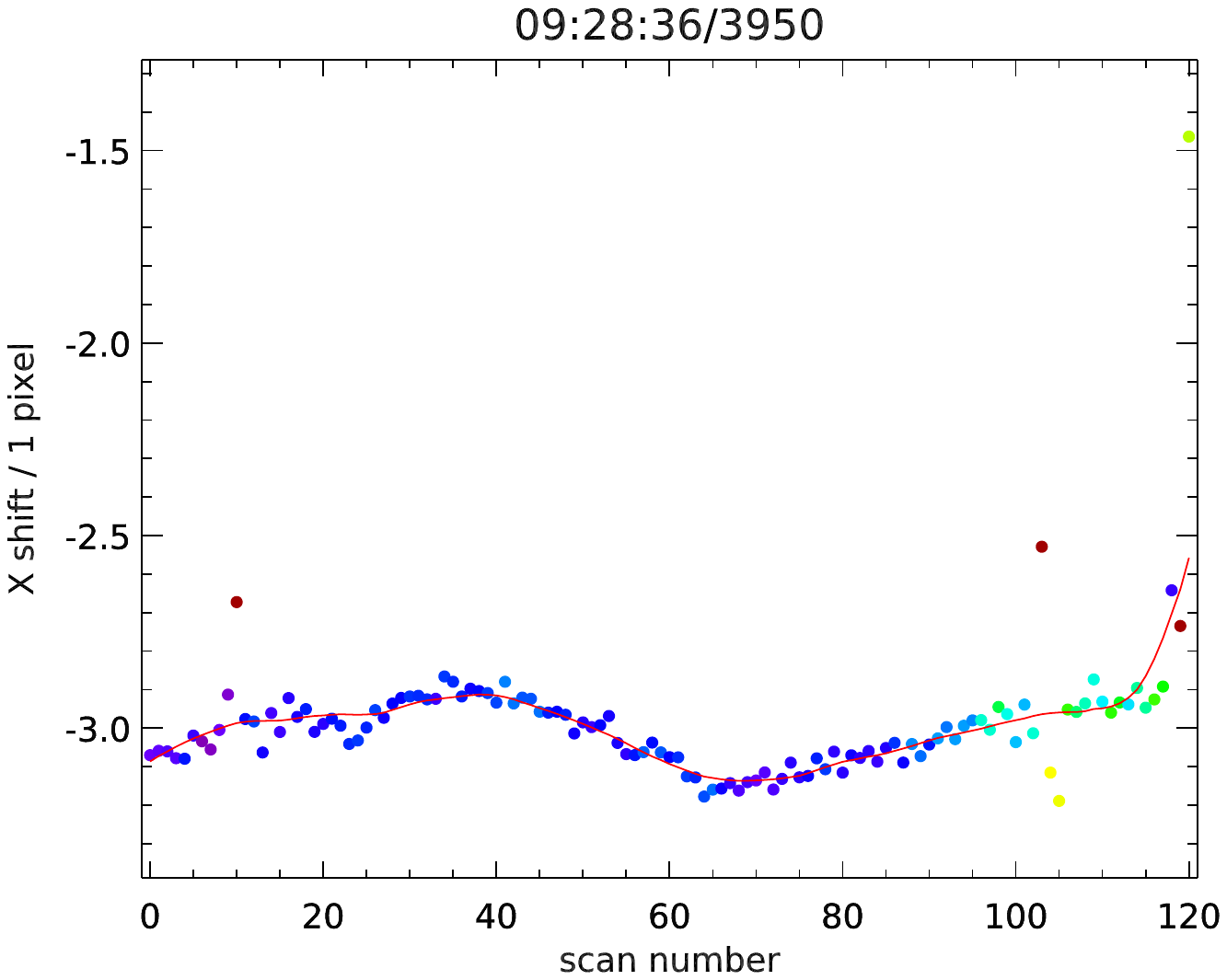}
\caption{X misalignment}
\label{fig:contalignX}
\end{subfigure}
\
\begin{subfigure}{0.32\linewidth}
\includegraphics[viewport=64 47 720 529,clip,width=\linewidth]{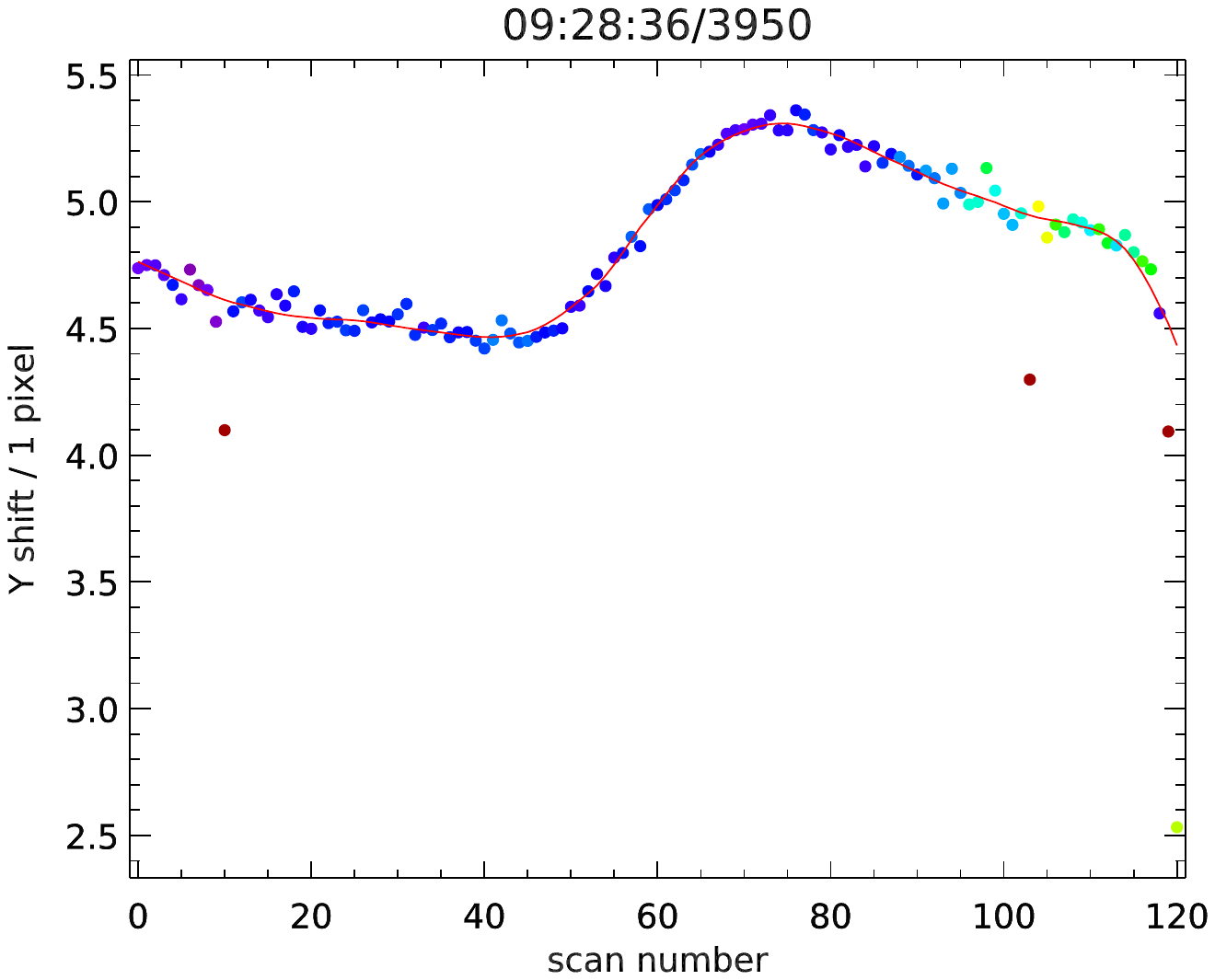}
\caption{Y misalignment}
\label{fig:contalignY}
\end{subfigure}
\caption{Continuum alignment diagnostic plots for 121 \ion{Ca}{ii}
scans collected from 09:28:36 on 2016 June 19, covering
approximately one period of the undulations shown in
Fig.~\ref{fig:contalign-temp}. \textbf{(a)} WB RMS contrast.
\textbf{(b)} Continuum/WB X misalignment. \textbf{(c)}
Continuum/WB Y misalignment. Filled circles in
\subref{fig:contalignX} and \subref{fig:contalignY}: Misalignment
measured by use of cross-correlation with MOMFBD restored data,
colors represent RMS contrast as shown in \subref{fig:contalignR}.
Lines in \subref{fig:contalignX} and \subref{fig:contalignY}:
Shifts in X and Y used for alignment, misalignment smoothed with a
heuristic algorithm that gives more weight to the high-contrast
data points. Diagnostic plots like these are automatically
produced by SSTRED to allow the user to verify how well the
contrast-aware smoothing worked.}
\label{fig:contalign}
\end{figure*}

Regardless of the source of the dispersion, we measured the
misalignment not only with WB and continuum, but also with images in
the \ion{Ca}{ii}~K blue wing and \ion{Ca}{ii}~H red wing (formed at
a similar height in the solar atmosphere) and found that it varies
linearly with wavelength over the relevant range.

SSTRED includes a procedure for measuring and correcting the
misalignment of the MOMFBD-restored \ion{Ca}{ii} scans. The process
needs to be repeated for each scan because of the temporal variations.
The misalignment between the WB and NB continuum images is measured by
cross-correlation in the MOMFBD-restored data. The misalignments at
the \ion{Ca}{ii}~H wavelengths are then calculated by linear
interpolation between the WB 395~nm and the NB continuum 399~nm.
Similarly, linear extrapolation gives the misalignment at the
\ion{Ca}{ii}~K wavelengths.

Because the cross-correlation measurements depend on the data
quality, they are smoothed with a method that ignores outliers
and weights data points with respect to image RMS contrast. The
variations around the smoothed line (disregarding the low-contrast
outliers) suggest that the precision is a few tenths of a pixel, see
Fig.~\ref{fig:contalign}.

Correcting this misalignment after MOMFBD restoration is much easier
and less time consuming than doing it in the raw data. The
misalignment is only a few pixels, which is small in comparison to the MOMFBD
subfield size. The MOMFBD assumption of a common wavefront in WB and
NB is therefore not significantly violated.

\subsection{Polarimetric cross-talk}
\label{sec:polar-cross-talk}

The telescope polarization model limits the accuracy of the
demodulation. In the resulting Stokes cubes, there is cross-talk
between the components. Due to the larger numbers in the $I$ component
compared to the differential $Q$, $U$, and $V$ components, the most
significant cross-talk is from $I$ to the other components.
This is corrected for similarly to the procedure used by
\citetads{1992ApJ...398..359S}.

If the FOV is free from polarization signal, any structures
that match $I$ in the differential components are cross-talk and the
rest is noise. With this assumption and $X \in \{Q,U,V\}$, the
correction for the $X$ component can be written as
\begin{equation}
\label{eq:14}
X \leftarrow X - w_X \cdot I
,\end{equation}
where the weight $w_X$ can be estimated with a least-squares fit to
the data in one or a few tuning states in (or as close as possible to)
the continuum. We calculate
\begin{equation}
w_X =  \langle I , X \rangle \bigm/ \langle I, I \rangle .
\label{eq:13}
\end{equation}
However, in our implementation, this is not exactly a least-squares
fit. In the usual $\ell^2$ inner product definition,
$ \langle A, B \rangle \propto \operatorname{mean}(A \cdot B) $, we
replaced the mean value with the histogram peak position, estimated
with a Gaussian fit. This made the procedure more robust with respect
to magnetic signal not completely removed by the masking, as well as
artifacts that often occur near the edges of the FOV.

In addition, when evaluating Eq.~(\ref{eq:13}), any spots, pores, or
other areas where a significant magnetic field can be expected are
masked from the FOV. A mask is automatically constructed based on the
bisquare weights (see the \texttt{biweight\_mean} function in the
IDL\-Astro library) of the pixels in a time-averaged Stokes $I$ image.
The user can then adjust this mask or construct one from scratch by
deselecting regions in a GUI.

The corrections in Eq.~\ref{eq:14} are then applied to all tuning
states. The weights are usually on the order $\pm10^{-3}$.

\subsection{Orientation and pointing}
\label{sec:spatial-coordinates-1}

The final science data cubes are by default oriented so that the pixel axis
directions are along the HPLN and HPLT (solar longitude and latitude)
coordinate axes. The second axis points toward solar North. The
spatial coordinates can be calibrated with SDO/HMI continuum images as
reference.

To calibrate the field rotation, we use datasets with some pores or
spots in the FOV, features for which the orientation can be
recognized in the lower resolution of HMI. See Fig.~8 in the CRISPRED
paper for a comparison of CRISP and HMI images of the same FOV.

The pointing information logged by the primary image guider
\citep[PIG; ][]{sliepen13primary} is estimated to be correct to within
5\arcsec{} if a four-limb calibration of the PIG/turret
system was performed close in time before the data were collected but
could otherwise be off by on the order of an arcminute. HMI continuum
images with spots and pores can also be used to refine the pointing
metadata in the spatial WCS information.

\section{Image restoration}
\label{sec:image-restoration}

\subsection{MFBD and MOMFBD}
\label{sec:mfbd-momfbd}

Isoplanatic image formation with optical aberrations can be modeled as
a convolution of an object with a PSF, where the PSF is determined by
the variation of the wavefront phase over the pupil. The wavefront can
be expanded into linear combinations of a finite set of modes that
span a subspace of all possible aberrations. Multi-frame blind
deconvolution (MFBD) methods, including phase diversity (PD, see
Sect.~\ref{sec:phase-diversity}), are based on estimating the
wavefront parameters by fitting the model to a data set, where the
assumption that multiple images depict the same object is a powerful
constraint (see, e.g., \citeads{2002SPIE.4792..146L}, and references
therein). With the assumption of additive white noise, the intensity
of the unknown object can be calculated by deconvolution of the data
so that the pixel values do not have to be estimated as independent model
parameters.

The wavefront aberrations are caused by turbulence in the Earth's
atmosphere, randomly mixing air with different temperature and
therefore different refractive index. We therefore have to make the
assumption that the exposures are short enough that the PSF does not
have time to change significantly. Due to turbulence at high
altitudes, the line of sight from the telescope to different parts of
the FOV passes through different turbulent structures, which means that the
image formation is not really isoplanatic. However, within
sufficiently small subfields, the assumption of isoplanatic image
formation is a good approximation. We solve the model fitting and
deconvolution problem independently within multiple subfields and form
restored versions of the full FOV by mosaicking the results from the
individual subfields.

\begin{figure}[ht]
\includegraphics[viewport=136 424 364 670, width=\linewidth]{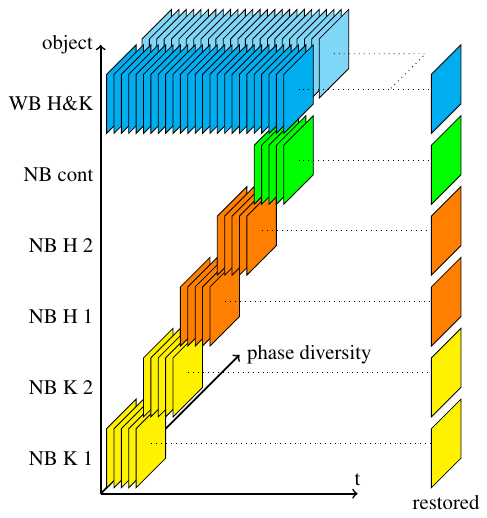}
\caption{Schematic representation of CHROMIS datasets for MOMFBD
processing. The depicted mock dataset has an NB continuum tuning
point (green), two tuning points each in an NB K filter (yellow)
and an NB H filter (orange). The H\&K WB filter (blue) anchors the
entire scan. This meas that it forces MOMFBD to align the restored images
of the scan.}
\label{fig:momfbd-setup-regular}
\end{figure}

For imaging spectro(polari)meters, each NB wavelength tuning and
polarization state produces its own cospatial ``object''. The number
of collected frames in each state is then determined by a trade-off:
we need many frames to boost the S/N, particularly in the core of deep
lines, but we wish to complete a full scan before the solar scene
evolves too much. Usually, only a few frames per state are collected,
too few for MFBD to effectively restore the images independently for
the individual NB states. In addition to the NB images, we also
collect WB images in synchronization with the NB images, with the
result that an NB line scan is always accompanied by a simultaneous WB
data set that spans the entire scan. The MOMFBD algorithm uses the WB
and NB data together to make a joint restoration
(\citeads{2002SPIE.4792..146L}; \citeads{2005SoPh..228..191V}). A
CHROMIS MOMFBD dataset is illustrated in
Fig.~\ref{fig:momfbd-setup-regular}, CRISP is similar but without a PD
channel (see Sect.~\ref{sec:phase-diversity}). The WB anchor images
aid the image restoration in two ways: 1) the large number of images
of the same object and many realizations of the random wavefronts
constrain the solution, and 2) the estimated wavefront tilt components
align all the WB images and therefore also the NB images to the WB.
The result is that the different NB images are also aligned to each
other.

Some particular challenges related to MOMFBD image restoration with
spectropolarimetric data are described by
\citetads{2008A&A...489..429V}, \citetads{2011A&A...534A..45S}, and in
Sect.~4.4 of the CRISPRED paper.
We use a MOMFBD code not published before (see
Appendix~\ref{sec:redux-code}) and a few developments in how the
MOMFBD processing is set up and run. The latter are described in the
following subsections.

\subsection{Phase diversity}
\label{sec:phase-diversity}

The PD wavefront-sensing and image-restoration technique is a form of
MFBD that is constrained by the intentional defocusing of one or more
images, corresponding to a known parabolic difference in phase over
the pupil. The technique was invented by
\citetads{1982OptEn..21..829G} and the theory was clarified and
extended to multiple diversities and multiple exposures by
\citeauthorads{1992JOSAA...9.1072P} (\citeyearads{1992JOSAA...9.1072P})
and (\citeyear{paxman92phase}). It was
independently developed for high-resolution solar data by
\citetads{1994A&AS..107..243L} and \citetads{1994SPIE.2302..268S}.
Both implementations were verified versus each other and versus speckle
interferometry by \citetads{1996ApJ...466.1087P}.
\citetads{2002SPIE.4792..146L} formulated it in terms of linear
equality constraints (LECs), thereby incorporating it into what later
became the MOMFBD method.

The diversity in phase does not have to be focus. However, this is
easily implemented and by far the most commonly used. The magnitude of
the phase difference also does not have to be known, although it does
constrain the solution much better if it is.

\begin{figure}[!htb]
\begin{subfigure}[t]{\linewidth}
\includegraphics[viewport=146 25 710 514,clip,width=\linewidth]{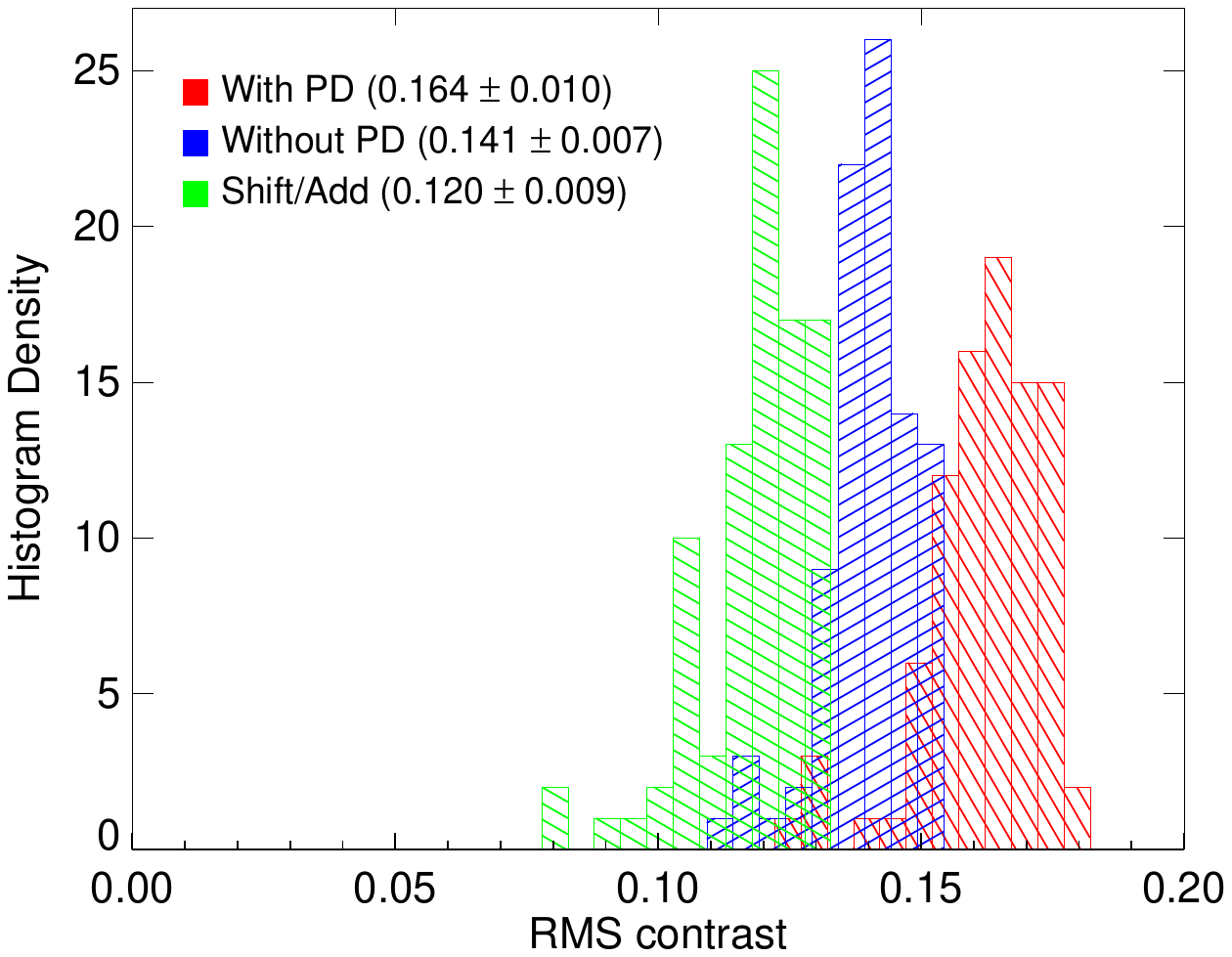}
\caption{RMS contrast histograms. The numbers in parentheses are
the median $\pm$ the (robust) standard deviations of the
contrasts.
\label{fig:pd-compare-plot}}
\end{subfigure}
\\[2mm]
\begin{subfigure}[t]{\linewidth}
\includegraphics[width=\linewidth]{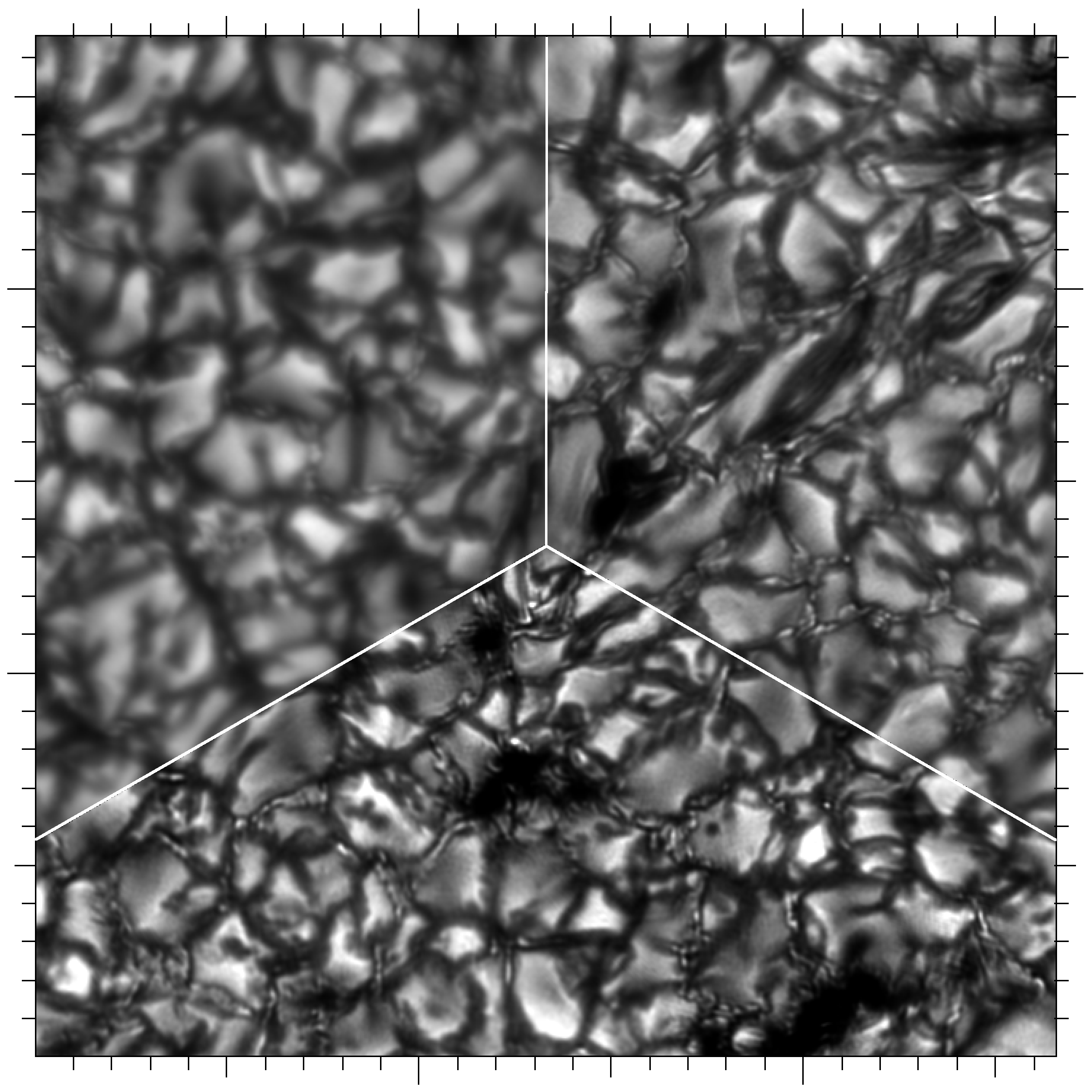}
\caption{One example of the FOV used in
(\subref{fig:pd-compare-plot}), comparing the three
different levels of processing. The tick marks are 1\arcsec{}
apart. Top left:~ Shift-and-add and MTF correction. Top
right:~60-mode MOMFBD. Bottom:~60-mode MOMFBD with PD.
\label{fig:pd-compare-image}}
\end{subfigure}
\caption{Contrast and image quality after different levels of MOMFBD
restoration of CHROMIS \ion{Ca}{ii} continuum images from 110
scans collected on 2016 September 19. The FOV is 700$\times$700 pixels
(26\farcs5$\times$26\farcs5). Processing levels: MOMFBD with and
without PD using 60~KL modes, and without PD using only tilt modes
(corresponding to shift-and-add by subfield and MTF correction).}
\label{fig:pd-compare}
\end{figure}

CHROMIS includes a WB PD camera. See Fig.~\ref{fig:pd-compare} for a
demonstration of the image quality obtained with three different
MOMFBD restorations of CHROMIS Ca H+K WB data: 60-mode MOMFBD with and
without PD, and MOMFBD without PD and only the two tip and tilt modes
corrected, corresponding to shift-and-add together with correction for
the theoretical, aberration-free modulation transfer function (MTF) of
the telescope. Both 60-mode restorations bring out fine structure that
is not visible in the MTF-corrected image, but PD improves the
contrast further, although it is still far from the expected
\ion{Ca}{ii} continuum granulation RMS contrast in excess of 27\%
\citepads{2019A&A...626A..55S}.

\subsection{Modes}
\label{sec:modes}

By default, we use Karhunen--Lo\`eve (KL) modes to parameterize the
unknown wavefronts to be estimated by the MOMFBD processing. Like
Zernike polynomials, they are orthogonal on a circular pupil. In
addition, they are statistically independent with respect to wavefronts
from atmospheric turbulence. As the SST adaptive optics (AO) monomorph
deformable mirror is designed to correct such atmospheric wavefronts,
the KL modes are also well suited to correcting residual wavefronts
from the AO.

The KL modes implemented are based on expansion in Zernike polynomials
\citepads{1990OptEn..29.1174R} and are therefore indexed as the dominating
Zernike polynomial as ordered by \citetads{1976JOSA...66..207N}. In
this order, the expected variance from turbulent seeing does not
decrease monotonically. To make the best subset selection, we therefore have
to order them first. Then we select the most significant subset by
truncating the list to the desired length.

We used to sort the selected subset of KL-modes back in index order,
for no better reason than to make the list look cleaner. We have since
realized that this is not optimal with respect to an important aspect
of the inner workings of the MOMFBD (and REDUX) program: In order to
stabilize the solution, a few iterations are run with first 5, then
10, then 15, etc., modes until all the specified modes are included.
The purpose is to allow the most significant modes to determine the
coarse shape of the wavefront before the finer features are fit.
This internal increase in the number of modes should therefore also be
performed in variance order, which means the selected set should be
specified in variance order.

\subsection{Residual alignment and destretching}
\label{sec:extra-wb-objects}

With perfectly estimated wavefronts, the anchor WB images should force
the restored images of all objects to be perfectly aligned. However,
model fitting with real data is rarely perfect, so that there are
small but noticeable residual misalignments between the NB images.
There is also residual stretching (from anisoplanatic wavefront
aberrations) on scales smaller than the subfields that MOMFBD cannot
compensate for.

\citetads{2012A&A...548A.114H} included a technique for a post-MOMFBD
dewarping step that compensates for the residual geometrical
differences between the NB images in a scan. This has since been used
routinely in the CRISPRED pipeline, see Sect.~4.4.2 of the CRISPRED
paper. The method is based on the MOMFBD program outputting additional
deconvolved WB images, in addition to the restored anchor WB image
based on all exposures in the scan. The additional WB images, one per
NB state, are generated by deconvolving only the subset of the raw WB
images that are simultaneous to the raw NB images of that state. The
additional deconvolved WB images are locally distorted in exactly the
same way as the deconvolved NB images and have the same residual
misalignment. Unlike NB images from different states, however, these
WB images can be successfully destretched against the anchor WB image
to measure the local warping, which is then applied to the NB images
to obtain a significantly better aligned data cube.

We modified the way to make the MOMFBD program generate these
additional WB images. We used to specify them as part of the MOMFBD
data set, using LECs to make their estimated wavefronts identical to
those of the simultaneous NB images. With zero weight in the error
metric, mathematically, they should not affect the solution. However,
numerically, it turns out that adding these additional objects, even
with zero weights, does have a small but measurable effect on the
converged solution. We attribute this to the changed LECs and to the
null spaces they generate.

The solution is modified slightly when zero-weight objects are added
to the LECs, but there is another downside to the previous approach.
The MOMFBD code treats the additional objects just as any other, that
is, it will load and process the WB data twice (once for the anchor
object and once for the additional WB objects), which leads to a
significant increase in both RAM usage and CPU load. For both CRISP
and CHROMIS, the number of data increases by 33\% by this method
(three cotemporal exposures are basically turned into four).

To overcome these issues, REDUX has a new mechanism in which the
additional WB objects are not specified with LECs. Instead, REDUX can
be configured to generate the desired images as a final step after the
wavefronts have been estimated. At this point, it deconvolves subsets
of the WB images that match the frame numbers of the NB images. As no
modifications are made to the problem itself, this will not interfere
with the solution at all.

In practice, although the converged solutions differ, the quality of
the restored data is similar. The real advantage of not including the
additional WB objects in the problem with LECs is that fewer gradients
of the error metric need to be calculated. For a typical CRISP or
CHROMIS data set with three cameras, this means a $\sim$25\% reduction
in the processing time.

\section{Metadata}
\label{sec:metadata}

For metadata, we follow the SOLARNET recommendations of
\citetads{2020arXiv201112139H}. The idea is that data should have
enough metadata for ``a Solar Virtual Observatory (SVO) [to] be able
to [\dots] find sets of successful observations matching a
hypothetical ideal observation proposal (if such observations
exist).'' The SOLARNET recommendations are an attempt to standardize
metadata for multiple telescopes and instruments so that searches can
be ``instrument agnostic''. The metadata should also facilitate
correct interpretation of data by a researcher who was not involved in
the observations or the pipeline processing. That is, they should be
complete enough that all important circumstances of the observations
are included. Moreover, the processing done by the pipeline should be
documented in the metadata, in particular when a processing step might
be run with different parameter settings.

We store science-ready data in FITS format
\citepads{2010A&A...524A..42P} with metadata information in the form
of FITS header keywords and extensions.
Some information is difficult to fit in standard FITS header keywords.
A number of conventions have therefore been developed to facilitate
storing particular types of information in FITS files. We use
the record-valued keywords of
\citet{calabretta04representations_distortions} to represent
coordinate distortions. We also use the JavaScript Object Notation
\citep[JSON;][]{RFC7159} to encode FITS keywords with structured
contents as strings, such as the SOLARNET \texttt{PRPARAn} keywords.
JSON was developed for JavaScript, but is now implemented in all major
programming languages. Finally, we use the SOLARNET variable keywords
(\citealp{haugan15metadata}; \citeyearads{2020arXiv201112139H}), a
mechanism for associating auxiliary data with the main data in a FITS
file for intensity statistics and some metadata from the log files of
auxiliary instruments. The SOLARNET variable keywords are stored in
FITS binary-table extensions and are associated with single-value
regular header keywords. The regular header keyword value is not
defined, but it is supposed to represent the whole cube in a way that
makes sense for the particular quantity.

\subsection{SOLARNET compliance and support}
\label{sec:solarnet-compliance}

A FITS file can be partially or fully SOLARNET compliant. For partial
compliance, it is enough to include four mandatory header keywords,
and if one of the keywords is different from the definitions in the
SOLARNET recommendations, a fifth keyword with a list of these
different keywords. For full compliance, a larger set of mandatory
keywords must be used depending on the nature of the observation. The
level of SOLARNET compliance itself is indicated with the mandatory
header keyword \texttt{SOLARNET} and values of 0.5 for partial
compliance and 1 for full compliance.

The SSTRED SOLARNET-compliant science data cubes are supported by
CRISPEX \citepads{2012ApJ...750...22V} version 1.7.4 (released in
January 2018) or later, see also Sect.~\ref{sec:crispex}. SOLARNET
compliance should make it easier for future data viewing and analysis
programs to support data from multiple sources.

In addition to SSTRED, several data pipelines are or will be preparing
their metadata in compliance with the SOLARNET recommendations. The
sTools pipeline is partially SOLARNET-compliant
\citepads{2018ApJS..236....5D}. The pipeline of SPICE produces
SOLARNET-compliant output. The Astronomical Institute of the Slovak
Academy of Sciences (AISAS) is updating their data pipelines,
developed for the Coronal Multi-channel Polarimeter for Slovakia
\citep[CoMP-S;][]{kucera10CoMP-S} and the Solar Chromospheric Detector
(SCD; \citeads{2015IAUGA..2246687K}) instruments of the Lomnick{\'y}
{\v S}t{\'i}t Observatory, according to the SOLARNET Metadata
Recommendations (J.~Ryb{\'a}k and P.~G{\"o}m{\"o}ry 2020, priv.
comm.). The recently funded Solar Activity Monitor NETwork (SAMNET)
project are planning to follow the SOLARNET recommendations for their
data structure (R.~Erd{\'e}lyi 2020, priv. comm.). The Royal
Observatory of Belgium (ROB) are aiming at producing, from early 2021,
image datasets from the Uccle Solar Equatorial Table (USET;
\citeads{2005AnGeo..23.3115B}) in white light, H$\alpha$, and
\ion{Ca}{ii}\,K with SOLARNET-compliant headers (R.~Vansintjan 2020,
priv. comm.). The Solar ALMA Pipeline (SoAP;
\citeads{2019asrc.confE.126S}) will follow the SOLARNET
recommendations \citepads{2020arXiv201112139H}.

This is evidence that the SOLARNET metadata recommendations have
gained momentum and do provide a means of metadata standardization.
This should significantly aid the analysis of data, in particular the
co-analysis of data from multiple instruments and telescopes.

\subsection{Observations}
\label{sec:observations}

Passing metadata from the observations through the SSTRED processing,
including the many calibration steps, requires access to metadata
pertaining to the details of the observations. These metadata have to be
collected from multiple sources, such as log files from auxiliary
instruments as well as the raw image data files themselves.

\subsubsection{Image data}
\label{sec:image-data}

The metadata collected from the raw data files include prefilter,
tuning, and polarimetry states, the detector that was used, time
stamps, and exposure times.
CHROMIS raw data are stored in multi-frame FITS files with metadata in
the headers, and CRISP raw data are stored one frame per file in an
ANA fz format. The latter format has less strictly structured headers
and many metadata are therefore encoded in the file names.

The reading of metadata from the raw files in a structured way is a
major bottleneck in SSTRED processing, in particular for CRISP, where
each frame has a file header to be read and a file name to be parsed.
We therefore started experimenting with ways of speeding this up.

A first attempt was to implement a dynamically loadable module (DLM)
that could be used to cache this information, so that SSTRED at least
did not have to read the same information more than once in the same
IDL session. This was mainly useful for development and testing, but
did not really help regular users, who generally only process the same
data once.

Recently, we implemented code that reads all the raw-file metadata
already at the time of the observations, and writes it in a database.
For users with access to the database, this is a very significant
speed-up in SSTRED processing that depends on the file system and the
current load. A few tests with reading metadata for 50,000 CRISP
frames (corresponding to 7.5 min of observations) took 20--45~min from
the files and 25--40~s from the database. For large datasets, we
therefore have a reduction from hours to minutes by reading metadata
from the database.

When during normal operations, metadata for older data are accessed
that are not already in the database, SSTRED automatically falls back
to reading the metadata from the data files and then adds them to the
database (effectively serving as a cache that is persistent between
IDL sessions).

The database runs on a MariaDB server. To avoid redundancy, it is
organized in hierarchical, linked tables with columns of metadata
appropriate for each level, see Table~\ref{tab:database-tables}. In
addition, there are auxiliary tables with information about image
detectors and prefilters.

\begin{table*}[!t]
\centering
\caption{Metadata database tables}
\begin{minipage}[t]{1.0\linewidth}
\begin{tabular}{lp{30mm}p{105mm}}
\hline
\hline
\noalign{\smallskip}
Name & Range & Types of metadata \\
\hline
\noalign{\smallskip}
\texttt{datasets} & Datasets & Date and time of a data-set
start, type,\tablefootmark{a}
instrument name\\
\texttt{configs} & Group of files & Camera name,\tablefootmark{b}
gain, exposure time, frame
dimensions, number of frames
per file, average
cadence etc.\\
\texttt{bursts} & Burst of exposures\tablefootmark{c} & Date and time of
beginning of burst,
prefilter,
FPI tuning,
scan number,
frame number of first frame,
detector temperature.\\
\texttt{chromis\_frames} & Individual frames & Time, frame
number, intensity statistics.\\
\texttt{crisp\_frames} & Individual frames  & Like \texttt{chromis\_frames} + liquid crystal state\\
\texttt{crisp\_polcal\_frames} &Individual frames & Like \texttt{crisp\_frames} +
quarter-wave
plate angle,
linear
polarizer angle.\\
\hline
\end{tabular}
\end{minipage}
\tablefoot{Hierarchical tables, parents on top.\\
\tablefoottext{a}{Types of datasets are darks, flats, science, etc.}
\tablefoottext{b}{The camera name here is the functional name of a
camera, like Crisp-R (CRISP NB camera in the beam reflected by
polarizing beamsplitter), Chromis-W (CHROMIS WB camera), etc.}
\tablefoottext{c}{By burst we refer to a group of image frames with
identical tuning and polarization states, collected during a short
period of time, to be processed with the assumption that the Sun
does not change. For CHROMIS this is a single file, and for CRISP
it is a group of files.}}
\label{tab:database-tables}
\end{table*}

The tables have \emph{unique} constraints to prevent writing the same
information in multiple rows. They also have \emph{foreign key}
constraints that prevent deletion of information from connected rows
in different tables.

As of this writing, the database is only available for users at the
Institute for Solar Physics in Stockholm. We plan to make it more
widely available after a testing period.

\subsubsection{Auxiliary instruments}
\label{sec:auxil-instr}

We extract some metadata from auxiliary instruments for the available
points in time within the span of the science data in a data cube.
These metadata are stored as SOLARNET variable keywords.

From the turret system log file, we read the elevation angle.
The AO log file provides the AO lock rate as well as two estimates of
the Fried parameter $r_0$. One of the $r_0$ values measures mostly
ground layer seeing, while the other also includes high-altitude
seeing \citepads{2019A&A...626A..55S}.

\subsection{World Coordinate System}
\label{sec:wcs}

Following the SOLARNET recommendations, we use the World Coordinate
System (WCS) to describe the observations. The WCS is part of the FITS
standard 3.0 (\citeads{2010A&A...524A..42P}, their Sect.~8), and it
allows the specification of coordinates for all (multi-dimensional)
pixels of a data cube. More details can be found in a series of papers
started by \citetads{2002A&A...395.1061G} and continued by authors
cited in the subsections below.

For our five-dimensional science-ready data cubes, the relevant
coordinates are spatial, temporal, spectral, and polarimetric. The WCS
part of a sample FITS header produced by SSTRED is shown in
Fig.~\ref{fig:sample-wcs-fits-header}. As specified with the
\texttt{PS$i\_$0} keywords, all our coordinates (except Stokes) are
tabulated in a FITS extension named \texttt{WCS-TAB}, in columns
numbered as given by the \texttt{PV$i\_$3} keywords.

\begin{figure*}[!t]
{ \small
\begin{verbatim}
PC1_1   =              1.00000 / No rotations
PC2_2   =              1.00000 / No rotations
PC3_3   =              1.00000 / No rotations
PC4_4   =              1.00000 / No rotations
PC5_5   =              1.00000 / No rotations

CTYPE1  = 'HPLN-TAB'           / SOLAR X
CUNIT1  = 'arcsec  '           / Unit along axis 1
CNAME1  = 'Spatial X'          /
PS1_0   = 'WCS-TAB '           / EXTNAME; EXTVER=EXTLEVEL=1 is default
PS1_1   = 'HPLN+HPLT+WAVE+TIME' / TTYPE for column w/coordinates
PS1_2   = 'HPLN-INDEX'         / TTYPE for INDEX
PV1_3   =                    1 / Coord. 1 tabulated coordinate number
CRPIX1  =                    0 / Unity transform
CRVAL1  =                    0 / Unity transform
CDELT1  =                    1 / Unity transform
CSYER1  =                   60 / Orientation unknown

CTYPE2  = 'HPLT-TAB'           / SOLAR Y
CUNIT2  = 'arcsec  '           / Unit along axis 2
CNAME2  = 'Spatial Y'          /
PS2_0   = 'WCS-TAB '           / EXTNAME; EXTVER=EXTLEVEL=1 is default
PS2_1   = 'HPLN+HPLT+WAVE+TIME' / TTYPE for column w/coordinates
PS2_2   = 'HPLT-INDEX'         / TTYPE for INDEX
PV2_3   =                    2 / Coord. 2 tabulated coordinate number
CRPIX2  =                    0 / Unity transform
CRVAL2  =                    0 / Unity transform
CDELT2  =                    1 / Unity transform
CSYER2  =                   60 / Orientation unknown

CTYPE3  = 'WAVE-TAB'           / Wavelength, function of tuning and scan number
CNAME3  = 'Wavelength'         /
CUNIT3  = 'nm      '           / Wavelength unit, tabulated for dim. 3 and 5
PS3_0   = 'WCS-TAB '           / EXTNAME; EXTVER=EXTLEVEL=1 is default
PS3_1   = 'HPLN+HPLT+WAVE+TIME' / TTYPE for column w/coordinates
PV3_3   =                    3 / Coord. 3 tabulated coordinate number
CRPIX3  =                    0 / Unity transform
CRVAL3  =                    0 / Unity transform
CDELT3  =                    1 / Unity transform

CTYPE4  = 'STOKES  '           / Stokes vector [I,Q,U,V]
CRPIX4  =                    1 / First (and only) quantity is I
CRVAL4  =                    1 / First (and only) quantity is I
CDELT4  =                    1 / [1,2,3,4] = [I,Q,U,V]

CTYPE5  = 'UTC--TAB'           / Time, function of tuning and scan number
CNAME5  = 'Time since DATEREF, increases with dim. 3 and 5' /
CUNIT5  = 's       '           /
PS5_0   = 'WCS-TAB '           / EXTNAME; EXTVER=EXTLEVEL=1 is default
PS5_1   = 'HPLN+HPLT+WAVE+TIME' / TTYPE for column w/coordinates
PV5_3   =                    4 / Coord. 5 tabulated coordinate number
CRPIX5  =                    0 / Unity transform
CRVAL5  =                    0 / Unity transform
CDELT5  =                    1 / Unity transform

TIMESYS = 'UTC     '           /
DATEREF = '2016-09-19T00:00:00.000000' / Reference time in ISO-8601
OBSGEO-X=              5327386 /  [m] SST location
OBSGEO-Y=             -1718721 /  [m] SST location
OBSGEO-Z=              3051720 /  [m] SST location
\end{verbatim}
}
\caption{WCS part of sample FITS header without polarimetry.}
\label{fig:sample-wcs-fits-header}
\end{figure*}

\subsubsection{Spatial coordinates}
\label{sec:spatial-coordinates}

The WCS for spatial coordinates was defined by
\citetads{2002A&A...395.1077C}.
\citeauthorads{2006A&A...449..791T}
(\citeyearads{2006A&A...449..791T}) and
(\citeyearads{2010A&A...515A..59T}) extended the notation with
coordinates relevant to solar observations.

Our spatial coordinates are obtained from PIG. PIG fits a circle to
the circumference of the primary image of the solar disk, as projected
on the bottom plate of the telescope vacuum tube. Knowing the location
of the exit window, it infers the pointing with $\sim$10\arcsec{}
accuracy and $\sim$1\arcsec{} precision and logs it every second.
However, the accuracy depends on a recent calibration, with several
positions on the limb as reference. Failing that, the pointing can be
1--2\arcmin{} off.

It is rare, but PIG sometimes loses track of the solar disk for
extended periods of time. When it does, SSTRED falls back to the less
accurate spatial coordinates from the turret system, logged every
30~s.

The SSTRED interpolates the logged coordinates to obtain the pointing
for any point in time.

All frames that belong to the same scan are aligned by the MOMFBD
processing so that they all have a common FOV. The temporal alignment
procedure follows the features in the photospheric WB images, trying
to keep them stationary in the FOV. This means that the spatial
coordinates are constant during a scan but can change with scan number
due to the solar rotation and possibly the motion of the tracked
feature over the solar surface. However, with the expected limited
accuracy of the pointing information, we initially set the spatial
coordinates to the median of the logged pointing coordinates. Varying
coordinates can be introduced later with calibrations to other
instruments.

We specify the spatial coordinates by tabulating them for the array
corner pixels for each scan. A capable WCS reader can then interpolate
to obtain the coordinates for any pixel.

Before May 2020, SSTRED did not have an absolute calibration of the
orientation of the cameras with respect to the solar coordinate
system. It can now do this (also for older data) by fitting SST images
of a suitable target to simultaneous SDO/HMI continuum images. Before,
only the average of the four corner coordinates could be trusted as
the coordinates of the center of the FOV. The lack of orientation
information is signaled by setting the systematic accuracy (keyword
\texttt{CSYER$n$}) for the spatial coordinates to 60\arcsec{} or worse
(comparable to the entire FOV). When the calibration is complete,
\texttt{CSYER$n$} is set to 5\arcsec{} or less.

\subsubsection{Spectral coordinates}
\label{sec:spectral-coordinates}

The spectral coordinates \citepads{2006A&A...446..747G} are initially
set by the tuning sequence decided by the observer. Because they are
in general not equidistant, they have to be tabulated.

Small adjustments of the wavelength scale are made based on a
calibration of disk center data versus an atlas spectrum (see
Sect.~\ref{sec:fitprefilter}). The cavity errors (see
Sect.~\ref{sec:cavity-errors}) are stored as distortions to the
wavelength coordinate by use of notation that extends the WCS
formalism, see Appendix \ref{sec:wavel-dist}.

Before we implemented a correction of the wavelengths based on the
spectral calibration (Sect.~\ref{sec:fitprefilter}), the wavelength of
the line core had to be calibrated for each target individually, and
it depends on the position of the observed target on the solar disk.
For example, \citetads{2011A&A...528A.113D} provided a calibration
method based on synthetic spectra derived from 3D MHD simulations of
the solar photosphere, and later studies by
\citeauthorads{2017A&A...607A..12L}
(\citeyearads{2017A&A...607A..12L}) and
(\citeyearads{2018A&A...611A...4L}) proposed to calibrate observations
using laser-based measurements.

Recent atmospheric model inversion codes such as NICOLE
\citepads{2015A&A...577A...7S} and the STockholm Inversion Code (STiC;
\citeads{2016ApJ...830L..30D}) need information about pixel- and
line-dependent wavelength shifts originating from cavity errors, for
instance. When the observations all have the same cavity errors, a
correction can be applied afterward to the estimated velocities. For
multi-line and multi-instrument data, however, the inversions need
instrumental profiles customized for each spectral region and pixel.

\subsubsection{Polarimetric coordinates}
\label{sec:polar-coord}

In the WCS, the polarimetric coordinates for Stokes vectors are
defined as indices $[1,2,3,4]$ for the $[I, Q, U, V]$ components
\citepads{2002A&A...395.1061G}. The Stokes $Q$, $U$, and $V$
components are defined with respect to a frame of reference, such as a
Cartesian coordinate system with axes $(x,y,z)$, where $Q>0$ ($Q<0$)
denotes linear polarization along the $x$-axis ($y$-axis), $U>0$
($U<0$) denotes linear polarization along the line $y=x$ ($y=-x$), and
$V>0$ ($V<0$) denotes right-handed (left-handed) circular polarization
around the $z$-axis.

Our reference system is oriented so that $x$ is parallel to the HPLT
axis (solar north) and $y$ is antiparallel to the HPLN axis, which
forms a right-handed system together with a $z$-axis pointing along
the propagation axis. However, the WCS surprisingly does not seem to
have a general notation to describe the orientation of the Stokes
reference system.

Two Stokes reference system conventions are in use in the night-time
community, both based on the equatorial coordinate system (Right
Ascension and Declination), but with the third axis pointing in
different directions (toward or away from the observer, respectively).
To break the ambiguity, \citet{gorski10healpix} defined the FITS
header keyword \texttt{POLCCONV} for the Hierarchical Equal Area
iso-Latitude Pixelization (HEALPix) data structure that is used for
sky maps from various sources. Their \texttt{POLCCONV} has two
possible string values, \texttt{'IAU'} and \texttt{'COSMO',} for these
two conventions.

Equatorial coordinates are not practical for high-resolution solar
observations, where we are more interested in coordinates on the solar
disk than in the Sun's position on the sky. The latest version of the
SOLARNET recommendations \citepads{2020arXiv201112139H} recommends a
right-handed coordinate system with the $z$-axis oriented toward the
observer, as for the IAU convention \citep{trans_IAU_15_commission40}.
The axes are specified with an extended use of \texttt{POLCCONV} in
the form \texttt{'(±XXXX,±YYYY,±ZZZZ)'}, where \texttt{XXXX},
\texttt{YYYY}, and \texttt{ZZZZ} are names of WCS coordinates, and a
plus (minus) means that the axis is parallel (antiparallel) to the
named coordinate.

Thus, we specify our reference system with \texttt{POLCCONV =
\allowbreak '(+HPLT,\allowbreak -HPLN,\allowbreak +HPRZ)'}, where
\texttt{HPRZ} is defined as the axis pointing toward the observer
\citepads{2006A&A...449..791T}.

This use of \texttt{POLCCONV} could serve as a general notation. For
example, the right-handed equatorial IAU system mentioned above can be
specified as \texttt{POLCCONV = \allowbreak '(+RA-{}-,\allowbreak
+DEC-,\allowbreak -DIST)'}, where \texttt{DIST}, defined as the
``radial distance from the observer'', points away from the observer.
The left-handed COSMO system is then \texttt{POLCCONV = \allowbreak
'(+RA-{}-,\allowbreak +DEC-,\allowbreak +DIST)'}.

When developing notation for the \texttt{POLCCONV} keyword, we
initially specified only the first two axes. For SST data, this means
that when the third axis is omitted, \texttt{+HPRZ} is implied.

\subsubsection{Temporal coordinates}
\label{sec:temporal-coordinates}

Temporal WCS coordinates are described by \citetads{2015A&A...574A..36R}.

Due to varying delays from FPI tuning and prefilter changes during
spectral scans, the temporal coordinates in our data
cubes are not equidistant and therefore have to be tabulated. Time is
read from the metadata of each raw-data frame and averaged for the
frames that were combined (by MOMFBD restoration) to make a frame in
the data cube.

Time is advanced not only from one scan to the next, but also from one
tuning position to the next. Polarimetric states are varied several
times per tuning state, so that the restored images are made from data
with overlapping time. These images are then mixed in approximately
equal proportions by demodulation into Stokes components. Because of
this, we assign the same time coordinate to all Stokes components of
the same tuning state.

\subsubsection{Telescope location}
\label{sec:telescope-location}

The telescope location is specified in three dimensions with the
keywords \texttt{OBSGEO-X}, \texttt{OBSGEO-Y}, \texttt{OBSGEO-Z}
(\citeads{2015A&A...574A..36R}, their Sect.~4.1.3). We calculated the
values of these keywords (as shown in
Fig.~\ref{fig:sample-wcs-fits-header}) following their recipe. As
input we used the geodetic location of the SST (latitude, longitude,
altitude) = ($28\fdg759693$, $-17\fdg880757$, 2380~m).

These numbers come from two services provided by Google. The latitude
and longitude is available with high precision by selection of a
location in Google
Maps.\footnote{\url{https://goo.gl/maps/XwGtEU6ueZv}} They also run a
separate elevation
service,\footnote{\url{https://developers.google.com/maps/documentation/javascript/examples/elevation-simple}}
by use of which the altitude of the SST location was determined to a
few meters above 2360~m. Allowing for the SST tower, the adopted
altitude value is 2380~m.
The accuracy is estimated to $\sim$10~m, which should be sufficient
for calculations of relative speeds versus the Sun.

\subsection{Statistics}
\label{sec:statistics}

The SOLARNET recommendations include the statistics metadata keywords
in Table~\ref{tab:stats-keywords}. One purpose of this is to enable
including intensity statistics in data archive search criteria.

\begin{table}[!tbh]
\centering
\caption{Statistics metadata keywords}
\begin{tabular}{lp{65mm}}
\hline
\hline
\noalign{\smallskip}
Keyword & Description \\
\hline
\noalign{\smallskip}
\texttt{DATAMIN} & The minimum data value. \\
\texttt{DATAMAX} & The maximum data value. \\
\texttt{DATAP}\textit{nn} & the \textit{nn} percentile, where
$\textit{nn} \in \{01, 02, 05, 10,$ $ 25, 75,90, 95, 98, 99\}$. \\
\texttt{DATAMEDN} & The median data value $\equiv$ the 50 percentile. \\
\texttt{DATAMEAN} & The average data value. \\
\texttt{DATANRMS} & The normalized RMS deviation from the mean. \\
\texttt{DATASKEW} & The skewness. \\
\texttt{DATAKURT} & The excess kurtosis. \\
\texttt{DATAMAD}  & The mean absolute deviation from the mean. \\
\texttt{NDATAPIX} & The number of data pixels, excluding padding
and other missing-data pixels. \\
\hline
\end{tabular}
\tablefoot{We use a subset of the statistics metadata keywords from
the SOLARNET recommendations.}
\label{tab:stats-keywords}
\end{table}

In addition to the keywords in Table~\ref{tab:stats-keywords}, there
is a \texttt{DATARMS} keyword in the SOLARNET recommendations.
However, because \texttt{DATANRMS} is just \texttt{DATARMS} normalized
with \texttt{DATAMEAN}, it would be redundant to include all three
quantities. In our opinion, \texttt{DATANRMS}, corresponding to what
is commonly known as the RMS contrast, is the more useful quantity.
Potential uses are searches in an SVO, for data quality in quiet-Sun
data, and perhaps in active regions to detect transient features such
as flares.

These statistical keywords are stored as SOLARNET variable keywords,
one value per frame in science data cubes. These values are easy to
calculate as the cube is written to file, frame by frame.

We also need regular header keyword values that accompany these
variable keywords and represent statistics for the whole cube. The
minimum and maximum values for the whole cube are easy to calculate,
they are just the minimum and maximum of the per-frame minimum and
maximum values. Moments and percentiles require special attention
because their calculation with built-in IDL commands requires that the
whole cubes are in memory. The NB cubes are potentially too large for
this (this is the reason CRISPEX needs the two differently ordered
versions of the cubes for fast access, see
Sect.~\ref{sec:data-cube-ordering}). We therefore need alternative
ways to compute them.

\citet{pebay16numerically} reported numerically stable one-pass
expressions for the mean, $\bar x = \frac{1}{n}\sum_{i=1}^nx_i$, as
well as the central moments,
$\mu_p=\frac{1}{n}\sum_{i=1}^n(x_i-\bar x)^p$ for any $p\ge2$, from
which RMS, skewness, and kurtosis can be calculated. However, we wish
to calculate the quantities for the whole cube from those of the
individual frames, therefore we instead use their ``pairwise''
formulas to combine the per-frame quantities as described in
Appendix~\ref{sec:moments}.

The cube percentiles cannot be calculated from the per-frame
percentiles. Instead, we calculate (approximate) percentile values
(including the median) from a cumulative histogram for the entire
cube, accumulated frame by frame, using a large number of bins (now
$2^{16}=65536$ bins). The percentile values can then be found as the
data values corresponding to the first bins that exceed the
percentiles. The accuracy of percentiles calculated in this way should
be on the order of the bin size (i.e., the range of values in the cube
divided by the number of bins). Assuming the data values are evenly
distributed within the bins, we further improve the accuracy by at
least an order of magnitude by use of interpolation.

The mean absolute deviation from the mean (MAD) was a late addition to
the included statistics. Like the percentiles, it cannot be calculated
for the whole cube from the per-frame values. This value is instead
built up during the histogram accumulation pass through the cube, the
mean being known from the first pass. The fact that we make two passes
removes one advantage with the pair-wise expressions in
Appendix~\ref{sec:moments}, as we could then also use the two passes
for the moments. If computing time is of the essence, a simple mean of
the per-frame values for the regular header keyword might be
considered.

\section{CRISPEX}
\label{sec:crispex}

An important part of the workflow with complex data products like the
science-ready data cubes produced by SSTRED is being able to view and
analyze them. Not a part of SSTRED per se, CRISPEX
\citepads{2012ApJ...750...22V} is an IDL tool that offers such data
cube browsing and analysis functionality. Details such as the
changelog, how to keep an up-to-date distribution of CRISPEX, and a
short usage tutorial can be found on the CRISPEX github
page.\footnote{See \url{https://github.com/grviss/crispex}.}

From version 1.7.4, released in January 2018, CRISPEX supports the
SOLARNET-compliant science data cubes output by SSTRED.
CRISPEX was originally developed for CRISP science data cubes stored
in LP format files. Along with the fz format, the LP (for ``La
Palma'') format has been used for SST/SVST data since at least the
early 1990s.
CRISPEX was earlier extended to support data from IRIS
\citepads{2014SoPh..289.2733D}, reading both its Level 3 spectrograph
(SG) FITS cube files and the Level~2 slit-jaw image (SJI)
files\footnote{See also the IRIS Technical Notes (ITN) 11 and 12 --
concerning data levels and header keywords, respectively -- on the
IRIS website: \url{https://iris.lmsal.com/documents.html}.}. In
fact, CRISPEX can handle any synthetic or observational data cube,
provided it has been formatted according to either the legacy LP
format, the SOLARNET-compliant FITS format, or the IRIS FITS format
(e.g., IRIS SJI-formatted SDO and Hinode data, or IRIS SG-formatted
Bifrost cubes).

The richer metadata in the SOLARNET cubes should result in an improved
user experience compared to the LP format cubes (and to some extent,
also compared to the IRIS format), both when calling the program and
during run-time. During run-time, the changes are visible mostly when
dealing with the WCS information, see Sect.~\ref{sec:wcs}. For
instance, while the LP and IRIS formats provide only scan-averaged
time, the SOLARNET format's WCS information contains image timing
information as function of wavelength tuning and modulation state (for
tuning instruments) or of spatial position and modulation state (for
(scanning) slit spectrographs). The time-dependent spatial position is
also available, which allows accounting for solar rotation during the
observations when returning the solar spatial coordinates. This is
particularly beneficial for browsing multi-instrument datasets (e.g.,
CRISP plus CHROMIS or CRISP/CHROMIS plus IRIS). First, because the
image to be displayed (determined through nearest-neighbor
interpolation in time) can be selected more closely in time when
considering tuning filter instruments or (rastering) spectrographs.
Second, because creating co-aligned cubes might in principle be
skipped, assuming the (time-dependent) spatial coordinates are well
defined: the pixel size difference, $xy$-translation, FOV rotation
with respect to solar north and solar rotation would automatically be
taken into account during run-time through the WCS information of the
respective files.

Moreover, to accommodate the larger CHROMIS images (larger than CRISP
and IRIS), the zooming functionality has been extended to allow zoom
true to size (i.e., 1:1 data-vs-monitor pixel scale) also on screens
that would normally not fit the image. This could in the future also
be useful for visualizing data from the Daniel~K.~Inouye Solar
Telescope (DKIST; \citeads{2020SoPh..295..172R}) or the European Solar
Telescope (EST; \citeads{2016SPIE.9908E..09M}), which are set to
deliver 4k$\times$4k images or larger.

The metadata needed, particularly the coordinate system of the data,
is provided in different ways for the different file formats. For
smooth display and for plotting of large data cubes, the data
themselves need to be ordered optimally for fast access. Details are
given in Appendix~\ref{sec:crispex-details}.

\section{Discussion}
\label{sec:discussion}

\subsection{Processing}
\label{sec:processing}

The SSTRED is a data processing pipeline that processes raw data from
CRISP and CHROMIS and outputs science data cubes in FITS files.

We identified two deficiencies with CHROMIS data that we had not
encountered in CRISP data. Wavelength- and time-dependent dispersion
in both atmosphere and optics caused misalignment within \ion{Ca}{ii}
scans. An alignment procedure was implemented as part of SSTRED.
Wavelength-dependent dispersion also appears to limit the accuracy of
the intensity calibration, which is measured in the WB but is applied to all
wavelengths.
The cause for these two issues is the wide wavelength range in
\ion{Ca}{ii} scans compared to other line scans collected with CRISP
and CHROMIS, together with the large wavelength gradient of the
dispersion in the 400~nm region.

Creating science cubes from MOMFBD-restored images involves multiple
spatial corrections, such as alignment, rotation, temporal
destretching to remove anisoplanatic distortions, and inter-tuning
destretching to remove MOMFBD-residual geometrical distortions. Some
of these operations require one correction to be applied before
another is measured, causing multiple consecutive interpolation
operations. As each interpolation is a blurring operation, we wish to
minimize the number of such operations. Instead of performing these
resampling operations consecutively, SSTRED can combine measured image
shifts, rotations, and spatial distortions into a single interpolation
operation, resulting in less blurring and fewer artifacts. Future
pipeline developers, designing code from scratch, should build in these
capabilities at a low level, perhaps by making images class
objects. The objects could include both original image data, a version
with operations so far performed on them, as well as a record of these
operations. When further operations are added, a new version
can be calculated in this way, using the original data and the combined operations.

Limb data are not as well supported as on-disk data.
This has to do with difficulties with alignment. The alignment between
WB and NB outside the limb is hampered by weak signals in the WB and
continuum. On disk, there can be problems with low contrast of
features on the disk and also with finding a rectangular subfield on
the disk on which to perform the cross correlation.

\subsection{Metadata}
\label{sec:metadata-1}

The SSTRED processes metadata, and the output science data cubes are
compliant with the SOLARNET metadata recommendations as formulated by
\citetads{2020arXiv201112139H}. This is meant to facilitate interpretations
as well as the search for science data in future Solar Virtual
Observatories (SVOs). As part of SOLARNET compliance, we specify the
coordinates of the output data cubes within the WCS system.

The number of pipelines that produce SOLARNET-compliant output is
growing. Any new instrument and pipeline should implement this from
the start. The metadata and SVO upload will facilitate the
findability, accessibility, interoperability, and reusability
\citepads[FAIR;][]{2016NatSD...360018W} of the data.

Metadata processing posed a few new challenges to the pipeline
development, leading to new developments.
The time-consuming reading of header data led to the implementation of
a database for raw-data metadata.
Cavity errors, pixel-dependent deviations from the nominal wavelength
coordinate, required an extension of an already proposed notation for
WCS coordinate distortions. While not (yet) part of WCS proper,
adoption of this notation by other pipelines would simplify the task
of writing software for the analysis of data from multiple
instruments.
By storing intensity statistics in the metadata, frame by frame as
variable keywords, we can calculate some of the statistics (the
moments) of data cubes from the statistics of the individual frames.
An SVO that implements this capability can facilitate searches by
quickly calculating statistics of subsets of data cubes from those of
the individual frames.

\subsection{Development}
\label{sec:development}

The SSTRED was first developed for CHROMIS data, generalizing the old
CRISPRED pipeline and adding metadata support. It should be fairly
easy to extend it to process data from other, similar instruments.

The development was made by a very small team after CHROMIS was
installed.
The development of the IDL parts of SSTRED has occupied one of us
(MGL) for a large fraction of his time since the installation of
CHROMIS. Another (TH) spent considerable time on REDUX. Together with
the time spent by the remaining authors, and counting also the
development of CRISPRED prior to the CHROMIS installation, this
represents an effort of several person-years. This is costly, but an
effort that is clearly needed to maximize the scientific output of
modern ground-based solar instrumentation.

Development of new calibrations has lead to new recommendations for
SST observations, such as collecting small datasets of disk-center,
quiet-Sun data every now and then throughout the observing period to
facilitate the intensity calibration. Artifacts are to be expected, and
it is important to identify and characterize them as soon as possible
and if possible, develop procedures for compensating for them.

The SOLARNET recommendations include versioning of data files. We take
this to also include intermediate data files, such as calibration data
that can perhaps be fitted in various ways or are based on parameters
that depend on the specific use case of the data. To make the final
data completely specified, versions of these data files need to be
specified in the metadata.
This is rather complicated and probably needs to be implemented at the
core level of the pipeline. We have chosen not to do this, but
next-generation pipelines should be designed with this in mind.

We rely on a few external IDL libraries that are self-contained and
available through git version control (IDLAstro and Coyote) or direct
download from a web site (mpfit). We made an early decision not to
require users to install SolarSoft \citepads{1998SoPh..182..497F} to run
SSTRED.
While this prevented us from using the WCS routines of
\citet{thompson10solarsoft} within the pipeline, we still wished
SolarSoft users to be able to use its WCS support when reading SSTRED
output. Testing showed that while our five-dimensional data cubes
complied with the WCS definitions, SolarSoft could not perform the
required interpolation in more than three coordinate dimensions with
routines included in IDL. We found the multi-dimensional interpolation
implementation of \citet{smith03ninterpolate}, which is now included
in SolarSoft.

The SST granulation images can be aligned to HMI images with very god
precision, as demonstrated by R.~Rutten (2020, priv. comm.), but this
requires HMI data with high enough cadence so that a frame can be
found that is close enough in time for the granulation pattern not to
have changed too much. Not using SolarSoft within SSTRED also limited
our use of HMI data to 15-minute cadence publicity jpeg images
available through predictable URLs. We did not find a way to
programmatically download HMI 45-second cadence science data without
SolarSoft. This is the reason that our pointing calibration versus HMI
data so far only works for data with spots or pores in the FOV.

\subsection{Data access}
\label{sec:data-access}

Access to science-ready CRISP and CHROMIS data depends on the data
policies of the owners of the data. Data that belong to the Institute
for Solar Physics are currently proprietary, but are shared with
researchers based in Sweden. Data collected within the SOLARNET access
program are proprietary for one year after successful pipeline
processing and are then released to the solar community. Access to data
collected by other research groups depends on their data policies.

The SVO (see Sect.~\ref{sec:svo}) database has access information and
can serve download URLs. Our own data, as well as data collected under
the second SOLARNET project access program are hosted by Stockholm
University, from where they can be downloaded, depending on the release
status. The data download is straightforward for data that are
released to the solar community, but download of proprietary data is
password-protected.

The Institute of Theoretical Astrophysics in Oslo and the Lockheed
Martin Solar and Astrophysics Laboratory have released co-aligned
SST/IRIS datasets elsewhere \citepads{2020A&A...641A.146R}.

\subsection{Concluding remarks}
\label{sec:final-comments}

We have described the SSTRED data-processing pipeline. We also
demonstrated what needs to go into the pipelines of a major
ground-based solar telescope and its instruments.

A pipeline that encodes optimized processing of data and calibrations
is absolutely necessary for a nontrivial instrument to reach its full
scientific potential. Known sources of error need to be at least
characterized and when possible compensated for.

Well-defined metadata are crucial for data to be usable for scientists
with little knowledge about the instruments and the circumstances of
the observations. We aim to follow and establish mechanisms and
notation to fully characterize the output from the pipeline.

The SSTRED was developed while the instruments and the pipeline were
used, which means that early versions of the pipeline were less
capable than the current version. For future telescopes, we must
improve on this by having full support very soon after commissioning
and also very well controlled data-collection procedures. A pipeline
can only work with the data that were collected, and the quality of
the processed data strongly depends on the calibration data that are
collected. We cannot and should not stop development of better
processing procedures, therefore we should collect more calibration
and environment data than we know we need. As an example, having a log
of the temperature sensor data from the telescope bottom plate made it
much easier to understand the wavelength- and time-dependent
misalignment of \ion{Ca}{ii} CHROMIS data.

All software must be expected to have bugs. Therefore it is extremely
important to archive not only reduced data, but also raw data,
including calibration data, at least for observations that are of
lasting value and/or are difficult to repeat (unusual events, part of
long time-sequences, etc.). While developing SSTRED, we have often
found reasons to reprocess data, sometimes multiple times, because of
corrected bugs or the addition of improved procedures. Retaining this
possibility is one way of ensuring the best-quality data for
scientific interpretation.

In our opinion, SSTRED represents the most ambitious data-processing
pipeline for solar imaging instruments. It can thus serve as a
reference for future similar developments.

\begin{acknowledgements}
Guus Sliepen implemented the new data acquisition system used for
CHROMIS.
We have had valuable discussions about metadata and FITS headers
with Mats Carlsson and Bill Thompson.
We are grateful to several users
-- in particular Vasco Henriques and Pradeep Chitta -- for patiently
testing SSTRED and reporting bugs.
Lewis Fox helped with the coding in the early stages of development.
This work was carried out as a part of the SOLARNET project
supported by the European Commission’s 7th Framework Programme under
grant agreement No. 312495.
This research has received financial support from the European
Union’s Horizon 2020 research and innovation program under grant
agreement No. 824135 (SOLARNET).
The Swedish 1-m Solar Telescope is operated on the island of La
Palma by the Institute for Solar Physics in the Spanish Observatorio
del Roque de los Muchachos of the Instituto de Astrof{\'\i}sica de
Canarias.
The Institute for Solar Physics is supported by a grant for research
infrastructures of national importance from the Swedish Research
Council (registration number 2017-00625).
JdlCR is supported by grants from the Swedish Research Council
(2015-03994), the Swedish National Space Board (128/15) and the
Swedish Civil Contingencies Agency (MSB). This project has received
funding from the European Research Council (ERC) under the European
Union's Horizon 2020 research and innovation programme (SUNMAG,
grant agreement 759548).

\end{acknowledgements}

\balance

\appendix

\section{Pipeline code}
\label{sec:pipeline-code}

The non-MOMFBD parts of SSTRED are coded in IDL, with some parts
implemented as DLMs coded in C/C++ for speed and/or to use existing
code. SSTRED works with IDL version 8.3 or later.

The SSTRED IDL code is available from
\url{https://github.com/ISP-SST/sstred}. The hash of the latest
commit, tagged 1.1.0-634, is \texttt{g3a676b9f2e5e}.

The SSTRED uses IDL objects, with the major steps in the processing
implemented as class methods. There are separate classes for CRISP and
CHROMIS, inheriting common code from a top class whenever possible and
practical.

Adding support for another, similar instrument should not be very
hard. It is a matter of entering information about detectors, file
formats, and filters, identifying common processing steps, and writing
a few new class methods that cover the differences. Identifying any
particular data imperfections for the new instrument and developing
calibrations and corrections for them could be time-consuming.

\subsection{Required IDL libraries}
\label{sec:required-libraries}

The following IDL code/libraries are required:
\begin{description}
\item[IDLAstro:] We use code from the IDL Astronomy User's Library
\citepads{1993ASPC...52..246L}, mainly to manipulate FITS headers
and extensions. This library is available through
git\footnote{\url{git://github.com/wlandsman/IDLAstro.git}}. The
hash of the latest commit is \texttt{0da32286d31d}.
\item[Coyote:] We make use of mainly plotting routines from the Coyote
library \citep{fanning11traditional}. The code is available in a
maintained version as a git
repository\footnote{\url{https://github.com/idl-coyote/coyote}}. The
hash of the latest commit is \texttt{ff720efbfa9a}.
\item[Mpfit:] Many steps in SSTRED require nonlinear model fitting.
For some of them, we use the Levenberg--Marquardt algorithm as
implemented in the IDL mpfit routines by
\citetads{2009ASPC..411..251M}, available for download from the
author's web
site\footnote{\url{http://www.physics.wisc.edu/~craigm/idl/down/mpfit.tar.gz}}.
The current version is 1.85, from 2017-01-03. (A C version of mpfit
is also incorporated in some of the C/C++ code.)
\end{description}

\subsection{DLMs}
\label{sec:dlms}

Some parts of the pipeline are implemented as DLMs coded in C/C++.
There are several reasons for this, primarily because it gives access
to lower-level C++ functionality, which allows for optimization. There
is also the benefit of obtaining access to high-level C++ libraries,
such as boost and OpenCV. In particular, the pinhole calibration
mentioned in \ref{sec:camera-alignment} utilizes camera calibration
tools that are readily available in OpenCV.

The DLM also allows direct access to code intrinsic to the
REDUX/MOMFBD software (see the Sect.~\ref{sec:redux-code} for access
and version information). This means that we can be certain that function
calls within IDL are doing the exact same thing as the REDUX code
does.

Sample tasks implemented as DLMs:
\begin{itemize}
\item Sum images.
\item Read and mosaic MOMFBD output.
\item Read and write fz format files.
\item Do pinhole alignment.
\item Geometric transform of images.
\item Measurements and application of stretch vectors.
\item Log file conversion.
\item Some metadata handling.
\item Access to some internal REDUX MOMFBD code.
\end{itemize}

\subsection{REDUX code}
\label{sec:redux-code}

CHROMIS and CRISP data are restored from optical aberrations caused by
turbulence in the atmopshere and partially corrected for by the SST AO.
We used the MOMFBD code of \citetads{2005SoPh..228..191V} as the
workhorse for SST image data for several years. One of us (TH) now
maintains a fork of that project (called \emph{REDUX}), where parts of
the original code are replaced with open-source libraries.

REDUX implements several improvements and new features compared to
MOMFBD. This includes a method for making sparser null-space matrices
from the LECs, leading to a speed-up of $\sim$33\%. Another is the
re-implementation of the additional WB objects used for dewarping (see
Sect.~\ref{sec:extra-wb-objects}) with an additional $\sim$25\%
speed-up.

As with the old MOMFBD code, pinhole calibration is used to specify
the relative geometry between the cameras involved in a data set to be
restored. The REDUX code can read the new projective transforms
described in Sect.~\ref{sec:camera-alignment}. (It still supports the
xoffs/yoffs files of the old MOMFBD program.)

The REDUX code, including the DLMs, is available from
github\footnote{\url{https://github.com/ISP-SST/redux}}. The hash of
the latest commit, tagged 1.0.18-13, is \texttt{g27ba623e9b22}. See
the Redux wiki
page\footnote{\url{https://dubshen.astro.su.se/wiki/index.php/Redux}}
for installation and usage instructions.

\section{Wavelength distortions}
\label{sec:wavel-dist}

This appendix describes how SSTRED stores cavity errors (see
Sect.~\ref{sec:cavity-errors}) as distortions in the wavelength
coordinate.

The etalons are fixed with respect to the detectors, so that the cavity
errors can be described as distortions that are a function of the
spatial pixel coordinates. The distorted coordinate is not a pixel
coordinate, however, it is the wavelength in physical units, that is, the
wavelength world coordinate.

\citet{calabretta04representations_distortions} defined a notation for
describing coordinate distortions as part of the WCS. (Unlike the
other WCS papers we refer to, this is published only in draft form.
However, the notation is in fact used for data from the Hubble Space
Telescope \citep{hack12distortion} and will be used for SPICE. It is
also implemented in the WCSLIB C library
\citep{wcslib76}.\footnote{See
\url{https://www.atnf.csiro.au/people/Mark.Calabretta/WCS/index.html}.})
However, while this notation can specify distortions that are a
function of pixel coordinates, the coordinate being distorted also has
to be a pixel coordinate.

This would not be a problem if the wavelength pixel-coordinate grid
were equidistant in wavelength. Then a distortion in the tuning pixel
coordinate would be equivalent to a distortion in the wavelength world
coordinate. In our case, however, the wavelength is tabulated without
restriction on the spacing between tuning points. This would lead to
a table look-up that has a discontinuous derivative, defined by the
spacing in wavelength between the two nearest tuning points.
Alternatively, if the distortion were defined in the tuning pixel
coordinate, it would have to be specified for each tuning, resulting
in an additional dimension in the distortion table, and this would still
cause problems at tuning points surrounded by different-length
wavelength intervals.

We use an extended notation, described by
\citeauthorads{2020arXiv201112139H}
(\citeyearads{2020arXiv201112139H}, version 1.5, in prep.). This
notation allows distortions to be associated with and applied to any
of a set of numbered stages in the conversion of pixel coordinates to
world coordinates in the WCS. In contrast to the cases considered by
\citet{calabretta04representations_distortions}, the associate and
apply stages are specified individually. Below we give a brief
description focused on the parts used for the cavity errors.

The new notation generalizes the record-valued
\texttt{DP$j$}/\texttt{DQ$i$} keywords and the associated
\texttt{CPDIS$j$}/\texttt{CQDIS$i$} keywords of
\citet{calabretta04representations_distortions} and defines keywords
with similar meaning, but with \texttt{W} substituted for \texttt{P}
and \texttt{Q}. The \texttt{DW$j$} (and $i$) keywords have a few
additional records compared to \texttt{DQ$i$} and \texttt{DP$j$},
namely \texttt{DW$j\cdot$ASSOCIATE} and \texttt{DW$j\cdot$APPLY}, the
values of which are the relevant stage numbers. See
Fig.~\ref{fig:sample-wcs-dist-fits-header} for the corresponding part
of a FITS header.

\begin{figure*}[!t]
{ \small
\begin{verbatim}
CWERR3  =            0.0105497 / [nm] Max total distortion
CWDIS3  = 'Lookup  '           / WAVE distortions in lookup table

DW3     = 'EXTVER: 1'          / Cavity error for 3934
DW3     = 'NAXES: 5'           / 3 axes in the lookup table
DW3     = 'AXIS1: 1'           / Spatial X
DW3     = 'AXIS2: 2'           / Spatial Y
DW3     = 'AXIS3: 3'           / Tuning
DW3     = 'AXIS4: 4'           / Stokes
DW3     = 'AXIS5: 5'           / Scan number
DW3     = 'OFFSET3: 11.0421'   / Tuning coordinates offset
DW3     = 'SCALE3: 0.0475238'  / Tuning coordinates scale
DW3     = 'CWERR: 0.0139631'   / [nm] Max distortion (this correction step)
DW3     = 'CWDIS.LOOKUP: 1'    / Distortions in lookup table
DW3     = 'ASSOCIATE: 1'       / Association stage (pixel coordinates)
DW3     = 'APPLY: 6'           / Application stage (world coordinates)

DW3     = 'EXTVER: 2'          / Cavity error for 3969
DW3     = 'NAXES: 5'           / 3 axes in the lookup table
DW3     = 'AXIS1: 1'           / Spatial X
DW3     = 'AXIS2: 2'           / Spatial Y
DW3     = 'AXIS3: 3'           / Tuning
DW3     = 'AXIS4: 4'           / Stokes
DW3     = 'AXIS5: 5'           / Scan number
DW3     = 'OFFSET3: -9.95792'  / Tuning coordinates offset
DW3     = 'SCALE3: 0.0475238'  / Tuning coordinates scale
DW3     = 'CWERR: 0.0105497'   / [nm] Max distortion (this correction step)
DW3     = 'CWDIS.LOOKUP: 1'    / Distortions in lookup table
DW3     = 'ASSOCIATE: 1'       / Association stage (pixel coordinates)
DW3     = 'APPLY: 6'           / Application stage (world coordinates)
\end{verbatim}
}
\caption{WCS distortions part of a sample FITS header with cavity
maps for two \ion{Ca}{ii} prefilters, 3934 and 3969. These lines
come after the \texttt{CDELT3} line in
Fig.~\ref{fig:sample-wcs-fits-header}.}
\label{fig:sample-wcs-dist-fits-header}
\end{figure*}

The cavity maps are a function of the spatial pixel coordinates at
stage~1 (the associate stage), but they are a distortion of the
spectral world coordinate at stage~6 (the applied stage).
Consequently, we set \texttt{DW3$\cdot$ASSOCIATE} to 1 and
\texttt{DW3$\cdot$APPLY} to 6.

Following \citet{calabretta04representations_distortions}, we store
the distortion array in a FITS image extension called
\texttt{WCSDVARR}. Multiple distortions, to be applied in sequence,
can be stored in multiple \texttt{WCSDVARR} extensions with different
version numbers (\texttt{EXTVER} keyword). Each distortion array can
be defined to apply to only a partial range of the apply coordinates.
For this, the \texttt{DW$j\cdot$SCALE$i$} and
\texttt{DW$j\cdot$OFFSET$i$} keywords are used (see
\citealp{calabretta04representations_distortions};
\citeads{2020arXiv201112139H} for details). We use this mechanism for
the multiple cavity maps resulting from CHROMIS line scans using
multiple prefilters (Sect.~\ref{sec:measurement}).

The SSTRED has code for reading at least the distortions written by SSTRED
itself. It should not be hard to port this to any language in which
the SSTRED output is to be read. Documentation can be found in the SST
wiki\footnote{\url{https://dubshen.astro.su.se/wiki/index.php/SSTRED}}.
The corrected wavelength coordinate for a data pixel,
$\lambda_\text{corr}(i_\text{x},i_\text{y},i_\text{tun},i_\text{pol},i_\text{scan})
= \lambda(i_\text{tun}) +
\delta\lambda(i_\text{x},i_\text{y},i_\text{tun})$, where
$\lambda(i_\text{tun})$ is the nominal tuning wavelength, and
$\delta\lambda$ is the contents of the image extension
\texttt{WCSDVARR}.

\section{CRISPEX-supported file formats}
\label{sec:crispex-details}

\begin{table*}[!t]
\caption{FITS
header keywords required by CRISPEX}
{\small
\begin{tabularx}{\textwidth}{lcccX}
\hline \hline \noalign{\smallskip}
Keyword  & SN & SG & SJ & Description and use \\
\hline \noalign{\smallskip}
\texttt{BITPIX} & X & X & X & {Number of bits per pixel; is
converted to IDL datatype and used to
initialize variables holding the data slices} \\
\texttt{NAXIS$i$} & X & X & X & {Numbered variables specifying the
size of each dimension} \\
\texttt{BTYPE} & X & X & X & {Description of data type, e.g.,
intensity, temperature, pressure, etc.; used for plot labeling
and user feedback} \\
\texttt{BUNIT} & X & X & X & {Data units, e.g.,
erg\,cm$^{-2}$\,s$^{-1}$\,sr$^{-1}$, K, m\,s$^{-1}$, etc.; used
for plot labeling and user feedback} \\
\texttt{BSCALE} & X &---& X & {Data scaling factor; used in
combination with \texttt{BZERO} to descale the data} \\
\texttt{BZERO} & X &---& X & {Data scaling offset; used in
combination with \texttt{BSCALE} to descale the data} \\
\texttt{CDELT$i$} & X & X & X & {Pixel size of each dimension in
terms of \texttt{CUNIT$i$}; used for WCS transformations
(pixel-to-WCS and  vice versa)} \\
\texttt{CRPIX$i$} & X & X & X & {Reference pixel of each
dimension; used for WCS transformations} \\
\texttt{CRVAL$i$} & X & X & X & {Reference pixel value of each
dimension in terms of \texttt{CUNIT$i$}; used for WCS
transformations} \\
\texttt{CTYPE$i$} & X & X & X & {Description of data type of each
dimension; used for plot labeling and user feedback (the
\texttt{CTYPE$i$} corresponding to the X-axis is also checked
for containing '\texttt{TAB}', to determine further processing)} \\
\texttt{CUNIT$i$} & X & X & X & {Data unit of each dimension
(e.g., arcsec, \AA ngstr\"om, seconds, etc.); used for plot
labeling and user feedback} \\
\texttt{PC$i\_j$} & X & X & X & {Elements of the PC-matrix; used
in WCS transformations} \\
\texttt{OBSID} &---& X & X & {IRIS observing program ID;
provided in user feedback} \\
\texttt{STARTOBS} & P & X & X & {Date and time of observations
start; used to determine the timing in UTC when combined with
second (first) auxiliary extension data of a SG (SJI) file.} \\
\texttt{DATE\_OBS} & X & X & X & {Date and time of observations
start; used if \texttt{STARTOBS} is not defined} \\
\texttt{INSTRUME} & X & X & P & {Instrument that produced the
data; used to (un)set the non-equidistant spectral warping
switch for IRIS SG data (unset if not equal to
'\texttt{IRIS}')} \\
\texttt{NWIN} &---& X &---& {Number of diagnostics
compressed in the wavelength dimension} \\
\texttt{WSTART$i$} &---& X &---& {Starting index of each
diagnostic; used in combination with \texttt{WWIDTH$i$} to
determine diagnostic boundaries within the spectral dimension} \\
\texttt{WWIDTH$i$} &---& X &---& {Width in pixels of each
diagnostic; used in combination with \texttt{WSTART$i$} to
determine diagnostic boundaries within the spectral dimension} \\
\texttt{WDESC$i$} &---& X &---& {Label of each diagnostic; used
in labeling plots and control panel selection options} \\
\texttt{TWAVE$i$} &---& X &---& {Central wavelength of each
diagnostic; used to determine the Doppler velocity} \\
\texttt{TDESC1} &---&---& X & {Label of detector; used for
control panel labeling of SJI functions} \\
\texttt{TELESCOP} & P &---& X & {Telescope name; used for
control panel labeling of SJI functions} \\
\hline \noalign{\smallskip}
\texttt{TIME} &---&---& X & {Time since {\tt STARTOBS}, keyword
indicates the row-index of the extension data array;
used to determine the timing in UTC} \\
\texttt{XCENIX} &---&---& X & {Time-dependent central FOV
X-value in {\tt CUNIT1} (row-index); used to determine the SJI
\texttt{CRVAL1}} \\
\texttt{YCENIX} &---&---& X & {Time-dependent central FOV
Y-value in \texttt{CUNIT2} (row-index); used to determine the
SJI \texttt{CRVAL2}} \\
\texttt{PC$i\_j$IX}&---&---& X & {Time-dependent PC-matrix
elements (row-index); used determine the SJI \texttt{CRPIX/2},
and by extension \texttt{CRVAL1/2}} \\
\texttt{SLTPX1IX}&---&---& X & {Time-dependent slit center
X-pixel (row-index); used to determine the SJI \texttt{CRPIX1}
and \texttt{CRVAL1}} \\
\texttt{SLTPX2IX}&---&---& X & {Time-dependent slit center
Y-pixel (row-index); used to determine the SJI \texttt{CRPIX2}
and \texttt{CRVAL2}} \\
\hline
\end{tabularx}}
\tablefoot{Columns 2--4 indicate whether a particular keyword is
required (marked X), present but not required / actively used
(marked P), or absent (marked ---) for the SOLARNET (SN), IRIS
spectrograph (SG), and slit-jaw image (SJ) files, respectively. The
keywords above the horizontal line are from the main data extension
header; those below the line correspond to the first auxiliary extension
header. For tabulated WCS coordinates, the keywords
\texttt{CDELT$i$}, \texttt{CRPIX$i$}, and \texttt{CRVAL$i$} are
defined differently, along with \texttt{PS$i$\_$j$} and
\texttt{PV$i$\_$j,$} they are used to find the relevant data in
binary extensions (see \citeads{2006A&A...446..747G}).}
\label{tab:crispex_fitshdr_keywords}
\end{table*}

\subsection{Metadata}

The CRISPEX program needs metadata to correctly handle and
display the data. In particular, it needs coordinates: spatial,
temporal, spectral, and polarimetric.

As the legacy LP format cubes come with a minimalistic header
(describing only the cube dimensions and the data type), the
CRISPEX IDL command takes a number of keywords, some are
Boolean flags (to obtain a particular behavior), and some are used to
supply auxiliary data files with additional information (e.g., time in
seconds and wavelength values).

For FITS cubes -- both of the IRIS and the SOLARNET compliant
varieties -- CRISPEX (un)sets these switches and populates
auxiliary information variables automatically from the file metadata,
thereby simplifying the call sequence.
Table~\ref{tab:crispex_fitshdr_keywords} lists which header keywords
are used (and how) by CRISPEX. Details are given in
Sects.~\ref{sec:iris-fits-files} and \ref{sec:solarn-compl-fits}
below.

\subsubsection{IRIS FITS files}
\label{sec:iris-fits-files}

Most header keywords are used to determine or read the spatial WCS
information that can then be processed by the SolarSoft IDL functions
\texttt{wcs\_get\_coord()} and \texttt{wcs\_get\_pixel()} to go
back and forth between cube pixel and data value, especially for the
coordinate transform between files of differently sized FOVs. Most
other header keywords are used to obtain correct labeling in plot
windows and control panel selection options, or as switches to
enable/disable certain behavior. The remainder are to deal with
multiple spectral windows for the particular case of IRIS Level 3
data, where the wavelength axis is usually a concatenation of
noncontiguous spectral diagnostic windows corresponding to the
various observed lines.

IRIS Level 3 files contain four extensions, the first three of which
are used by CRISPEX: (0) the main data (in \texttt{BUNIT}),
(1) the wavelength array (in \texttt{CUNIT3}), and (2) the timing
array (in \texttt{CUNIT4} since \texttt{STARTOBS}).
IRIS Level 2 SJI files contain three extensions, of which CRISPEX only
uses the first two. The main extension again contains the main data.
The first auxiliary extension holds the time- and raster
step-dependent information on the slit.

\subsubsection{SOLARNET-compliant FITS files}
\label{sec:solarn-compl-fits}

The SOLARNET-compliant FITS cubes use WCS for all coordinates, see
Sect.~\ref{sec:wcs}. The WCS information is in main header keywords
and/or binary extensions, depending on whether it is on regular grids
or has to be tabulated.

CRISPEX does not access the WCS headers and extensions directly, but
rather retrieves the coordinate information through the SolarSoft
functions \texttt{fitshead2wcs()} and \texttt{wcs\_get\_coord()}.
CRISPEX ensures loading a modified version of
\texttt{wcs\_proj\_tab} as well as auxiliary routines (all provided
with the CRISPEX distribution) that allow interpolation in the
five-dimensional coordinate look-up table when accessed through
\texttt{wcs\_get\_coord()}.

\subsection{Data cube ordering}
\label{sec:data-cube-ordering}

Regardless of the file format (LP, IRIS-style FITS, or SOLARNET
FITS), CRISPEX expects a certain data-cube ordering for it to
correctly access the data when the cursor is moved or changed, for
example, the frame number or wavelength tuning position.
CRISPEX cubes can be written as three-, four- or five-dimensional cubes, but are
upon access considered to be three-dimensional, a sequence of
two-dimensional frames with any higher dimension combined, or
``folded'' into a third dimension.

At this time, the ordering in the image cubes and spectral cubes is
hard-coded. Should the need arise, CRISPEX could be updated to obtain the
ordering information from WCS \texttt{CTYPE}\textit{n} header
keywords.

\subsubsection{Image cubes}
\label{sec:image-cubes}

From the point of view of CRISPEX, the basic data cube is a
sequence of two-dimensional images. With notation from
Sect.~\ref{sec:post-momfbd}, its dimensions are
$[N_\text{x},N_\text{y},N_3]$, where
$N_3=N_\text{tun} \cdot N_\text{pol} \cdot N_\text{scan}$.
CRISPEX will then subscript the third dimension with index
$i_3 = i_\text{scan} \cdot N_\text{tun} \cdot N_\text{pol} + i_\text{pol} \cdot N_\text{tun} + i_\text{tun} $
to obtain the $xy$-image at spectrotemporal position
$(i_\text{tun},i_\text{pol},i_\text{scan})$.

\begin{figure}[tb]
\includegraphics[viewport=0 5 595 284,clip,width=\linewidth]{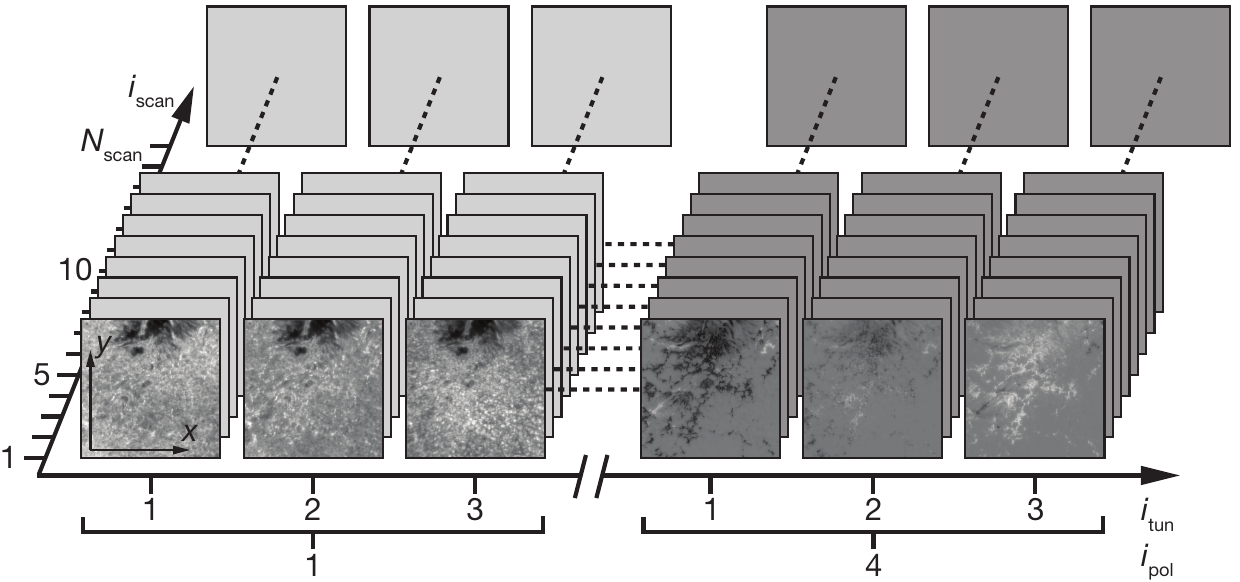}
\caption{Schematic representation of the CRISPEX data ordering of a
spectrotemporal Stokes cube with three tuning positions. The third
dimension has been ``unfolded'' into a tuning/polarimetry axis and
a scan axis, showing that the data are ordered sequentially as
wavelength scans for each Stokes parameter separately before
proceeding to the next scan. On the polarimetry axis,
$i_\text{pol}=[1,2,3,4]$
$\Leftrightarrow \text{Stokes}\ [I,Q,U,V]$ ($Q$ and $U$ not shown
in the figure).}
\label{fig:crispex_dataorder}
\end{figure}

In words, intensity images are stacked according to wavelength tuning
first, Stokes parameter second, and scan number (i.e., repetition)
third. See Fig.~\ref{fig:crispex_dataorder} for an illustration
involving polarimetric data.

\subsubsection{Spectral cubes}
\label{sec:spectral-files}

For single-wavelength time-series or single-wavelength scans, all
dimensions present can be accessed quickly enough so that an image cube
on its own suffices. However, for multi-dimensional cubes (e.g.,
time series of wavelength scans, or time series of imaging
spectropolarimetry), traversing the image cube to extract the local
spectrum or construct the spectrum as function of time for any pixel
during run-time would take a considerable amount of time, and the
recommended procedure therefore is to create a reordered, so-called
``spectral'', data-cube file for swifter access to the time-dependent
spectra.

A CRISPEX spectral cube has dimensions
$[N_\text{tun}, N_\text{scan}, N_3]$, where
$N_3 = N_\text{x} \cdot N_\text{y} \cdot N_\text{pol}$. The
spectrum-time data frame at position
$(i_\text{x},i_\text{y},i_\text{pol})$ is obtained by indexing the
third dimension with
$i_3 = i_\text{y} \cdot N_\text{x} \cdot N_\text{pol} + i_\text{x} \cdot N_\text{pol} + i_\text{pol}$,
that is, spectrum-time diagrams are stacked according to Stokes parameter
first, followed by the two spatial dimensions, $x$-coordinate second,
and $y$-coordinate third.

\section{Calculation of mean and moments}
\label{sec:moments}

In this appendix we detail how we calculate the mean, variance (and
its positive square root, the standard deviation from the mean),
skewness, and excess kurtosis for the SOLARNET statistical keywords.
We wish to do this for a data cube or a subset thereof, without having
the entire cube in memory at the same time, by combining the same
statistics calculated for the individual frames.
This would potentially also be useful for an SVO that supports serving
subsets of data defined on the fly.

Following \citet{pebay16numerically} we write the mean of a set of $n$
data points as
\begin{equation}
\bar x = \sum_{i=1}^n x_i\end{equation}
and define the central moments for any $p\ge2$ as
\begin{equation}
\mu_p=M_p/n,
\label{eq:22}
\end{equation}
where
\begin{equation}
M_p = \sum_{i=1}^n (x_i - \bar x)^p.
\label{eq:23}
\end{equation}
We partition the $n$ data points into two subsets $\mathcal A$ and
$\mathcal B$ with $n_{\mathcal A}$ and $n_{\mathcal B}$ data points,
respectively, and define $\bar x_{\mathcal A}$ ($\bar x_{\mathcal B}$)
and $M_p^{\mathcal A}$ ($M_p^{\mathcal B}$) as $\bar x$ and $M_p$,
but for the subset $\mathcal A$ ($\mathcal B$). We also define
\begin{equation}
\delta_{\mathcal B,A} = \bar x_{\mathcal B} - \bar x_{\mathcal A}.
\label{eq:26}
\end{equation}

With these definitions, \citet{pebay16numerically} write ``pairwise
and update formulas'' for the mean value of the $n$ data points as the
following combination of the means for the two subsets,
\begin{equation}
\bar x = \bar x_{\mathcal A} + \frac{n_{\mathcal B}}{n} \delta_{\mathcal B,A},
\label{eq:27}
\end{equation}
and $n$ times the central moments as
\begin{align}
M_p &= M_p^{\mathcal A} + M_p^{\mathcal B}
+ n_{\mathcal A}\left( \frac{-n_{\mathcal B}}{n} \delta_{\mathcal B,A} \right)^p
+ n_{\mathcal B}\left( \frac{ n_{\mathcal A}}{n} \delta_{\mathcal B,A} \right)^p \notag\\
& \quad + \sum_{k=1}^{p-2} \binom{p}{k} \; \delta_{\mathcal B,A}^k
\left[ M_{p-k}^{\mathcal A}\left(\frac{-n_{\mathcal B}}{n}\right)^k
+ M_{p-k}^{\mathcal B}\left(\frac{n_{\mathcal A}}{n}\right)^k\right].
\label{eq:25}
\end{align}

We use equations (\ref{eq:27}) and (\ref{eq:25}) repeatedly to
incrementally calculate the statistics for the whole cube one frame at
a time.
The mean, $\bar x$, as defined above for a data cube with $n$ pixels,
is what we need. The variance, skewness, and excess kurtosis are
not the same as the central moments, $\mu_p$, however, although the conversions
are easy enough. By definition, the unbiased estimate of the variance
can be written as
\begin{equation}
\sigma^2 = \frac{1}{n-1}\sum_i (x_{i} - \bar x_m)^2 \equiv \mu_2 \frac{n}{n-1}
\label{eq:3}
\end{equation}
and the standard deviation is its positive square root, $\sigma$. The
Fisher--Pearson coefficient of skewness is defined as
\begin{align}
s = \frac{1}{n}\sum_i \left(
\frac{x_{i} - \bar x}{\sigma}
\right)^3  \equiv \mu_3 \sigma^{-3}.
\label{eq:6}
\end{align}
The kurtosis for a normal distribution is 3. It is common to subtract
3 to obtain 0 for the normal distribution. The result is referred to
as the excess kurtosis,
\begin{align}
k = \frac{1}{n}\sum_i \left(
\frac{x_{i} - \bar x}{\sigma}\right)^4  - 3 \equiv \mu_4\sigma^{-4} - 3.
\label{eq:9}
\end{align}
This is the quantity returned by the IDL \texttt{kurtosis()} and
\texttt{moment()} functions, as well as by default by the python
\texttt{scipy.stats.kurtosis()} function. \footnote{See
\url{www.harrisgeospatial.com/docs/moment.html} and
\url{docs.scipy.org/doc/scipy/reference/generated/scipy.stats.kurtosis.html}.}


\begin{thebibliography}{88}
\expandafter\ifx\csname natexlab\endcsname\relax\def\natexlab#1{#1}\fi

\bibitem[{{Berghmans} {et~al.}(2005){Berghmans}, {van der Linden}, {Vanlommel},
{Warnant}, {Zhukov}, {Robbrecht}, {Clette}, {Podladchikova}, {Nicula},
{Hochedez}, {Wauters}, \& {Willems}}]{2005AnGeo..23.3115B}
{Berghmans}, D., {van der Linden}, R.~A.~M., {Vanlommel}, P., {et~al.} 2005,
Annales Geophysicae, 23, 3115

\bibitem[{{Bose} {et~al.}(2021){Bose}, {Joshi}, {Henriques}, \& {Rouppe van der
Voort}}]{2021A&A...647A.147B}
{Bose}, S., {Joshi}, J., {Henriques}, V. M.~J., \& {Rouppe van der Voort}, L.
2021, \aap, 647, A147%

\bibitem[{Brault \& Neckel(1987)}]{brault87spectal}
Brault, J.~W. \& Neckel, H. 1987, {Spectral Atlas of Solar Absolute
Diskaveraged and Disk-Center Intensity from 3290 to 12510 \AA}%

\bibitem[{Bray(2014)}]{RFC7159}
Bray, T. 2014, The JavaScript Object Notation (JSON) Data Interchange Format,
RFC 7159, RFC Editor%

\bibitem[{Buehler {et~al.}(2019)Buehler, Esteban~Pozuelo, de~la Cruz~Rodriguez,
\& Scharmer}]{2019ApJ...876...47B}
Buehler, D., Esteban~Pozuelo, S., de~la Cruz~Rodriguez, J., \& Scharmer, G.~B.
2019, \apj, 876, 47%

\bibitem[{Calabretta(2021)}]{wcslib76}
Calabretta, M. 2021, WCSLIB 7.6, Australia Telescope National Facility, CSIRO%

\bibitem[{{Calabretta} \& {Greisen}(2002)}]{2002A&A...395.1077C}
{Calabretta}, M.~R. \& {Greisen}, E.~W. 2002, \aap, 395, 1077%

\bibitem[{Calabretta {et~al.}(2004)Calabretta, Valdes, Greisen, \&
Allen}]{calabretta04representations_distortions}
Calabretta, M.~R., Valdes, F.~G., Greisen, E.~W., \& Allen, S.~L. 2004,
Representations of distortions in FITS world coordinate systems,
\url{https://fits.gsfc.nasa.gov/wcs/dcs_20040422.pdf}, {(draft dated
2004-04-22)}

\bibitem[{{Collados}(2017)}]{2017psio.confE...1C}
{Collados}, M. 2017, in SOLARNET IV: The Physics of the Sun from the Interior
to the Outer Atmosphere, 1%

\bibitem[{{Couvidat} {et~al.}(2016){Couvidat}, {Schou}, {Hoeksema}, {Bogart},
{Bush}, {Duvall}, {Liu}, {Norton}, \& {Scherrer}}]{2016SoPh..291.1887C}
{Couvidat}, S., {Schou}, J., {Hoeksema}, J.~T., {et~al.} 2016, \solphys, 291,
1887

\bibitem[{Criscuoli \& Tritschler(2014)}]{criscouli14IBIS}
Criscuoli, S. \& Tritschler, A. 2014, IBIS Data Reduction Notes, NSO5%

\bibitem[{de~la Cruz~Rodr{\'{\i}}guez {et~al.}(2011)de~la
Cruz~Rodr{\'{\i}}guez, Kiselman, \& M.}]{2011A&A...528A.113D}
de~la Cruz~Rodr{\'{\i}}guez, J., Kiselman, D., \& M., C. 2011, \aap, 528, A113%

\bibitem[{{de la Cruz Rodr{\'{\i}}guez} {et~al.}(2016){de la Cruz
Rodr{\'{\i}}guez}, {Leenaarts}, \& {Asensio Ramos}}]{2016ApJ...830L..30D}
{de la Cruz Rodr{\'{\i}}guez}, J., {Leenaarts}, J., \& {Asensio Ramos}, A.
2016, \apjl, 830, L30%

\bibitem[{de~la Cruz~Rodr{\'i}guez {et~al.}(2015)de~la Cruz~Rodr{\'i}guez,
L{\"o}fdahl, S{\"u}tterlin, Hillberg, \& Rouppe van~der
Voort}]{2015A&A...573A..40D}
de~la Cruz~Rodr{\'i}guez, J., L{\"o}fdahl, M., S{\"u}tterlin, P., Hillberg, T.,
\& Rouppe van~der Voort, L. 2015, \aap, 573, A40%

\bibitem[{{De Pontieu} {et~al.}(2014){De Pontieu}, {Title}, {Lemen}, {Kushner},
{Akin}, {Allard}, {Berger}, {Boerner}, {Cheung}, {Chou}, {Drake}, {Duncan},
{Freeland}, {Heyman}, {Hoffman}, {Hurlburt}, {Lindgren}, {Mathur}, {Rehse},
{Sabolish}, {Seguin}, {Schrijver}, {Tarbell}, {W{\"u}lser}, {Wolfson},
{Yanari}, {Mudge}, {Nguyen-Phuc}, {Timmons}, {van Bezooijen}, {Weingrod},
{Brookner}, {Butcher}, {Dougherty}, {Eder}, {Knagenhjelm}, {Larsen},
{Mansir}, {Phan}, {Boyle}, {Cheimets}, {DeLuca}, {Golub}, {Gates}, {Hertz},
{McKillop}, {Park}, {Perry}, {Podgorski}, {Reeves}, {Saar}, {Testa}, {Tian},
{Weber}, {Dunn}, {Eccles}, {Jaeggli}, {Kankelborg}, {Mashburn}, {Pust},
{Springer}, {Carvalho}, {Kleint}, {Marmie}, {Mazmanian}, {Pereira}, {Sawyer},
{Strong}, {Worden}, {Carlsson}, {Hansteen}, {Leenaarts}, {Wiesmann},
{Aloise}, {Chu}, {Bush}, {Scherrer}, {Brekke}, {Martinez-Sykora}, {Lites},
{McIntosh}, {Uitenbroek}, {Okamoto}, {Gummin}, {Auker}, {Jerram}, {Pool}, \&
{Waltham}}]{2014SoPh..289.2733D}
{De Pontieu}, B., {Title}, A.~M., {Lemen}, J.~R., {et~al.} 2014, \solphys, 289,
2733

\bibitem[{{de Wijn} {et~al.}(2021){de Wijn}, {de la Cruz Rodr{\'\i}guez},
{Scharmer}, {Sliepen}, \& {S{\"u}tterlin}}]{2021AJ....161...89D}
{de Wijn}, A.~G., {de la Cruz Rodr{\'\i}guez}, J., {Scharmer}, G.~B.,
{Sliepen}, G., \& {S{\"u}tterlin}, P. 2021, The Astronomical Journal, 161,
89%

\bibitem[{{Denker} {et~al.}(2018){Denker}, {Kuckein}, {Verma}, {Gonz{\'a}lez
Manrique}, {Diercke}, {Enke}, {Klar}, {Balthasar}, {Louis}, \&
{Dineva}}]{2018ApJS..236....5D}
{Denker}, C., {Kuckein}, C., {Verma}, M., {et~al.} 2018, \apjs, 236, 5

\bibitem[{Díaz~Baso(2018)}]{diaz_baso18analysis}
Díaz~Baso, C.~J. 2018, PhD thesis, Universidad de La Laguna%

\bibitem[{{Esteban Pozuelo} {et~al.}(2019){Esteban Pozuelo}, {de la Cruz
Rodriguez}, {Drews}, {Rouppe van der Voort}, {Scharmer}, \&
{Carlsson}}]{2019ApJ...870...88E}
{Esteban Pozuelo}, S., {de la Cruz Rodriguez}, J., {Drews}, A., {et~al.} 2019,
\apj, 870, 88%

\bibitem[{Fanning(2011)}]{fanning11traditional}
Fanning, D.~W. 2011, {Coyote's Guide to Traditional IDL Graphics} (Coyote Book
Publishing)

\bibitem[{{Freeland} \& {Handy}(1998)}]{1998SoPh..182..497F}
{Freeland}, S.~L. \& {Handy}, B.~N. 1998, \solphys, 182, 497

\bibitem[{Gonsalves(1982)}]{1982OptEn..21..829G}
Gonsalves, R.~A. 1982, Opt. Eng., 21, 829

\bibitem[{G{\'o}rski {et~al.}(2010)G{\'o}rski, Wandelt, Hivon, Hansen, \&
Banday}]{gorski10healpix}
G{\'o}rski, K.~M., Wandelt, B.~D., Hivon, E., Hansen, F.~K., \& Banday, A.~J.
2010, {The HEALPix Primer}, Jet Propulsion Laboratory, v. 2.15a%

\bibitem[{{Greco} {et~al.}(2019){Greco}, {Sordini}, {Cauzzi}, {Reardon}, \&
{Cavallini}}]{2019A&A...626A..43G}
{Greco}, V., {Sordini}, A., {Cauzzi}, G., {Reardon}, K., \& {Cavallini}, F.
2019, \aap, 626, A43

\bibitem[{{Greisen} \& {Calabretta}(2002)}]{2002A&A...395.1061G}
{Greisen}, E.~W. \& {Calabretta}, M.~R. 2002, \aap, 395, 1061%

\bibitem[{{Greisen} {et~al.}(2006){Greisen}, {Calabretta}, {Valdes}, \&
{Allen}}]{2006A&A...446..747G}
{Greisen}, E.~W., {Calabretta}, M.~R., {Valdes}, F.~G., \& {Allen}, S.~L. 2006,
\aap, 446, 747%

\bibitem[{Hack {et~al.}(2012)Hack, Dencheva, Fruchter, \&
Greenfield}]{hack12distortion}
Hack, W., Dencheva, N., Fruchter, A., \& Greenfield, P. 2012, Distortion
Correction in HST FITS Files, Tech. Rep. TSR 2012-01, Space Telescope Science
Institute%

\bibitem[{Hartley \& Zisserman(2000)}]{hartley00multiple}
Hartley, R. \& Zisserman, A. 2000, Multiple View Geometry in Computer Vision,
2nd edn. (Cambridge University Press)

\bibitem[{Haugan \& Fredvik(2015)}]{haugan15metadata}
Haugan, S. V.~H. \& Fredvik, T. 2015, {Document on Standards for Data Archiving
and VO}, Deliverable D20.4, SOLARNET (EC 7th FP grant 312495)

\bibitem[{Haugan \& Fredvik(2020)}]{2020arXiv201112139H}
Haugan, S. V.~H. \& Fredvik, T. 2020 %
[\eprint[arXiv]{2011.12139v1}]

\bibitem[{Heeschen {et~al.}(1973)Heeschen, Howard,
{et~al.}}]{trans_IAU_15_commission40}
Heeschen, D.~S., Howard, W.~E., {et~al.} 1973, Transactions of the IAU, 15A(2),
165

\bibitem[{{Henriques}(2012)}]{2012A&A...548A.114H}
{Henriques}, V.~M.~J. 2012, \aap, 548, A114%

\bibitem[{Jess \& Keys(2017)}]{jess17rosapipeline}
Jess, D. \& Keys, P. 2017, ROSA data reduction pipeline, Queen's University
Belfast Astrophysics Research Centre%

\bibitem[{{Joshi} {et~al.}(2020){Joshi}, {Rouppe van der Voort}, \& {de la Cruz
Rodr{\'\i}guez}}]{2020A&A...641L...5J}
{Joshi}, J., {Rouppe van der Voort}, L. H.~M., \& {de la Cruz Rodr{\'\i}guez},
J. 2020, \aap, 641, L5%

\bibitem[{Kasten \& Young(1989)}]{1989ApOpt..28.4735K}
Kasten, F. \& Young, A.~T. 1989, Appl. Opt., 28, 4735

\bibitem[{{Kianfar} {et~al.}(2020){Kianfar}, {Leenaarts}, {Danilovic}, {de la
Cruz Rodr{\'\i}guez}, \& {Jos{\'e} D{\'\i}az Baso}}]{2020A&A...637A...1K}
{Kianfar}, S., {Leenaarts}, J., {Danilovic}, S., {de la Cruz Rodr{\'\i}guez},
J., \& {Jos{\'e} D{\'\i}az Baso}, C. 2020, \aap, 637, A1%

\bibitem[{{Kosugi} {et~al.}(2007){Kosugi}, {Matsuzaki}, {Sakao}, {Shimizu},
{Sone}, {Tachikawa}, {Hashimoto}, {Minesugi}, {Ohnishi}, {Yamada}, {Tsuneta},
{Hara}, {Ichimoto}, {Suematsu}, {Shimojo}, {Watanabe}, {Shimada}, {Davis},
{Hill}, {Owens}, {Title}, {Culhane}, {Harra}, {Doschek}, \&
{Golub}}]{2007SoPh..243....3K}
{Kosugi}, T., {Matsuzaki}, K., {Sakao}, T., {et~al.} 2007, \solphys, 243, 3

\bibitem[{Ku{\v c}era {et~al.}(2010)Ku{\v c}era, Ambr{\'o}z, G{\"o}m{\"o}ry,
Koz{\'a}k, \& Ryb{\'a}k}]{kucera10CoMP-S}
Ku{\v c}era, A., Ambr{\'o}z, J., G{\"o}m{\"o}ry, P., Koz{\'a}k, M., \&
Ryb{\'a}k, J. 2010, Contrib. Astron. Obs. Skalnat{\'e} Pleso, 40, 135%

\bibitem[{Ku{\v c}era {et~al.}(2015)Ku{\v c}era, Tomczyk, Ryb{\'a}k, Sewell,
G{\"o}m{\"o}ry, Schwartz, Ambroz, \& Koz{\'a}k}]{2015IAUGA..2246687K}
Ku{\v c}era, A., Tomczyk, S., Ryb{\'a}k, J., {et~al.} 2015, in IAU General
Assembly, Vol.~29, 2246687

\bibitem[{{Kuckein} {et~al.}(2017){Kuckein}, {Denker}, {Verma}, {Balthasar},
{Gonz{\'a}lez Manrique}, {Louis}, \& {Diercke}}]{2017IAUS..327...20K}
{Kuckein}, C., {Denker}, C., {Verma}, M., {et~al.} 2017, in Fine Structure and
Dynamics of the Solar Atmosphere, ed. S.~Vargas~Dom{\'i}nguez, A.~G.
Kosovichev, L.~Harra, \& P.~Antolin, Proc. IAUS No. 327%

\bibitem[{{Kuridze} {et~al.}(2019){Kuridze}, {Mathioudakis}, {Morgan},
{Oliver}, {Kleint}, {Zaqarashvili}, {Reid}, {Koza}, {L{\"o}fdahl},
{Hillberg}, {Kukhianidze}, \& {Hanslmeier}}]{2019ApJ...874..126K}
{Kuridze}, D., {Mathioudakis}, M., {Morgan}, H., {et~al.} 2019, \apj, 874, 126%

\bibitem[{{Landsman}(1993)}]{1993ASPC...52..246L}
{Landsman}, W.~B. 1993, in ASP Conf. Ser., Vol.~52, Astronomical Data Analysis
Software and Systems II, ed. R.~J. Hanisch, R.~J.~V. Brissenden, \&
J.~Barnes, 246%

\bibitem[{{Leenaarts} {et~al.}(2018){Leenaarts}, {de la Cruz Rodr{\'{\i}}guez},
{Danilovic}, {Scharmer}, \& {Carlsson}}]{2018A&A...612A..28L}
{Leenaarts}, J., {de la Cruz Rodr{\'{\i}}guez}, J., {Danilovic}, S.,
{Scharmer}, G., \& {Carlsson}, M. 2018, \aap, 612%

\bibitem[{L{\"o}fdahl(2002)}]{2002SPIE.4792..146L}
L{\"o}fdahl, M.~G. 2002, in Proc. SPIE, Vol. 4792, Image Reconstruction from
Incomplete Data II, ed. P.~J. Bones, M.~A. Fiddy, \& R.~P. Millane, 146%

\bibitem[{{L{\"o}fdahl} {et~al.}(2018){L{\"o}fdahl}, {Hillberg}, {de la Cruz
Rodr{\'i}guez}, {Vissers}, {Scharmer}, {Haugan}, \&
{Fredvik}}]{lofdahl_data-processing_v1}
{L{\"o}fdahl}, M.~G., {Hillberg}, T., {de la Cruz Rodr{\'i}guez}, J., {et~al.}
2018, In prep. [\eprint[arXiv]{1804.03030v1}]

\bibitem[{L{\"o}fdahl \& Scharmer(1994)}]{1994A&AS..107..243L}
L{\"o}fdahl, M.~G. \& Scharmer, G.~B. 1994, \aaps, 107, 243

\bibitem[{{L{\"o}hner-B{\"o}ttcher} {et~al.}(2017){L{\"o}hner-B{\"o}ttcher},
{Schmidt}, {Doerr}, {Kentischer}, {Steinmetz}, {Probst}, \&
{Holzwarth}}]{2017A&A...607A..12L}
{L{\"o}hner-B{\"o}ttcher}, J., {Schmidt}, W., {Doerr}, H.-P., {et~al.} 2017,
\aap, 607, A12

\bibitem[{{L{\"o}hner-B{\"o}ttcher} {et~al.}(2018){L{\"o}hner-B{\"o}ttcher},
{Schmidt}, {Stief}, {Steinmetz}, \& {Holzwarth}}]{2018A&A...611A...4L}
{L{\"o}hner-B{\"o}ttcher}, J., {Schmidt}, W., {Stief}, F., {Steinmetz}, T., \&
{Holzwarth}, R. 2018, \aap, 611, A4

\bibitem[{Mampaey {et~al.}(2017)Mampaey, Vansintjan, \&
Delouille}]{2017psio.confE..91V}
Mampaey, B., Vansintjan, R., \& Delouille, V. 2017, in SOLARNET IV: The Physics
of the Sun from the Interior to the Outer Atmosphere, 91%

\bibitem[{{Markwardt}(2009)}]{2009ASPC..411..251M}
{Markwardt}, C.~B. 2009, in ASP Conf. Ser., Vol. 411, Astronomical Data
Analysis Software and Systems XVIII, ed. D.~A. Bohlender, D.~Durand, \&
P.~Dowler, 251

\bibitem[{{Matthews} {et~al.}(2016){Matthews}, {Collados}, {Mathioudakis},
{Erdelyi}, \& {the EST Team}}]{2016SPIE.9908E..09M}
{Matthews}, S.~A., {Collados}, M., {Mathioudakis}, M., {Erdelyi}, R., \& {the
EST Team}. 2016, in Proc. SPIE, Vol. 9908, Ground-based and Airborne
Instrumentation for Astronomy VI, ed. C.~J. {Evans}, L.~{Simard}, \&
H.~{Takami}, 990809

\bibitem[{{Neckel}(1999)}]{1999SoPh..184..421.}
{Neckel}, H. 1999, \solphys, 184, 421

\bibitem[{{Neckel}(2005)}]{2005SoPh..229...13N}
{Neckel}, H. 2005, \solphys, 229, 13

\bibitem[{{Neckel} \& {Labs}(1994)}]{1994SoPh..153...91N}
{Neckel}, H. \& {Labs}, D. 1994, \solphys, 153, 91

\bibitem[{Noll(1976)}]{1976JOSA...66..207N}
Noll, R.~J. 1976, J. Opt. Soc. Am., 66, 207

\bibitem[{Paxman {et~al.}(1992{\natexlab{a}})Paxman, Schulz, \&
Fienup}]{1992JOSAA...9.1072P}
Paxman, R.~G., Schulz, T.~J., \& Fienup, J.~R. 1992{\natexlab{a}}, J. Opt. Soc.
Am. A, 9, 1072

\bibitem[{Paxman {et~al.}(1992{\natexlab{b}})Paxman, Schulz, \&
Fienup}]{paxman92phase}
Paxman, R.~G., Schulz, T.~J., \& Fienup, J.~R. 1992{\natexlab{b}}, in Technical
Digest Series, Vol.~11, Signal Recovery and Synthesis {IV}, Optical Society
of America, 5%

\bibitem[{Paxman {et~al.}(1996)Paxman, Seldin, L{\"o}fdahl, Scharmer, \&
Keller}]{1996ApJ...466.1087P}
Paxman, R.~G., Seldin, J.~H., L{\"o}fdahl, M.~G., Scharmer, G.~B., \& Keller,
C.~U. 1996, \apj, 466, 1087

\bibitem[{P\'ebay {et~al.}(2016)P\'ebay, Terriberry, Kolla, \&
Bennett}]{pebay16numerically}
P\'ebay, P., Terriberry, T.~B., Kolla, H., \& Bennett, J. 2016, Computational
Statistics, 31, 1305

\bibitem[{{Pence} {et~al.}(2010){Pence}, {Chiappetti}, {Page}, {Shaw}, \&
{Stobie}}]{2010A&A...524A..42P}
{Pence}, W.~D., {Chiappetti}, L., {Page}, C.~G., {Shaw}, R.~A., \& {Stobie}, E.
2010, \aap, 524, A42

\bibitem[{{Pietrow} {et~al.}(2020){Pietrow}, {Kiselman}, {de la Cruz
Rodr{\'\i}guez}, {D{\'\i}az Baso}, {Pastor Yabar}, \&
{Yadav}}]{2020A&A...644A..43P}
{Pietrow}, A.~G.~M., {Kiselman}, D., {de la Cruz Rodr{\'\i}guez}, J., {et~al.}
2020, \aap, 644, A43%

\bibitem[{{Rimmele} {et~al.}(2020){Rimmele}, {Warner}, {Keil}, {Goode},
{Kn{\"o}lker}, {Kuhn}, {Rosner}, {McMullin}, {Casini}, {Lin}, {W{\"o}ger},
{von der L{\"u}he}, {Tritschler}, {Davey}, {de Wijn}, {Elmore}, {Fehlmann},
{Harrington}, {Jaeggli}, {Rast}, {Schad}, {Schmidt}, {Mathioudakis},
{Mickey}, {Anan}, {Beck}, {Marshall}, {Jeffers}, {Oschmann}, {Beard},
{Berst}, {Cowan}, {Craig}, {Cross}, {Cummings}, {Donnelly}, {de Vanssay},
{Eigenbrot}, {Ferayorni}, {Foster}, {Galapon}, {Gedrites}, {Gonzales},
{Goodrich}, {Gregory}, {Guzman}, {Guzzo}, {Hegwer}, {Hubbard}, {Hubbard},
{Johansson}, {Johnson}, {Liang}, {Liang}, {McQuillen}, {Mayer}, {Newman},
{Onodera}, {Phelps}, {Puentes}, {Richards}, {Rimmele}, {Sekulic}, {Shimko},
{Simison}, {Smith}, {Starman}, {Sueoka}, {Summers}, {Szabo}, {Szabo},
{Wampler}, {Williams}, \& {White}}]{2020SoPh..295..172R}
{Rimmele}, T.~R., {Warner}, M., {Keil}, S.~L., {et~al.} 2020, \solphys, 295,
172

\bibitem[{Roddier(1990)}]{1990OptEn..29.1174R}
Roddier, N. 1990, Opt. Eng., 29, 1174

\bibitem[{{Rots} {et~al.}(2015){Rots}, {Bunclark}, {Calabretta}, {Allen},
{Manchester}, \& {Thompson}}]{2015A&A...574A..36R}
{Rots}, A.~H., {Bunclark}, P.~S., {Calabretta}, M.~R., {et~al.} 2015, \aap,
574, A36

\bibitem[{{Rouppe van der Voort} {et~al.}(2017){Rouppe van der Voort}, {De
Pontieu}, {Scharmer}, {de la Cruz Rodriguez}, {Martinez-Sykora},
{Nobrega-Siverio}, {Guo}, {Jafarzadeh}, {Pereira}, {Hansteen}, {Carlsson}, \&
{Vissers}}]{2017ApJ...851L...6R}
{Rouppe van der Voort}, L., {De Pontieu}, B., {Scharmer}, G.~B., {et~al.} 2017,
\apjl, 851, L6%

\bibitem[{{Rouppe van der Voort} {et~al.}(2020){Rouppe van der Voort}, {De
Pontieu}, {Carlsson}, {de la Cruz Rodr{\'\i}guez}, {Bose}, {Chintzoglou},
{Drews}, {Froment}, {Go{\v{s}}i{\'c}}, {Graham}, {Hansteen}, {Henriques},
{Jafarzadeh}, {Joshi}, {Kleint}, {Kohutova}, {Leifsen},
{Mart{\'\i}nez-Sykora}, {N{\'o}brega-Siverio}, {Ortiz}, {Pereira}, {Popovas},
{Quintero Noda}, {Sainz Dalda}, {Scharmer}, {Schmit}, {Scullion}, {Skogsrud},
{Szydlarski}, {Timmons}, {Vissers}, {Woods}, \&
{Zacharias}}]{2020A&A...641A.146R}
{Rouppe van der Voort}, L.~H.~M., {De Pontieu}, B., {Carlsson}, M., {et~al.}
2020, \aap, 641, A146%

\bibitem[{{Rouppe van der Voort} {et~al.}(2021){Rouppe van der Voort}, {Joshi},
{Henriques}, \& {Bose}}]{2021A&A...648A..54R}
{Rouppe van der Voort}, L. H.~M., {Joshi}, J., {Henriques}, V. M.~J., \&
{Bose}, S. 2021, \aap, 648, A54%

\bibitem[{{Sanchez Almeida} \& {Lites}(1992)}]{1992ApJ...398..359S}
{Sanchez Almeida}, J. \& {Lites}, B.~W. 1992, \apj, 398, 359

\bibitem[{Scharmer(2017)}]{2017psio.confE..85S}
Scharmer, G. 2017, in SOLARNET IV: The Physics of the Sun from the Interior to
the Outer Atmosphere, 85

\bibitem[{Scharmer(2006)}]{2006A&A...447.1111S}
Scharmer, G.~B. 2006, \aap, 447, 1111

\bibitem[{Scharmer {et~al.}(2003)Scharmer, Bjelksj{\"o}, Korhonen, Lindberg, \&
Pettersson}]{2003SPIE.4853..341S}
Scharmer, G.~B., Bjelksj{\"o}, K., Korhonen, T.~K., Lindberg, B., \&
Pettersson, B. 2003, in Proc. SPIE, Vol. 4853, Innovative Telescopes and
Instrumentation for Solar Astrophysics, ed. S.~Keil \& S.~Avakyan, 341%

\bibitem[{Scharmer {et~al.}(2019)Scharmer, L{\"o}fdahl, Sliepen, \& de~la
Cruz~Rodr{\'i}guez}]{2019A&A...626A..55S}
Scharmer, G.~B., L{\"o}fdahl, M.~G., Sliepen, G., \& de~la Cruz~Rodr{\'i}guez,
J. 2019, \aap, 626, A55%

\bibitem[{Scharmer {et~al.}(2008)Scharmer, Narayan, Hillberg, de~la
Cruz~Rodriguez, S{\"u}tterlin, L{\"o}fdahl, van Noort, Kiselman, \&
Lagg}]{2008ApJ...689L..69S}
Scharmer, G.~B., Narayan, G., Hillberg, T., {et~al.} 2008, \apjl, 689, L69

\bibitem[{Schnerr {et~al.}(2011)Schnerr, de~la Cruz~Rodr{\'{\i}}guez, \& van
Noort}]{2011A&A...534A..45S}
Schnerr, R., de~la Cruz~Rodr{\'{\i}}guez, J., \& van Noort, M. 2011, \aap, 534,
A45%

\bibitem[{Selbing(2005)}]{2010arXiv1010.4142S}
Selbing, J. 2005, Master's thesis, Stockholm University%

\bibitem[{Seldin \& Paxman(1994)}]{1994SPIE.2302..268S}
Seldin, J.~H. \& Paxman, R.~G. 1994, in Proc. SPIE, Vol. 2302, Image
reconstruction and restoration, ed. T.~J. Schultz \& D.~L. Snyder, 268%

\bibitem[{Sliepen \& S{\"u}tterlin(2013)}]{sliepen13primary}
Sliepen, G. \& S{\"u}tterlin, P. 2013, in Synergies Between Ground and Space
Based Solar Research, 1st SOLARNET -- 3rd EAST/ATST meeting

\bibitem[{Smith(2003)}]{smith03ninterpolate}
Smith, J.~D. 2003, Ninterpolate function,
\url{https://tir.astro.utoledo.edu/idl/ninterpolate.pro}, see also
``Multidimensional Interpolation'' thread from 2003 in
\url{news:comp.lang.idl-pvwave}.

\bibitem[{{Socas-Navarro} {et~al.}(2015){Socas-Navarro}, {de la Cruz
Rodr{\'i}guez}, {Asensio Ramos}, {Trujillo Bueno}, \& {Ruiz
Cobo}}]{2015A&A...577A...7S}
{Socas-Navarro}, H., {de la Cruz Rodr{\'i}guez}, J., {Asensio Ramos}, A.,
{Trujillo Bueno}, J., \& {Ruiz Cobo}, B. 2015, \aap, 577, A7%

\bibitem[{{SPICE Consortium} {et~al.}(2020){SPICE Consortium}, {Anderson},
{Appourchaux}, {Auch{\`e}re}, {Aznar Cuadrado}, {Barbay}, {Baudin},
{Beardsley}, {Bocchialini}, {Borgo}, {Bruzzi}, {Buchlin}, {Burton},
{B{\"u}chel}, {Caldwell}, {Caminade}, {Carlsson}, {Curdt}, {Davenne},
{Davila}, {Deforest}, {Del Zanna}, {Drummond}, {Dubau}, {Dumesnil}, {Dunn},
{Eccleston}, {Fludra}, {Fredvik}, {Gabriel}, {Giunta}, {Gottwald}, {Griffin},
{Grundy}, {Guest}, {Gyo}, {Haberreiter}, {Hansteen}, {Harrison}, {Hassler},
{Haugan}, {Howe}, {Janvier}, {Klein}, {Koller}, {Kucera}, {Kouliche},
{Marsch}, {Marshall}, {Marshall}, {Matthews}, {McQuirk}, {Meining},
{Mercier}, {Morris}, {Morse}, {Munro}, {Parenti}, {Pastor-Santos}, {Peter},
{Pfiffner}, {Phelan}, {Philippon}, {Richards}, {Rogers}, {Sawyer},
{Schlatter}, {Schmutz}, {Sch{\"u}hle}, {Shaughnessy}, {Sidher}, {Solanki},
{Speight}, {Spescha}, {Szwec}, {Tamiatto}, {Teriaca}, {Thompson}, {Tosh},
{Tustain}, {Vial}, {Walls}, {Waltham}, {Wimmer-Schweingruber}, {Woodward},
{Young}, {de Groof}, {Pacros}, {Williams}, \&
{M{\"u}ller}}]{2020A&A...642A..14S}
{SPICE Consortium}, {Anderson}, M., {Appourchaux}, T., {et~al.} 2020, \aap,
642, A14

\bibitem[{{Szydlarski}(2019)}]{2019asrc.confE.126S}
{Szydlarski}, M. 2019, in ALMA2019: Science Results and Cross-Facility
Synergies, 126

\bibitem[{Thompson(2010{\natexlab{a}})}]{thompson10solarsoft}
Thompson, W. 2010{\natexlab{a}}, The SolarSoft WCS Routines: A Tutorial, Adnet
Systems, Inc., NASA Goddard Space Flight Center%

\bibitem[{{Thompson}(2006)}]{2006A&A...449..791T}
{Thompson}, W.~T. 2006, \aap, 449, 791

\bibitem[{Thompson(2010{\natexlab{b}})}]{2010A&A...515A..59T}
Thompson, W.~T. 2010{\natexlab{b}}, \aap, 515, A59

\bibitem[{van Noort {et~al.}(2005)van Noort, Rouppe van~der Voort, \&
L{\"o}fdahl}]{2005SoPh..228..191V}
van Noort, M., Rouppe van~der Voort, L., \& L{\"o}fdahl, M.~G. 2005, \solphys,
228, 191

\bibitem[{{van Noort} \& {Rouppe van der Voort}(2008)}]{2008A&A...489..429V}
{van Noort}, M.~J. \& {Rouppe van der Voort}, L.~H.~M. 2008, \aap, 489, 429

\bibitem[{{Vissers} \& {Rouppe van der Voort}(2012)}]{2012ApJ...750...22V}
{Vissers}, G. \& {Rouppe van der Voort}, L. 2012, \apj, 750, 22%

\bibitem[{{Vissers} {et~al.}(2019){Vissers}, {de la Cruz Rodriguez},
{Libbrecht}, {Rouppe van der Voort}, {Scharmer}, \&
{Carlsson}}]{2019A&A...627A.101V}
{Vissers}, G.~J.~M., {de la Cruz Rodriguez}, J., {Libbrecht}, T., {et~al.}
2019, \aap, 627, A101%

\bibitem[{{Wilkinson} {et~al.}(2016){Wilkinson}, {Dumontier}, {Aalbersberg},
{Appleton}, {Axton}, {Baak}, {Blomberg}, {Boiten}, {da Silva Santos},
{Bourne}, {Bouwman}, {Brookes}, {Clark}, {Crosas}, {Dillo}, {Dumon},
{Edmunds}, {Evelo}, {Finkers}, {Gonzalez-Beltran}, {Gray}, {Groth}, {Goble},
{Grethe}, {Heringa}, {'T Hoen}, {Hooft}, {Kuhn}, {Kok}, {Kok}, {Lusher},
{Martone}, {Mons}, {Packer}, {Persson}, {Rocca-Serra}, {Roos}, {van Schaik},
{Sansone}, {Schultes}, {Sengstag}, {Slater}, {Strawn}, {Swertz}, {Thompson},
{van der Lei}, {van Mulligen}, {Velterop}, {Waagmeester}, {Wittenburg},
{Wolstencroft}, {Zhao}, \& {Mons}}]{2016NatSD...360018W}
{Wilkinson}, M.~D., {Dumontier}, M., {Aalbersberg}, I.~J., {et~al.} 2016,
Scientific Data, 3, 160018

\end{thebibliography}
\end{document}